\documentclass[hidelinks,12pt]{article}
\usepackage{amsmath}

\usepackage[margin=1in]{geometry}

\usepackage{amssymb}

\usepackage{tipa}

\usepackage{epsfig}
\usepackage{subcaption}
\usepackage{float}

\usepackage{natbib}

\usepackage{etoc}

\bibpunct{(}{)}{;}{a}{,}{,}
\usepackage{setspace}

\newcommand{\blind}{1}

\usepackage{mathabx}

\usepackage{amsthm}

\theoremstyle{definition}
\newtheorem{definition}{Definition}[section]

\theoremstyle{remark}

\DeclareMathOperator*{\argmin}{arg\,min}

\newcommand{\tr}[1]{\operatorname{tr}\left(#1\right)}

{\end{eqnarray}\end{subequations}\hskip-4.0pt}

\def\fatnorm#1{|\kern-.2ex|\kern-.2ex| #1 |\kern-.2ex|\kern-.2ex|}

\newcommand{\V}{\mathcal{V}}

\newcommand{\N}{{\mathcal N}}

\newcommand{\cov}{\textsf{Cov}}

\def\conv{\mathop{\text{\rm conv}\kern.2ex}}

\newcommand{\silent}[1]{}

\def\qed{\hskip1pt $\;\;\scriptstyle\Box$}
\def\Ber{\mathop{\text{Bernoulli}\kern.2ex}}
\def\supp{\mathop{\text{supp}\kern.2ex}}
\def\corr{\mathop{\text{corr}\kern.2ex}}
\def\prec{\mathop{\text{precision}\kern.2ex}}
\def\recall{\mathop{\text{recall}\kern.2ex}}
\def\cov{\mathop{\text{Cov}\kern.2ex}}
\def\mnorm{\mathcal{N}_{f,m}\kern.2ex}
\def\var{\mathop{\text{Var}\kern.2ex}}
\def\ess{\mathop{\text{ess}\kern.2ex}}
\def\dom{\mathop{\text{dom}\kern.2ex}}
\def\lin{\mathop{\text{lin}\kern.2ex}}

\def\supp{\mathop{\text{\rm supp}\kern.2ex}}
\def\argmin{\mathop{\text{arg\,min}\kern.2ex}}

\newcommand{\beq}{\begin{equation}}
\newcommand{\eeq}{\end{equation}}
\newcommand{\ben}{\begin{eqnarray}}
\newcommand{\een}{\end{eqnarray}}
\newcommand{\bnum}{\begin{enumerate}}
\newcommand{\enum}{\end{enumerate}}
\newcommand{\bit}{\begin{itemize}}
\newcommand{\eit}{\end{itemize}}
\newcommand{\bens}{\begin{eqnarray*}}
\newcommand{\eens}{\end{eqnarray*}}

\def\qed{\hskip1pt $\;\;\scriptstyle\Box$}

\begin{document}

\if1\blind
{
  \title{Tensor models for linguistics pitch curve data of native speakers of Afrikaans
\footnote{
}}
  \author{Michael Hornstein\textsuperscript{1}, Shuheng Zhou\textsuperscript{2}, Kerby Shedden\textsuperscript{1}, 
\hspace{.2cm} \\
 \textsuperscript{1}Department of Statistics, University of Michigan, MI \\
 \textsuperscript{2}Department of Statistics, University of California, Riverside, CA}
  \maketitle
} \fi

\if0\blind
{
  \bigskip
  \bigskip
  \bigskip
  \begin{center}
    {\LARGE\bf Tensor models for linguistics pitch curve data of native speakers of Afrikaans}
\end{center}
  \medskip
} \fi

\def\spacingset#1{\renewcommand{\baselinestretch}%
{#1}\small\normalsize} \spacingset{1}

\bigskip
\begin{abstract} \normalsize
We use tensor analysis techniques for high-dimensional data to gain insight into pitch curves, which play an important role in linguistics research.  In particular, we propose that demeaned phonetics pitch curve data can be modeled as having a Kronecker product inverse covariance structure with sparse factors corresponding to words and time.  Using data from a study of native Afrikaans speakers, we show that by targeting conditional independence through a graphical model, we reveal relationships associated with natural properties of words as studied by linguists.  We find that words with long vowels cluster based on whether the vowel is pronounced at the front or back of the mouth, and words with short vowels have strong edges associated with the initial consonant.  
\end{abstract}

\noindent%
\vfill

\newpage
\spacingset{1.5} 

\section{Introduction}

Pitch curve data in phonetics is used to address a wide variety of research questions, including studying the cognitive processes of sound perception and production \citep{chetouani2009time, shami2007evaluation, sheikhan2013modular}, population variation in speech patterns, and software generation of speech \citep{hirst2011analysis, houde1998adaptation, iriondo2009automatic}.  To obtain pitch curve data, human speech is recorded, and pitch (in Hz) is extracted at a dense sequence of time points.  Linguists are interested in the relationship between pitch and other attributes of spoken words, which gives insight into how meaning is conveyed through acoustical properties of speech.  Extensive research has documented associations between pitch curves and acoustical characteristics of words such as voicing \citep{hanson2009effects}.  Such pitch curve analysis has traditionally employed parametric models such as mixed effects models with random effects for speaker and word \citep{baayen2008mixed}.  By contrast, we use a more flexible matrix-variate model that focuses on word-word and time-time covariances.  By analyzing data from \citet{coetzee2018plosive}, we show that word-word correlations are associated with word attributes of interest to linguists.  



\begin{figure}[h]
 \centering 
\begin{subfigure}{0.55\textwidth}
\includegraphics[width=\textwidth]{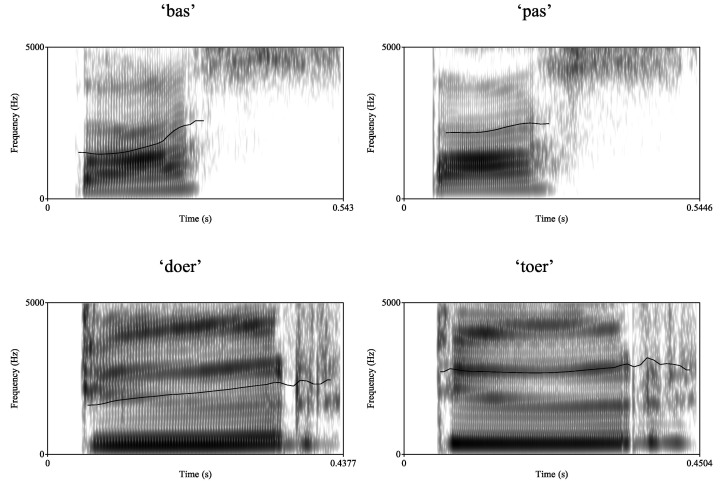} 
\caption{Spectrograms from Coetzee et. al.\ 2017.}
\label{longShortVowels_Coetzee}
\end{subfigure}
\begin{subfigure}{0.4\textwidth}
\includegraphics[width=\textwidth]{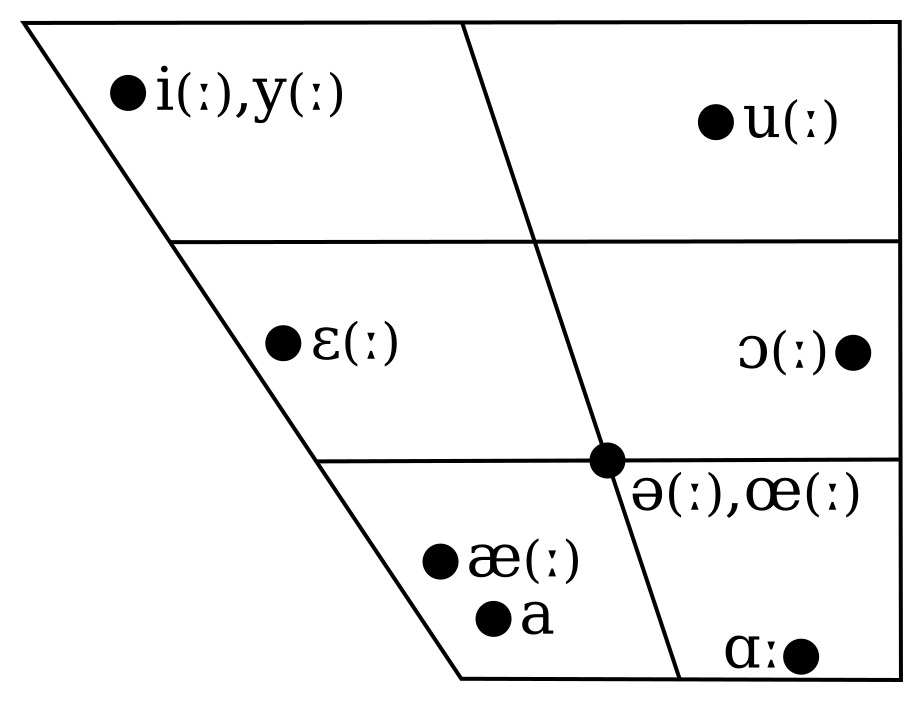}
\caption{Vowel formant diagram from \citet{wissing2012integrasie}.}
\label{AfrikaansVowelFormantDiagram}
\end{subfigure} 
\caption{Phonetics data in Afrikaans.}
\label{fig:image2}
\end{figure}

\begin{figure}[H]
 \centering 
 \begin{subfigure}{\textwidth}
\includegraphics[width=0.8\textwidth]{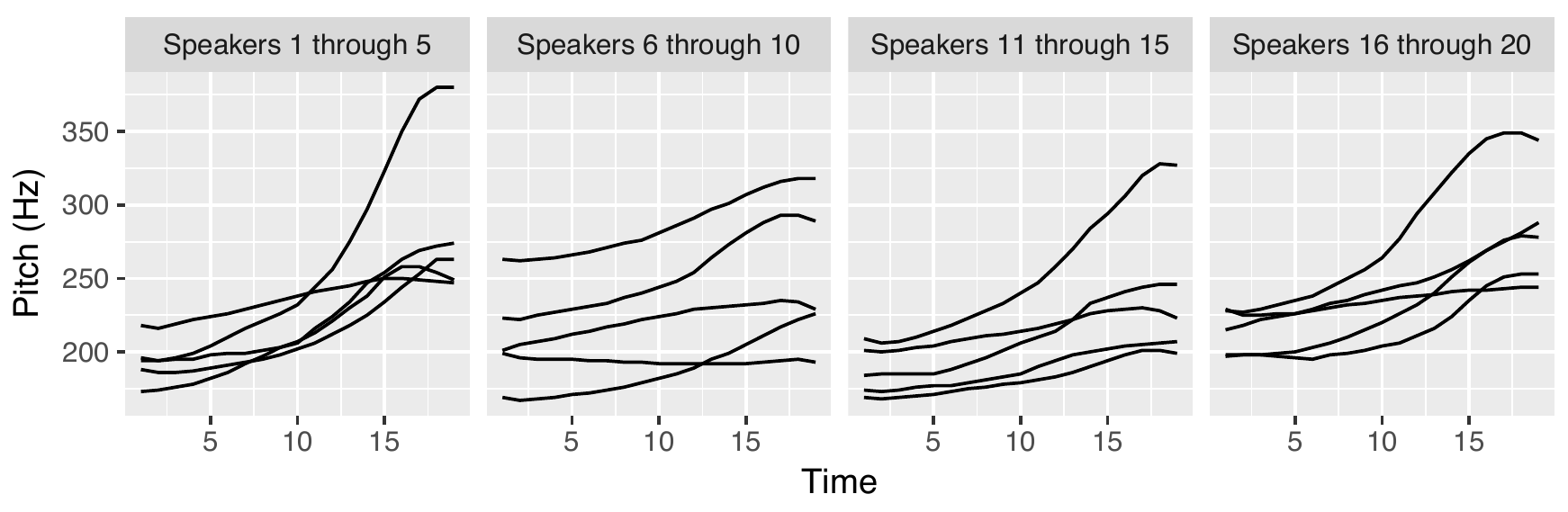}
\caption{Pitch curves for the word ``met.''  For readability, the pitch curves are displayed in four panels. }
\label{firstUtteranceAllPeople_met_noNames}
\end{subfigure}
\begin{subfigure}{\textwidth}
\includegraphics[width=0.8\textwidth]{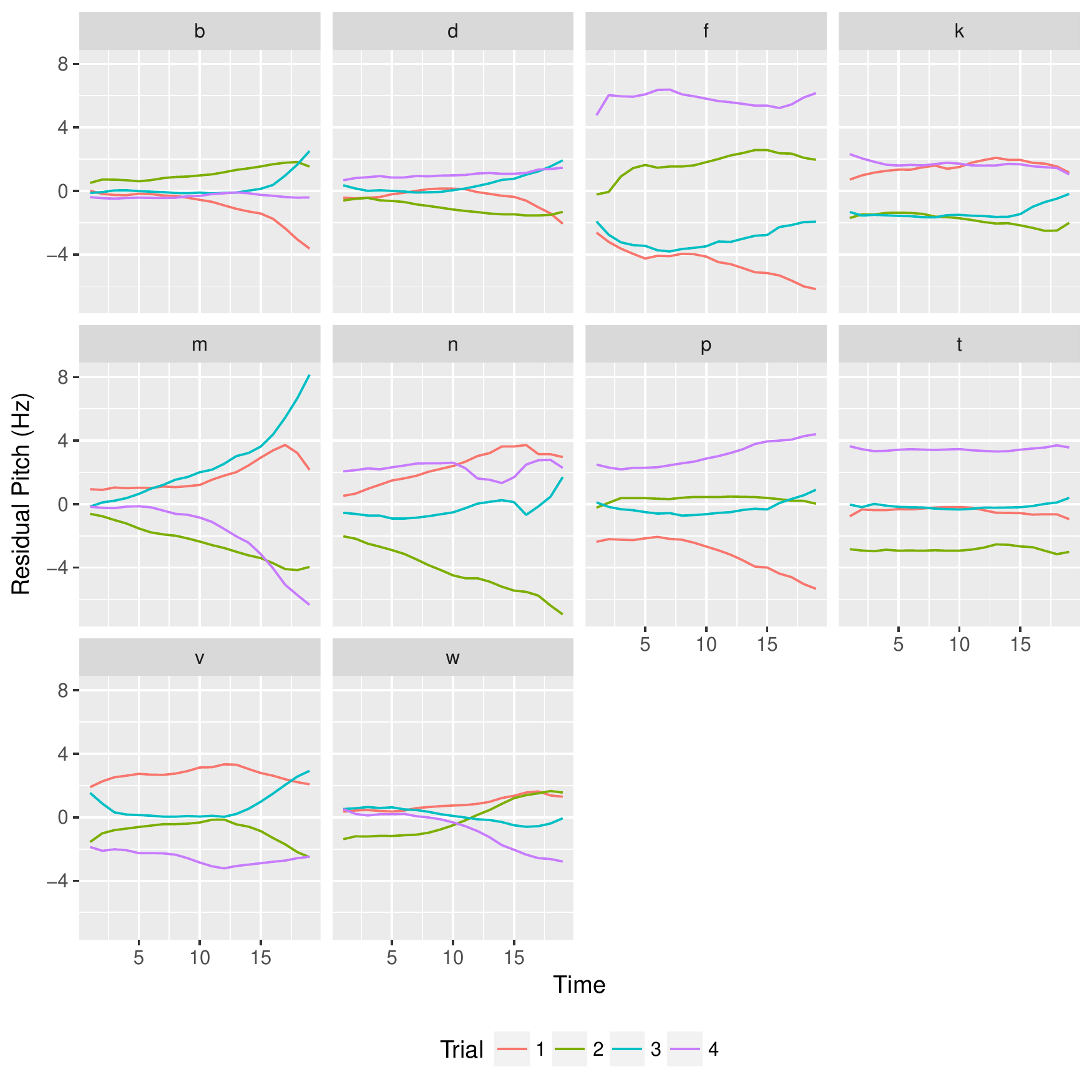}
\caption{Trial residualized pitch curves for each of the four trials, averaged over speaker and word for each initial consonant.}
\label{trialResid_personAvg_allInitialConsonants}
\end{subfigure} 
\caption{Phonetics pitch curve data in Afrikaans.}
\label{fig:image2}
\end{figure}

Here we propose a Kronecker product model and de-meaning procedure for pitch curve data, yielding correlation and precision information along two axes--words and time points.  Using estimated precision matrices and associated graphical models, we show that word-word correlations are driven by natural properties of words studied in phonetics, including onset (initial consonant), coda (final consonant), and mouth position.  In particular, we find words with long vowels naturally cluster according to the region of the mouth in which the vowel is pronounced (front of mouth vs. back of mouth).  We furthermore show that time-time covariance matrices are similar across subsets of words, which justifies pooling words to estimate a common time-time covariance matrix.  


To estimate covariance matrices, inverse covariance matrices, and graphical models, we use statistical methods with rigorous theoretical guarantees \citep{Zhou14a}.  In particular, we use Gemini and nodewise regression with theoretically guided penalty parameters.  The data contains trial replicates, with four utterances of each word by each speaker; we make use of trial residualization to center the data prior to estimating the covariance matrices.  Due to stable inter-individual differences in speech patterns, unadjusted covariance matrices are highly ill-conditioned.  We therefore developed a non-parametric demeaning procedure such that the residuals were suitable for graphical analysis.  We anticipate that this approach will be useful in other types of data with trial replicates, including linguistics and neuroscience data. 

	
Figure \ref{longShortVowels_Coetzee} displays the spectrogram of four words in Afrikaans, in which the darkness represents amplitude as a function of frequency and time.  To provide a basic illustration of pitch curve data, in Figure \ref{firstUtteranceAllPeople_met_noNames} we display the first utterance of the word ``met'' for each of the speakers.  For ease of visualization, the pitch curves for the speakers are displayed in four panels.  These are individual raw pitch curves that have not been centered or averaged.  The pitch curves are obtained by segmenting the word to extract the vowel, then extracting pitch measurements at $19$ equally spaced time points within the vowel, using \emph{Praat} ~\citep{boersma1993accurate}.  There is substantial heterogeneity among speakers pronouncing a given word.   As discussed below, we center the data using trial residualization, subtracting from each individual pitch curve the corresponding point-wise mean pitch curve over each subject $\times$ word, taken over the four trials.  To illustrate, in Figure \ref{trialResid_personAvg_allInitialConsonants}, we display four trials, centered by first removing the speaker $\times$ word mean, and then averaging these residuals over speakers and words within initial consonant (with a separate panel shown for each initial consonant).  The trials are centered as discussed below in equation \eqref{trialCentering}.

The raw data we consider here \citep{coetzee2018plosive} can be represented as a four-index array.  Specifically, let $X(i, j, r, t)$ denote the pitch measurement for speaker $i$, word $j$, trial $r$, and time $t$.  Let $n(s)$, $n(w)$, $n(r)$, and $n(t)$ denote the respective dimension along each axis.  We describe a matrix-variate model that captures word-word and time-time correlations, treating the trials as replicates nested within speakers by words.  Let $X(i, r) \in \mathbb{R}^{n(w) \times n(t)}$ denote the data for speaker $i$, trial $r$.  For the mean structure, we assume that for each speaker $i$, a common mean matrix $M(i) \in \mathbb{R}^{n(w) \times n(t)}$ is shared across the four trials.   Denote the sample mean estimate of $M(i)$ as 
\begin{equation} 
\overline{X}(i) = \frac{1}{n(r)} \sum_{r = 1}^{n(r)} X(i, r), 
\end{equation} 
with $M(i) = \mathbb{E} \left[ \overline{X}(i) \right]$.  For the covariance structure, we specify a Kronecker product model.  Hence the mean and covariance structure can be expressed, respectively, as 
\begin{equation} \label{trialCentering}
\mathbb{E}\left[ X(i, r) - \overline{X}(i) \right] = 0 \qquad \text{ and } \qquad  \text{Cov}(\text{vec}(X(i, r))) = A \otimes B,
\end{equation} 
where $A$ is a time-time covariance matrix and $B$ is a word-word covariance matrix.   

 We use estimation procedures with known convergence properties to recover $A$ and $B$ from the data, then use the corresponding estimated graph structures to probe word-word and time-time associations.    
 
The following definitions, taken verbatim from \citet{Zhou14a}, concerns Gaussian graphical models for random vectors and matrices, respectively.  
%
\begin{definition}
\label{def::ggm}
Let $V = (V_1, \ldots,V_n)^T$ be a random Gaussian vector,
which we represent by an undirected graph $G = (\mathcal{V}, F)$. The
vertex set
$\mathcal{V} := \{1, \ldots, n\}$ has one vertex for each component
of the vector $V$. The edge set
$F$ consists of pairs $(j, k)$ that are joined by an edge. If $V_j$ is
independent of $V_k$ given the other variables, then $(j, k) \notin F$.
\end{definition}
\begin{definition}
\label{def::matrixVariateGraphical}
Let $\V=\{1, \ldots, n\}$ be an index set which enumerates
rows of $X$ according to a fixed order.
For all $i=1, \ldots, m$, we assign to each variable of a column vector
$x^i$ exactly one element of the set $\V$ by a rule of
correspondence $g:x^i \to\V$ such that $g(x^i_j) = j, j=1,\ldots, n$.
The graphs $G_i(\V, F)$ constructed
for each random column vector $x^i, i=1, \ldots, m$ according to
Definition~\ref{def::ggm}
will share an identical edge set $F$,
because the normalized column vectors ${x^{1}}/{\sqrt{a_{11}}},
\ldots,
x^{m}/\sqrt{a_{mm}}$ follow the same multivariate normal distribution
$\N_n(0, B)$, where $a_{jj}$ denotes the $j$th diagonal entry of $A$ as defined in \eqref{trialCentering}.
Hence, graphs $G_1, \ldots, G_m$ are isomorphic and we write $G_i
\simeq G_j, \forall i, j$.
Due to the isomorphism,
we use $G(\V, F)$ to represent the family of graphs $G_1, \ldots, G_m$.
Hence, a pair $(\ell, k)$ which is absent in $F$ encodes
conditional independence between the ${\ell}${th} row and the
$k$th row given all other rows.
Similarly, let $\Gamma=\{1, \ldots, m\}$ be the index set
which enumerates columns of $X$ according to a fixed order.
We use $H(\Gamma, E)$ to represent the family of graphs $H_1, \ldots,
H_f$, where $H_i$ is constructed for row vector $y^i$, and
$H_i \simeq H_j, \forall i, j$.
Now $H(\Gamma, E)$ is a graph with adjacency matrix $\Upsilon(H) =
\Upsilon(A^{-1})$ as edges in $E$ encode nonzeros in $A^{-1}$, and $G(\V, F)$ is a graph with adjacency matrix
$\Upsilon(G) = \Upsilon(B^{-1})$.
\end{definition}

The bigraphical lasso \citep{KLLZ13} models the graph as Cartesian product of $G(\V, F)$ and $H(\Gamma, E)$.

\section{Methods} \label{sec:PitchCurveMethods}

We define the sample covariance matrices based on trial residualized data.  We then define the Gemini estimators and nodewise regression.  

\subsection{Covariance and precision matrices for time points and words}


In analyzing data from \citet{coetzee2018plosive}, we treat the $20$ subjects and four trials in the Afrikaans study as $20 \times 4 = 80$ independent random arrays with a common covariance structure.  Each such array is a $n(w) \times n(t)$  matrix which has been centered over the trials as discussed above in \eqref{trialCentering}.  We then apply the Glasso method with theoretically guided penalty parameters to the word and time Gram matrices.  This approach gives us graph structures among the words and among the time points.  

We now define the sample covariance matrices based on trial residualization; the corresponding sample correlation matrices are used as inputs to Gemini and nodewise regression.  Let $S_A$ denote the word-word sample covariance matrix,
\begin{equation} \label{sampleCovWordTrialResid}
	S_A = \frac{1}{n(t) n(s) n(r)} \sum_{i = 1}^{n(s)} \sum_{r = 1}^{n(r)}  \left[ X(i, r) - \overline{X}(i) \right] \left[ X(i, r) - \overline{X}(i) \right]^T 
\end{equation}
and let $S_B$ denote the time-time sample covariance matrix, 
\begin{equation} \label{sampleCovTimeTrialResid}
	S_B = \frac{1}{n(w) n(s) n(r)} \sum_{i = 1}^{n(s)} \sum_{r = 1}^{n(r)}  \left[ X(i, r) - \overline{X}(i) \right]^T \left[ X(i, r) - \overline{X}(i) \right].
\end{equation}
Note that in this formulation, speakers and trials are taken as replicates, so each Gram matrix is an average of $n(s) \cdot n(r)$ Gram matrices.  



\noindent{\bf Gemini estimators.}  Let the sample correlation matrices corresponding to \eqref{sampleCovWordTrialResid} and \eqref{sampleCovTimeTrialResid} be defined as
\begin{equation}  \label{defGammaAGammaB}
	\widehat{\Gamma}_{ij}(A) = \frac{(S_A)_{ij} }{ \sqrt{(S_A)_{ii}(S_A)_{jj}}} \quad \text{ and } \quad \widehat{\Gamma}_{ij}(B) = \frac{(S_B)_{ij} }{ \sqrt{(S_B)_{ii}(S_B)_{jj}}}.
\end{equation} 
The Gemini estimators apply graphical lasso with adjusted penalty to account for correlation along the other axis.  The Gemini inverse correlation estimators ~\citep{Zhou14a} are defined as follows,
using a pair of penalized estimators for the correlation matrices $\rho(A) = (a_{ij}/\sqrt{a_{ii} a_{jj}})$
and $\rho(B) = (b_{ij}/\sqrt{b_{ii} b_{jj}})$,
\begin{align} \label{geminiObjectiveFnA}
\widehat{A}_\rho &= \argmin_{A_\rho \succ 0} \left\{ \tr{\widehat{\Gamma}(A) A_\rho^{-1}}
+ \log |A_\rho| + \lambda_B |A_\rho^{-1}|_{1, \text{off}}\right\} \quad \text{ and } \\
\label{geminiObjectiveFnB}
\widehat{B}_\rho &= 
\argmin_{B_\rho \succ 0} \left\{ \tr{\widehat{\Gamma}(B) B_\rho^{-1}} 
  + \log |B_\rho| + \lambda_A |B_\rho^{-1}|_{1, \text{off}} \right\}, 
\end{align}
where the inputs are a pair of sample correlation matrices as defined in \eqref{defGammaAGammaB}.  The theoretically guided penalty parameters used in \eqref{geminiObjectiveFnA} and \eqref{geminiObjectiveFnB} are 
\begin{equation} \label{glassoWordTimeTuningParam}
\lambda_A= \sqrt{\frac{\log(n(w))}{n(s) \cdot n(r) \cdot n(w)}} \qquad \text{and} \qquad  \lambda_B = \sqrt{\frac{\log(n(w))}{n(s) \cdot n(r) \cdot n_{\text{eff}}(t) }},
\end{equation} 
where $n(w)$ is the number of words, $n(s)$ is the number of speakers, $n(r)$ is the number of replicates, and $n_{\text{eff}}(t)$ is the number of ``effective'' time points.  Note that the effective time points per utterance is smaller than $19$, because the pitch curves are smooth curves, so adjacent points are dependent.  Due to the stretched time scale over short vowels versus the long vowels, we believe that $n_{\text{eff}}(t)$ for short vowels is smaller than that for long ones; hence we recommend using larger penalty when we interpret the graphs over short vowels.  Figures \ref{OlympicVowelsWithLong_Glasso_shortVowel_pen0p36_thresh0} through \ref{OlympicVowelsWithLong_Glasso_shortVowel_pen0p52_thresh0} in the Appendix illustrate that edges persist at higher penalty values for words with short vowels than for words with long vowels.


\subsection{Nodewise regression} \label{sec:NodewiseRegression}

In addition to using Glasso, we also estimate edges using nodewise regression.  \citet{MB06} proposed variable selection via nodewise regression, in which each variable is regressed on each other variable via $\ell_1$ penalized regression.  The edges correspond to the nonzero entries of the regression coefficients (i.e. an edge exists between vertices $i$ and $j$ if either the regression coefficient of variable $i$ on $j$ is nonzero, or the regression coefficient of variable $j$ on $i$ is nonzero).  \citet{MB06} proved variable selection consistency of nodewise regression.  

We define Pearson correlation.  For random variables $X, Y \in \mathbb{R}$, the population Pearson correlation is defined as $\rho(X, Y) = \text{Cov}(X, Y) / \sqrt{\text{Var}(X) \text{Var}(Y)}$.  For example, the Pearson correlation between ``bate'' and ``maak'' is $0.50$,  with trial residualized pitch curves displayed in Figure \ref{pitchCurves_trialResid_bate_maak} in the appendix.  Let $\widehat{\Gamma}$ denote a sample Pearson correlation matrix as defined in \eqref{defGammaAGammaB}.

 
We now explain nodewise regression in more detail.  Let $\widehat{\Gamma}^{(i)} \in \mathbb{R}^{(m - 1) \times (m - 1)}$ denote the submatrix of $\widehat{\Gamma}$ obtained by excluding the $i$th column and $i$th row.  Let $\widehat{\gamma}^{(i)}$ denote the $i$th column of $\widehat{\Gamma}$ excluding the diagonal entry.  The regression coefficient for the $i$th variable is obtained by solving the $\ell_1$ penalized least squares problem, 
\begin{equation} \label{nodewiseLasso}
\widehat{\beta}^i = \text{arg min}_{\beta: \beta \in \mathbb{R}^{m - 1}} \left \{ \frac{1}{2} \beta^T \widehat{\Gamma}^{(i)} \beta - \langle \widehat{\gamma}^{(i)}, \beta \rangle + \lambda \lVert \beta \rVert_1 \right \}.
\end{equation} 
Afterwards, the inverse correlation matrix is reconstructed by first obtaining a matrix $\widetilde{\Theta}$, 
\begin{equation} 
\widetilde{\Theta}_{-j, -j} = -(\widehat{\Gamma}_{jj} - \widehat{\Gamma}_{j, -j} \widehat{\beta}^j)^{-1} \widehat{\beta}^j, \quad \text{and} \quad \widetilde{\Theta}_{jj} = (\widehat{\Gamma}_{jj} - \widehat{\Gamma}_{j, -j} \widehat{\beta}^j)^{-1},
\end{equation} 
then projecting $\widetilde{\Theta}$ onto the space of symmetric matrices.  

Using nodewise regression with a refit to obtain an estimate of the inverse covariance matrix was proposed by \citet{Yuan10, LW12}.  In \citet{zhou2011high}, they combine a multiple regression approach with ideas of thresholding and refitting: first they infer a sparse undirected graphical model structure via thresholding of each among many $\ell_1$-norm penalized regression functions of \eqref{nodewiseLasso}.  They show that under suitable conditions, this approach yields consistent estimation in terms of graphical structure and fast convergence rates with respect to the operator and Frobenius norm for the covariance matrix and its inverse. In our data analysis, using the same penalty parameters for nodewise regression as defined in  \eqref{glassoWordTimeTuningParam} follows the theoretical guidance of \citet{zhou2011high} and yields similar edge structures and sparsity levels to Glasso.  

Our nodewise regression with thresholding procedure follows from  ideas of \citet{zhou2011high} and \citet{Zhou10}.  Since our input matrix is positive semidefinite, the methods of \citet{LW12}, \citet{Yuan10}, and \citet{ZRXB11} would all work to obtain $\Theta$.

\section{Data Analysis}


\subsection{Long vowel conditional dependence is driven by front/back of mouth pronunciation}

We show that for words with long vowels, a natural clustering exists in the edges of the graphical model according to whether the vowel is pronounced toward the front of the mouth or the back of the mouth.  In Figure \ref{AfrikaansVowelFormantDiagram}, vowels on the left are pronounced in the front of the mouth (\textipa{i}, \textepsilon), whereas vowels on the right are pronounced in the back of the mouth (\textscripta, \textopeno, \textipa{u}).  Figure \ref{OlympicVowelsWithLong_Glasso_longVowels_cluster_pen0p30_thresh0} in the Appendix displays the estimated inverse covariance graph for words with long vowels, using Glasso with a penalty of $0.3$.  Based on the estimated effective sample size ($n(r) = 3$, $n_{\text{eff}}(t)=3$ or $4$, $n(s) = 20$) and theoretical guidance from \citet{Zhou14a},
we believe the theoretical penalty should be in the range of $[0.11, 0.13]$, so a penalty of $0.3$ displays the strongest edges.  Based on the edges, the words naturally cluster according to whether the vowels are spoken at the front or back of the mouth.  The sum of the absolute values of within-cluster Glasso-estimated edge weights is $0.76$, whereas the sum of between-cluster edge weights is $0.18$ (i.e. the total edge weight of edges that cross the graph cut is substantially larger). 	   

\begin{figure}[h!]
\includegraphics[width=\linewidth, keepaspectratio]{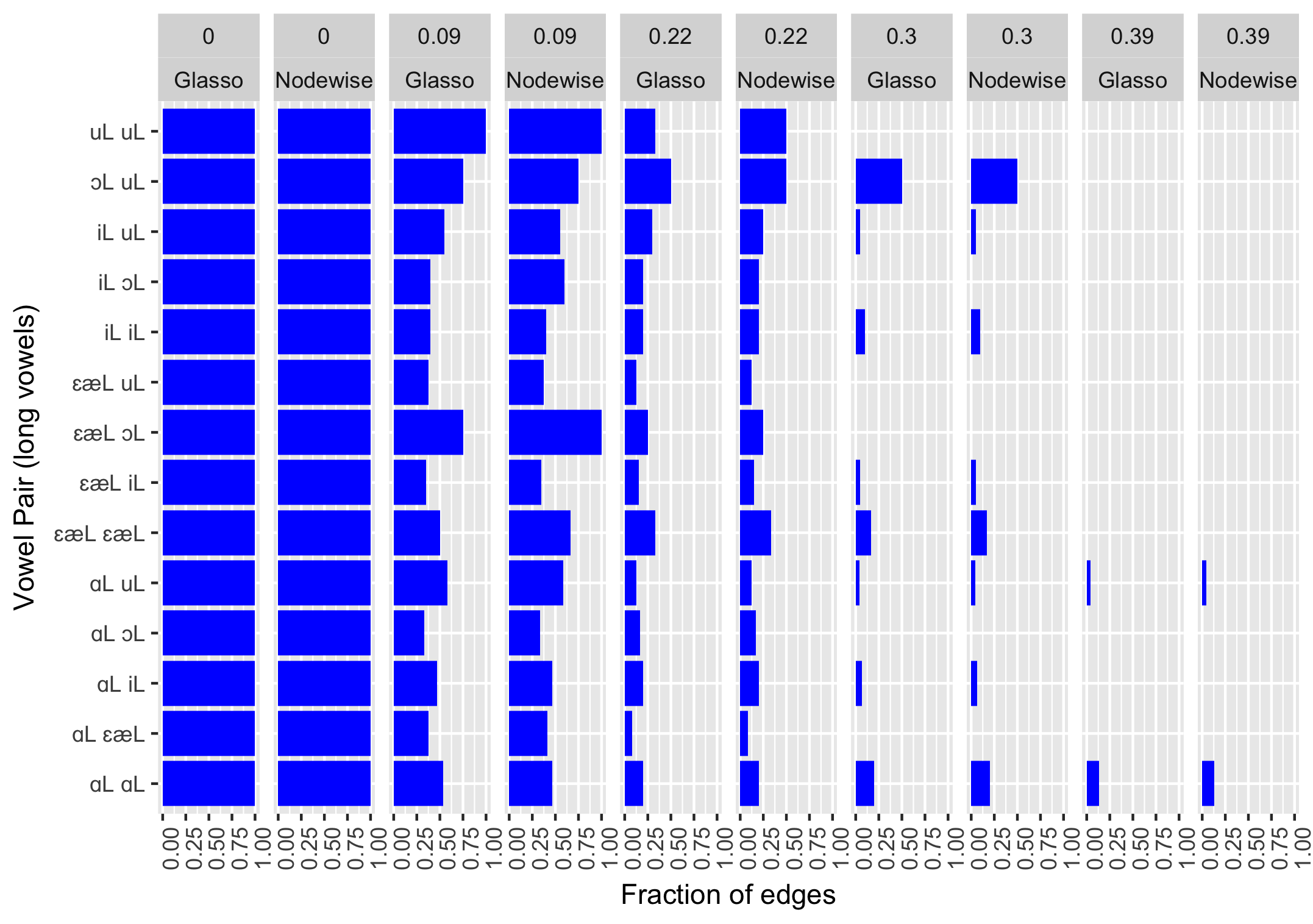} \caption{Bar chart of fraction of edges for long vowels, estimated using Glasso and nodewise regression.}  \label{hist_vowelWithLongMerged_longVowels_glasso_nodewise_fracedges}
\end{figure}

Figure \ref{hist_vowelWithLongMerged_longVowels_glasso_nodewise_fracedges} displays a bar chart of the fraction of edges present among each pair of long vowels.  The edges are estimated using a sequence of penalty parameters for Glasso and nodewise regression.  Note that when the penalty is zero, the Glasso estimate reduces to the inverse sample correlation, which is a fully dense matrix, so the fraction of edges is equal to one.  As the penalty increases, the fraction of edges decreases more rapidly for some vowel pairs than for others.  For word pairs that have larger Pearson correlation but smaller penalized inverse correlation, the words are marginally correlated, but not conditionally correlated given the other words; that is, the relationship between those words is explained by other words.  As seen in Figure \ref{hist_vowelWithLongMerged_longVowels_glasso_nodewise_fracedges}, the long vowel pairs \textscripta-\textscripta \,and \textscripta-u, both of which are spoken at the front of the mouth, persist to a penalty of $0.4$.  By contrast, the between-cluster edges drop out first as the penalty increases.  For example, the \textepsilon/\ae-\textopeno vowel pairs have many edges at smaller penalty parameters, but no edges at a penalty of $0.3$.  Figure \ref{OlympicVowelsWithLong_Nodewise_longThenShortVowels_pen_thresh_pen0p16_thresh0p08} displays the estimated inverse covariance graph for words with long vowels, using nodewise regression with a penalty of $0.3$.  

For each pair of long vowels, Figure \ref{hist_vowelWithLongMerged_longVowels_glasso_nodewise_absSampleCorr} displays the average absolute values of the Pearson correlation entries among edges.  Note that the edges are obtained via the precision matrix, but the average is taken using entries of the sample correlation matrix.  For example, let $E(A, A)$ denote the set of edges between words with a long ``\textscripta'' vowel, and let $|E(A, A)|$ denote the number of edges between words with long ``\textscripta'' vowels.  Then we calculate
\begin{equation} 
\frac{1}{|E(A, A)|} \sum_{(i, j) \in E(A, A)} |S_{ij}|.
\end{equation} 
Note that as the penalty increases, the number of edges decreases, so the average Pearson correlation is taken over the stronger edges that remain, resulting in a larger value.  At the highest penalty shown, three edges remain: bate-maak, maak-kaas, and bate-toer.  Pearson correlations between word pairs with strong edges are shown in Table \ref{wordWordPearson}. 

Figure \ref{pitchCurves_trialResid_maak_kaas} displays the trial residual pitch curves for maak and kaas.  For multiple speakers, the variability increases towards the end of the word, flaring out over time.  The Pearson correlation between two words is high if corresponding utterances within speakers predominantly have the same sign (e.g. if the first utterance of maak is positive for the same time points as the first utterance of kaas, the second utterance of maak is positive for the same time points as the second utterance of kaas, etc., and if this pattern holds across speakers).  Analgously, Figure \ref{pitchCurves_trialResid_bate_maak} shows the trial residual pitch curves for bate and maak.  Figure \ref{pitchCurves_trialResid_bate_toer} shows the trial residual pitch curves for bate and toer.


\begin{figure}[h!]
\includegraphics[width=\linewidth, keepaspectratio]{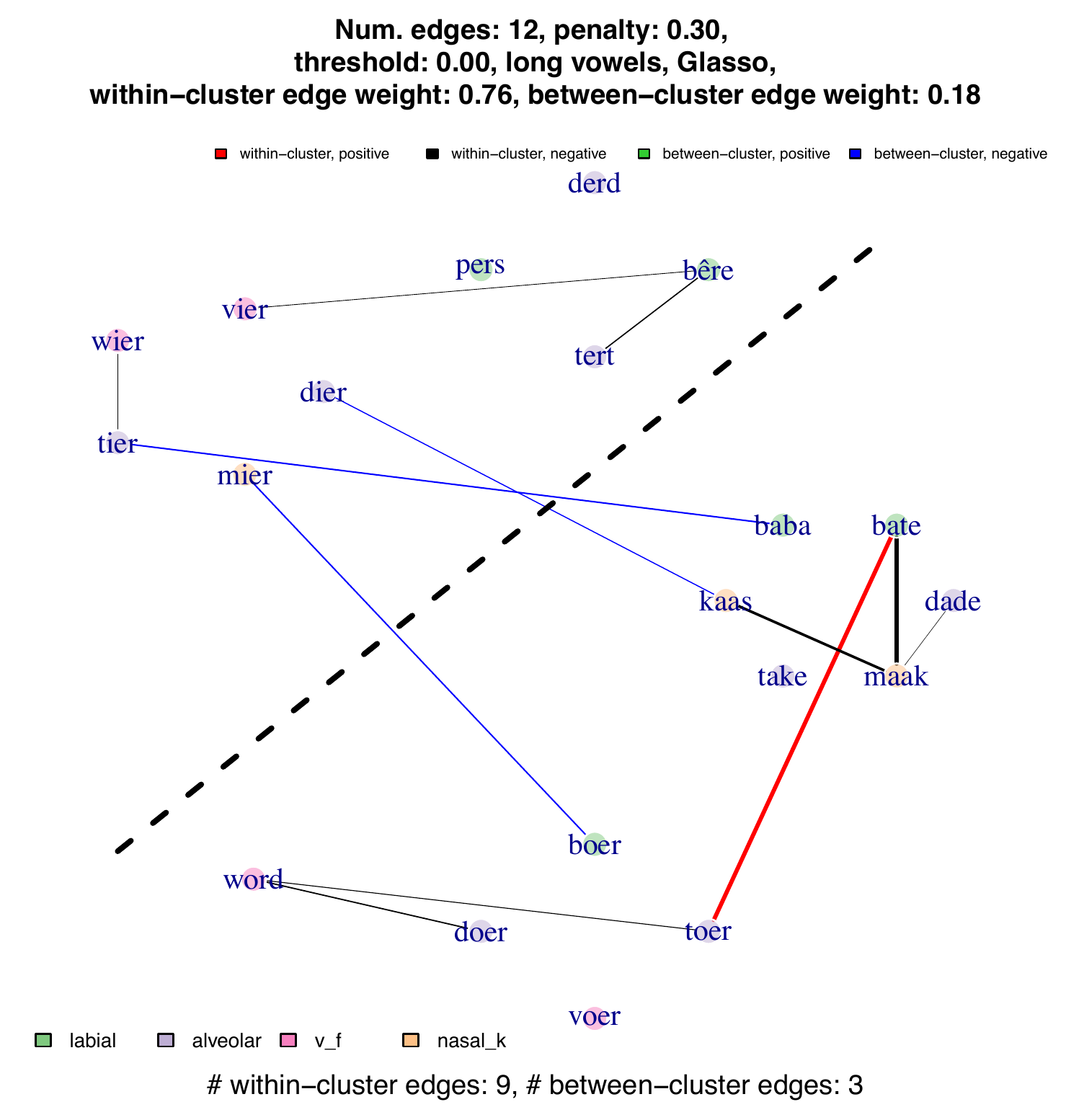} 
\caption{Inverse correlation edge graph for words with long vowels.  The words are organized by vowel, with each circle of words sharing a common vowel.  The words cluster based on whether the vowel is pronounced in the front of the mouth (upper cluster) or back of the mouth (lower cluster); see Figure \ref{AfrikaansVowelFormantDiagram}.}  \label{OlympicVowelsWithLong_Glasso_longVowels_cluster_pen0p30_thresh0}
\end{figure}

\subsection{Short vowel conditional dependence is associated with initial and final consonants}
Among words with short vowels, edges are associated with word attributes including onset (initial consonant) and coda (consonant after the vowel).  Among edges that persist to a penalty of $0.52$ are words beginning with n and w (both continuously voiced), m and w (also continuously voiced), and n and t (both alveolar).  Figure \ref{barChartShortVowelOnsetCoda} (and Section \ref{sec:ShortVowel} in the appendix) displays bar charts of the average sample correlation corresponding to edges between words with short vowels for each pair of onsets, first codas, and last codas.   

  \begin{figure} 
  \begin{subfigure}{0.8\textwidth}
  \includegraphics[width=\textwidth]{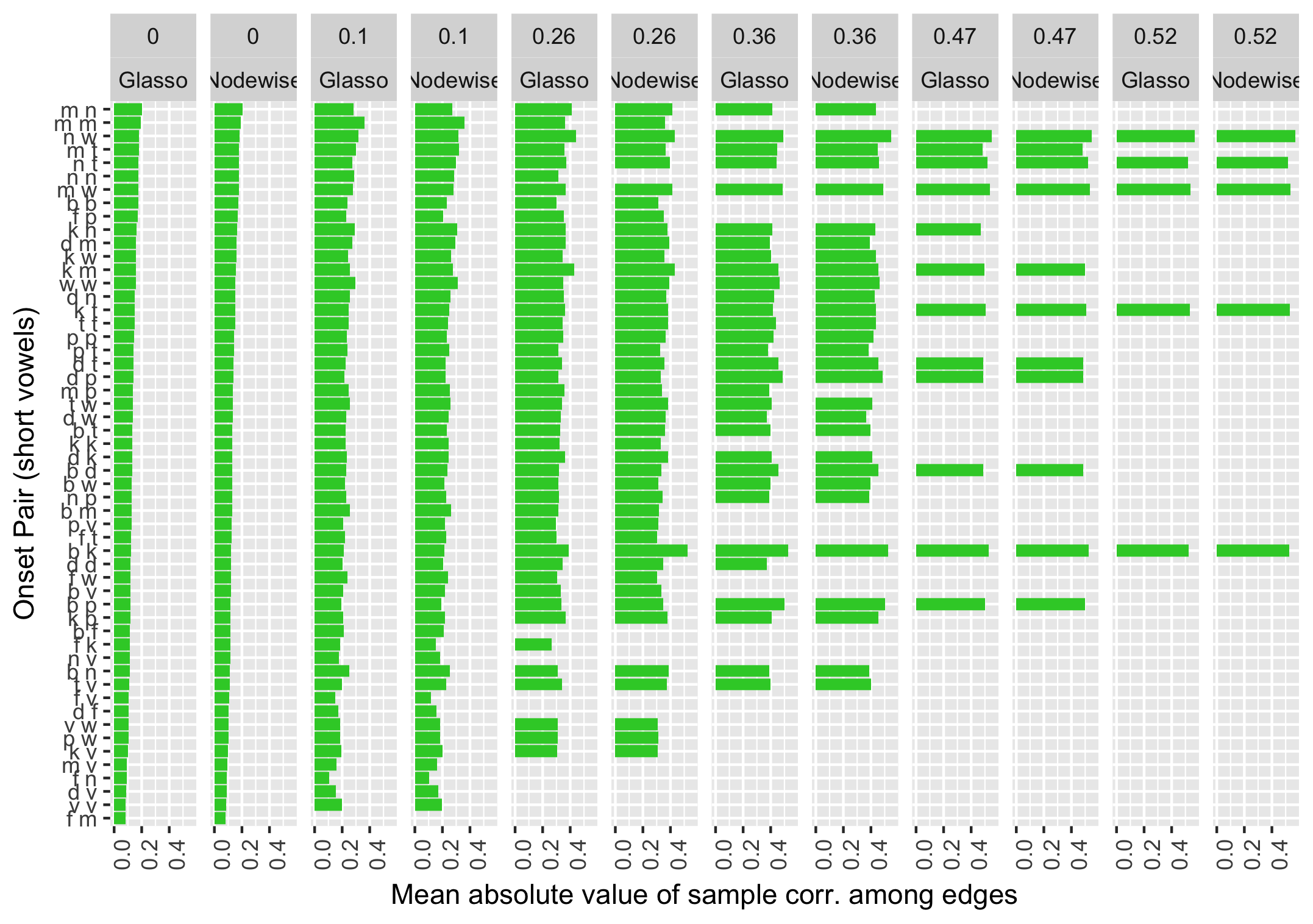}
  \caption{Bar chart of average sample correlation between words with edges for each pair of onsets, estimated using Glasso and nodewise regression.}
        \end{subfigure} 
  \begin{subfigure}{0.8\textwidth}
  \includegraphics[width=\textwidth]{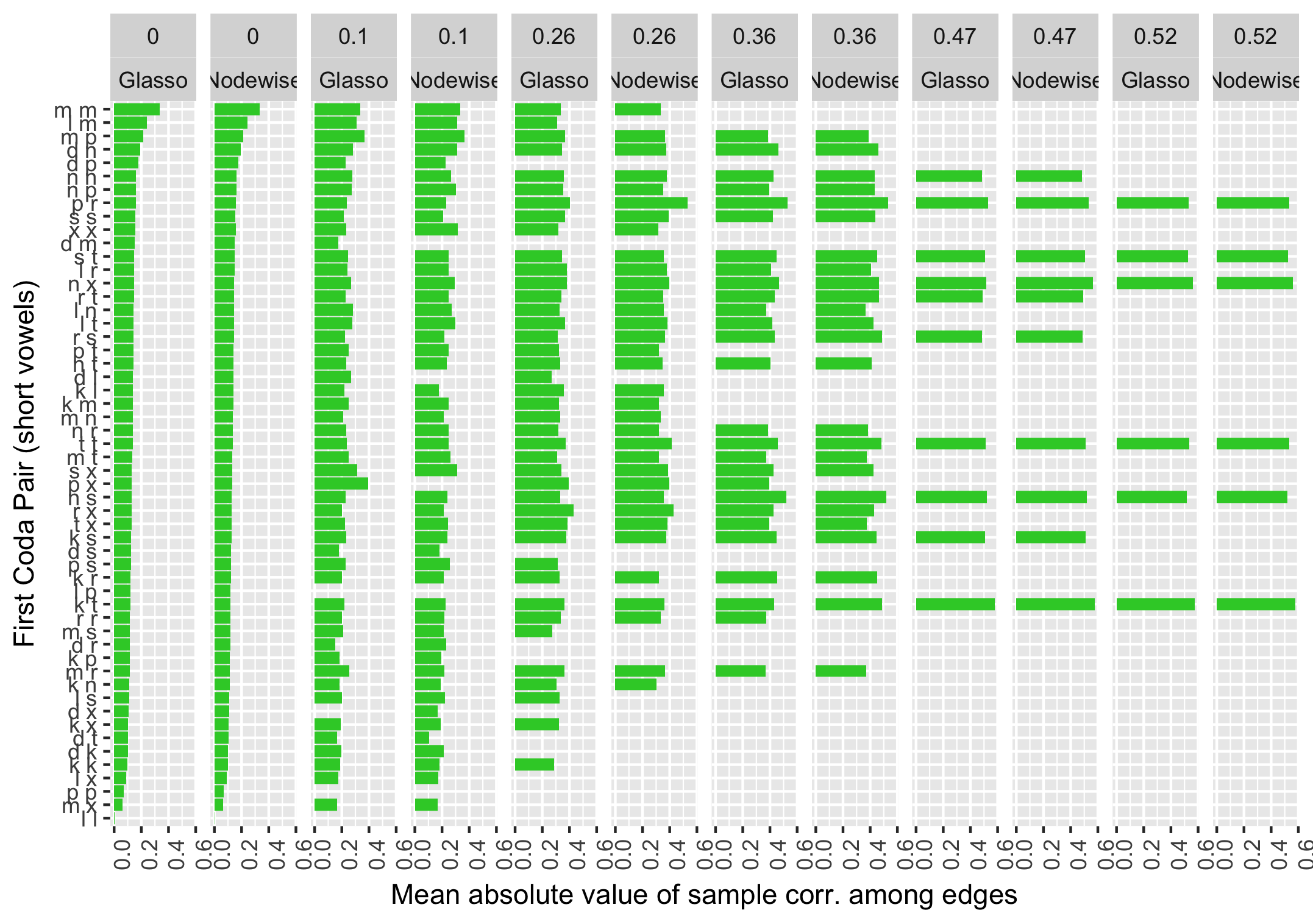}
  \caption{Bar chart of average sample correlation between words with edges for each pair of first codas, estimated using Glasso and nodewise regression.}
        \end{subfigure} 
         \caption{Bar charts for short vowels.} \label{barChartShortVowelOnsetCoda}
  \end{figure}

  
  Figure \ref{hist_vowelWithLongMerged_shortVowels_glasso_nodewise_fracedges} in the Appendix displays a bar chart of the fraction of edges between each pair of short vowels.    By contrast to Figure \ref{hist_vowelWithLongMerged_longVowels_glasso_nodewise_fracedges}, the edges appear to be uniformly distributed among vowel pairs.  Figure \ref{hist_vowelWithLongMerged_shortVowels_glasso_nodewise_absSampleCorr} displays a bar chart for short vowels analogous to Figure \ref{hist_vowelWithLongMerged_longVowels_glasso_nodewise_absSampleCorr}.  Hence for words with short vowels,  edges appear to be driven more by the initial consonant than the vowel.     

\subsection{Conditional independence reveals initial consonants associate with vowel pronunciation}

\begin{figure}[h!]
\includegraphics[width=\linewidth, keepaspectratio]{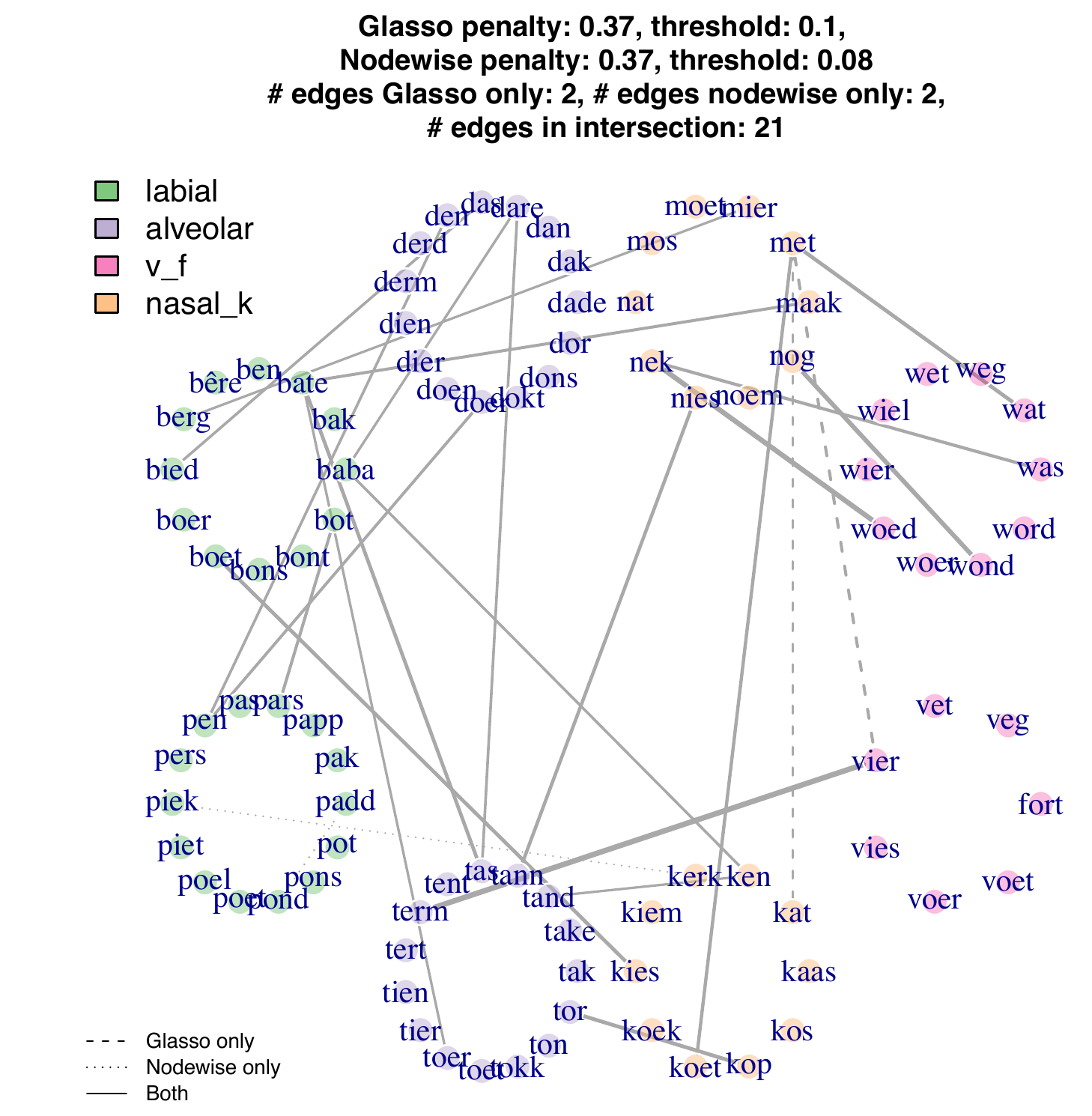} \caption{Inverse covariance graph of all words, comparing Glasso edges with nodewise regression edges.}  \label{OlympicInitCons_compare_Glasso_pen0p37_thresh0p1_Nodewise_pen0p37_thresh_0p08}
\end{figure}

\begin{figure}[h!]
\includegraphics[width=\linewidth, keepaspectratio]{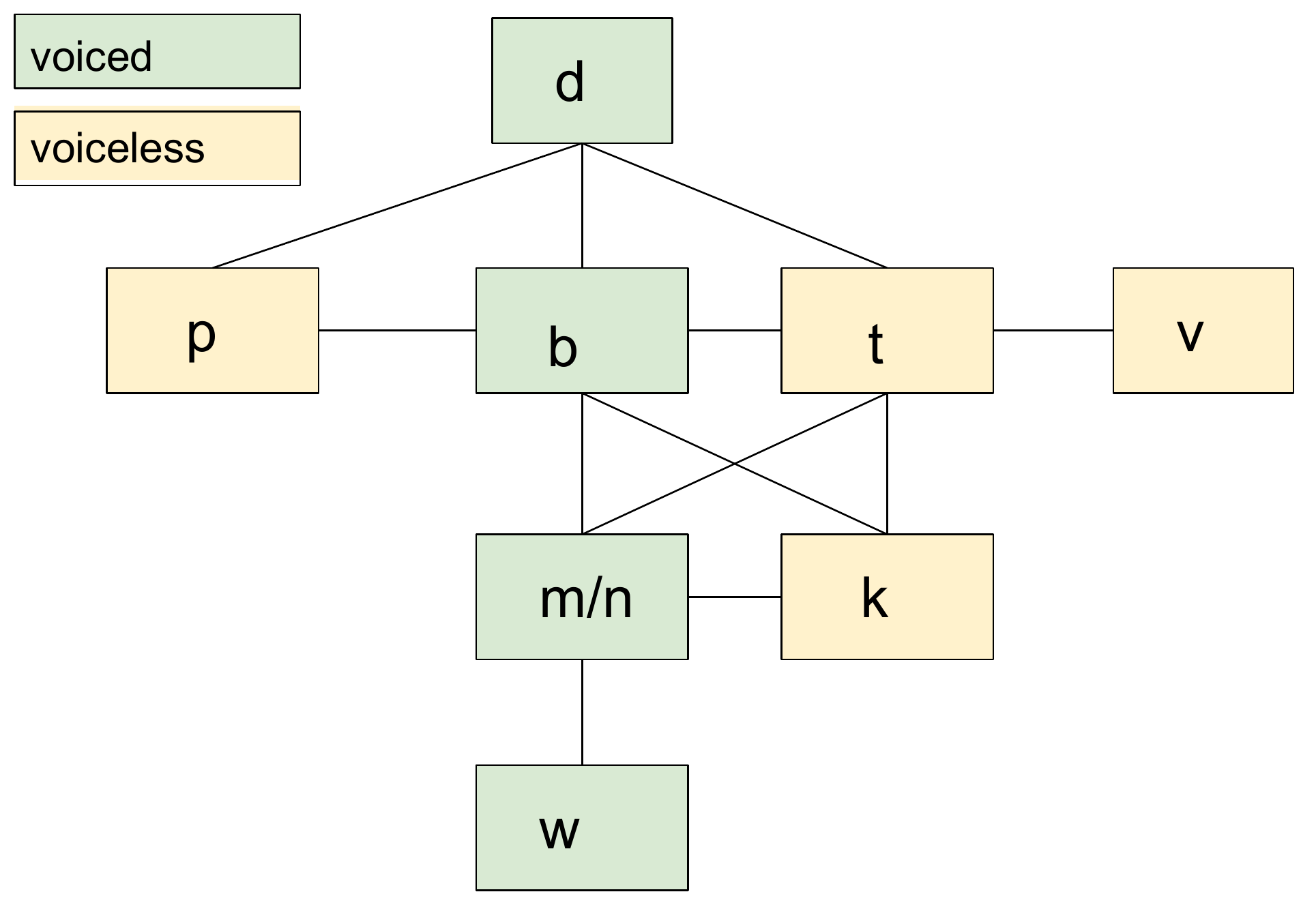} \caption{Diagram displaying connectivity among consonants, providing a higher-level representation of Figure \ref{OlympicInitCons_compare_Glasso_pen0p37_thresh0p1_Nodewise_pen0p37_thresh_0p08} by combining nodes within a consonant type into ``supernodes.''  Two nodes are connected in this diagram if there is an edge estimated by both Glasso and nodewise regression between words with the corresponding initial consonants in Figure \ref{OlympicInitCons_compare_Glasso_pen0p37_thresh0p1_Nodewise_pen0p37_thresh_0p08}.}  \label{consonantConnectivityDiagram_v}
\end{figure}

In Figure \ref{OlympicInitCons_compare_Glasso_pen0p37_thresh0p1_Nodewise_pen0p37_thresh_0p08}, we display the inverse correlation graph for all words, organized by initial consonant.  The nodes are colored by category of the initial consonant: labial (pronounced with the lips), alveolar (pronounced with the tongue on the ridge behind the teeth), nasal (pronounced using the sinus cavity), and v/f (fricative, pronounced by partially obstructing the air).  The Glasso penalty is $0.37$, followed by a threshold of $0.1$, and the nodewise regression penalty is $0.37$, followed by a threshold of $0.08$.  The words are organized by initial consonant.  Almost all of the edges are between group rather than within group; that is, almost all edges are between words starting with different consonants.  In Figure \ref{consonantConnectivityDiagram_v}, we present a high-level summary of this edge graph, by aggregating words with the same consonant into ``supernodes.''  Two supernodes are connected if there is an edge in Figure \ref{OlympicInitCons_compare_Glasso_pen0p37_thresh0p1_Nodewise_pen0p37_thresh_0p08} between two words with the corresponding consonants, estimated by both Glasso and nodewise regression.  This diagram holds for a particular choice of penalty and threshold.  Figure \ref{pearsonCorrShortVowels} displays the Pearson correlations for word pairs that have edges in Figure \ref{OlympicInitCons_compare_Glasso_pen0p37_thresh0p1_Nodewise_pen0p37_thresh_0p08}. 

The edges in the supernode diagram are associated with properties of words studied by linguists.  Both m/n and w are continuously voiced consonants (i.e.\ they can be pronounced in a sustained way, with continuous vocal fold vibration).  Figure \ref{hist_initialConsonantMerged_glasso_fracEdges} in the appendix displays a bar chart of the fraction of edges between each pair of onsets (initial consonants), for a sequence of Glasso penalty parameters.  Even at a penatly of $0.43$, edges persist between words beginning with m/n and words beginning with w.  Figure \ref{OlympicVowels_labialAlveolarMerged_initialConsonant_seqPen_pen0p30_thresh0p10} in the appendix displays the inverse covariance graph for words with labial and alveolar initial consonants, estimated using Glasso with penalty $0.3$.  At this penalty, the words starting with p and t are disconnected, whereas many edges remain between words starting with the other labial and alveolar consonants.  The edges that remain are the same edges that are present in the supernode diagram, Figure \ref{consonantConnectivityDiagram_v}.  The leaf nodes are the fricative consonants v and w, whereas the labial b and alveolar t have more connections.    



 \begin{figure}
\begin{subfigure}{0.4\textwidth}  
\resizebox{\textwidth}{!}{\begin{tabular}{c|c|c}
     Word One & Word Two & Pearson Correlation  \\ 
     \hline
     kop & tor & $0.53$ \\
     nog & wond & $0.56$ \\
     den & pen & $0.49$ \\
     baba & ken & $-0.49$ \\
     bate & maak & $0.50$ \\
     bate & tas & $0.52$ \\
     bate & toer & $-0.48$ \\
     berg & mier & $0.48$ \\
     bied & das & $0.49$ \\
     boet & kies & $0.60$  \\
     bot & pars & $0.50$ \\
     dare & baba & $0.48$
\end{tabular}}
\end{subfigure}
\begin{subfigure}{0.4\textwidth}  
\resizebox{\textwidth}{!}{\begin{tabular}{c|c|c}
     Word One & Word Two & Pearson Correlation  \\ 
     \hline
     doer & pen & $0.50$ \\
     kat & met & $-0.48$ \\
     ken & tand & $0.48$ \\
     kerk & piek & $0.45$ \\
     koet & met & $0.51$ \\
     met & vier & $0.52$ \\
     met & wat & $0.53$ \\
     nek & was & $0.51$ \\
     nek & woed & $0.58$ \\
     padd & pond & $0.46$ \\
     term & vier & $0.60$ \\
     & & 
\end{tabular}}
\end{subfigure}
\caption{Word-word pearson correlations for words with edges in Figure \ref{OlympicInitCons_compare_Glasso_pen0p37_thresh0p1_Nodewise_pen0p37_thresh_0p08}.}  \label{pearsonCorrShortVowels}
\end{figure}

\subsection{Heteroscedasticity and nonstationary autocorrelation indicate heterogeneity over time}

Since the pitch curves are smooth, strong local correlations along the time axis are expected.  The time-time dependence structure is informative in that it provides a characterization of the variance of the pitch curves as a function of time, and reveals the extent to which local dependencies decay.   

The time-time covariance matrices for each word group (labial, alveolar, nasal, fricative) shown in Section \ref{timeGlassoOutput} of the Appendix indicate that the variance increases over time; that is, the pitch exhibits greater variability at the end of the word utterance than at the beginning.  This indicates that speech may be more constrained at the beginning of a word token than at the end.  The correlation matrices are approximately banded, and essentially all pairwise correlations are above $0.5$. In some cases the correlations decay faster at the end of the utterance than at the beginning.    

The diagonal entries of the inverse covariance matrix reflect the residual variances of each time point when regressed on the other other time points; a small diagonal entry corresponds to large residual variance.  For each of the word groups, the diagonal entries of the precision matrix are decreasing in time, also consistent with the early portion of the utterance being more constrained and predictable than the later portion of the utterance.     


In Table \ref{timeCorrMetrics}, we report metrics related to the Glasso estimate of the time-time correlation matrix.  Based on the estimated effective sample using all words ($n(r) =4$, $n(w)=93$, $n(s) = 20$), using the identity matrix for the word-word covariance, the theoretical penalty is $\sqrt{\log(n(w)) / (n(s) * n(r) * n(w))} = 0.03$.  In practice, due to dependence on the other axis, one should use a larger penalty when estimating the time-time inverse covariance.  The similar time-time edge structure for the word groups motivates pooling across the word groups to estimate a common time-time inverse correlation structure.   

\section{Conclusion}

Using recently developed methods for tensor data with multi-way dependence, we show that word-word edges in an Afrikaans pitch curve data set are associated with natural word attributes of interest to linguists.  This data analysis suggests that the notion of conditional independence and tools of graphical modeling, which have proved useful in other scientific fields are also useful in the field of phonetics.  In particular, long vowel edges are driven by whether the vowel is pronounced at the front of the mouth or back of the mouth.  A future research question of interest is whether the patterns observed here hold in other languages.  

\section*{Acknolwedgments} We thank Will Styler for very delightful discussions of our initial findings as detailed in this report.

\bibliographystyle{ims}
\bibliography{subgaussian}{}

\appendix




%

\section{Outline of supplement}


Section \ref{sec:IPA} displays the International Phonetic Alphabet (IPA) representation for Afrikaans words used in \cite{coetzee2018plosive}.  Section \ref{supplementVisualizationEdges} discusses word-word inverse covariance estimation and graphical modeling.  Section \ref{sec:SupplementLongVowels} presents an inverse covariance graph and trial residualized pitch curves for words with long vowels, as well as bar charts and Pearson correlation tables summarizing the edges for long and short vowels.  Section \ref{sec:LabialAlveolar} demonstrates that for words starting with labial and alveolar consonants, ``p'' and ``t'' become disconnected in the graphical model for sufficiently high penalty parameter.  Section \ref{supplementOnsetBarCharts} displays bar charts of the fraction of edges and Pearson correlation between each pair of initial consonants.  Section \ref{sec:ShortVowel} shows that edges for words with short vowels are associated with initial consonant (onset) and final consonant (coda).  Further visualizing these edges, Section \ref{sec:BicyclePlotsAllWordsSmallerPen} displays the graphical model for all $93$ words, with each plot containing a subgraph consisting of two word categories (among labial, alveolar, nasal/k, and fricative) with a penalty of $0.32$ and threshold of $0.16$, and Section \ref{bicyclePlotsAppendix} displays analogous plots with a penalty of $0.26$ and threshold $0.08$.  Section \ref{timeGlassoOutput} discusses estimation of the time-time covariance and inverse covariance matrices, and compares time-time edges estimated using different subsets of words. 

\subsection{IPA representation of Afrikaans words} \label{sec:IPA}
  \begin{figure}[H]
  \begin{subfigure}{\textwidth}
  \includegraphics[width=0.86\textwidth]{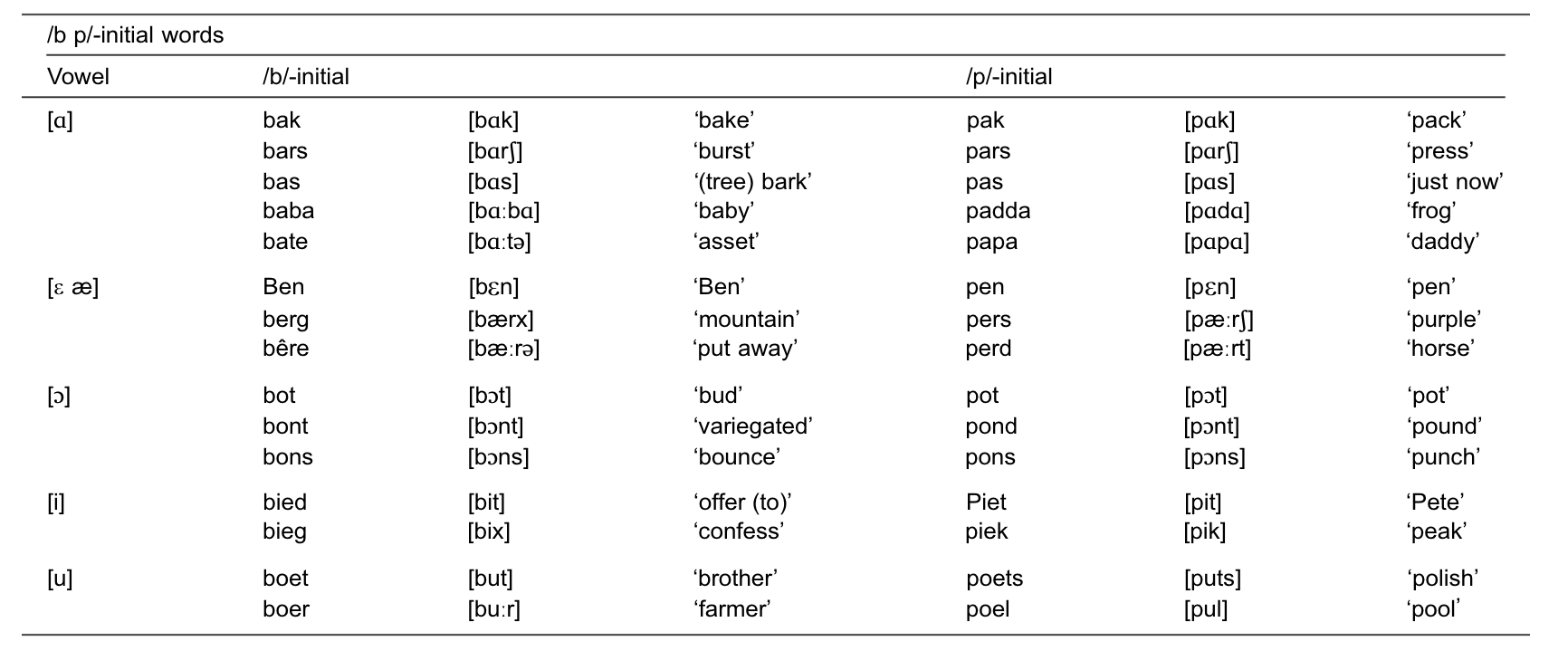}
  \end{subfigure}
  \begin{subfigure}{\textwidth}
  \includegraphics[width=0.86\textwidth]{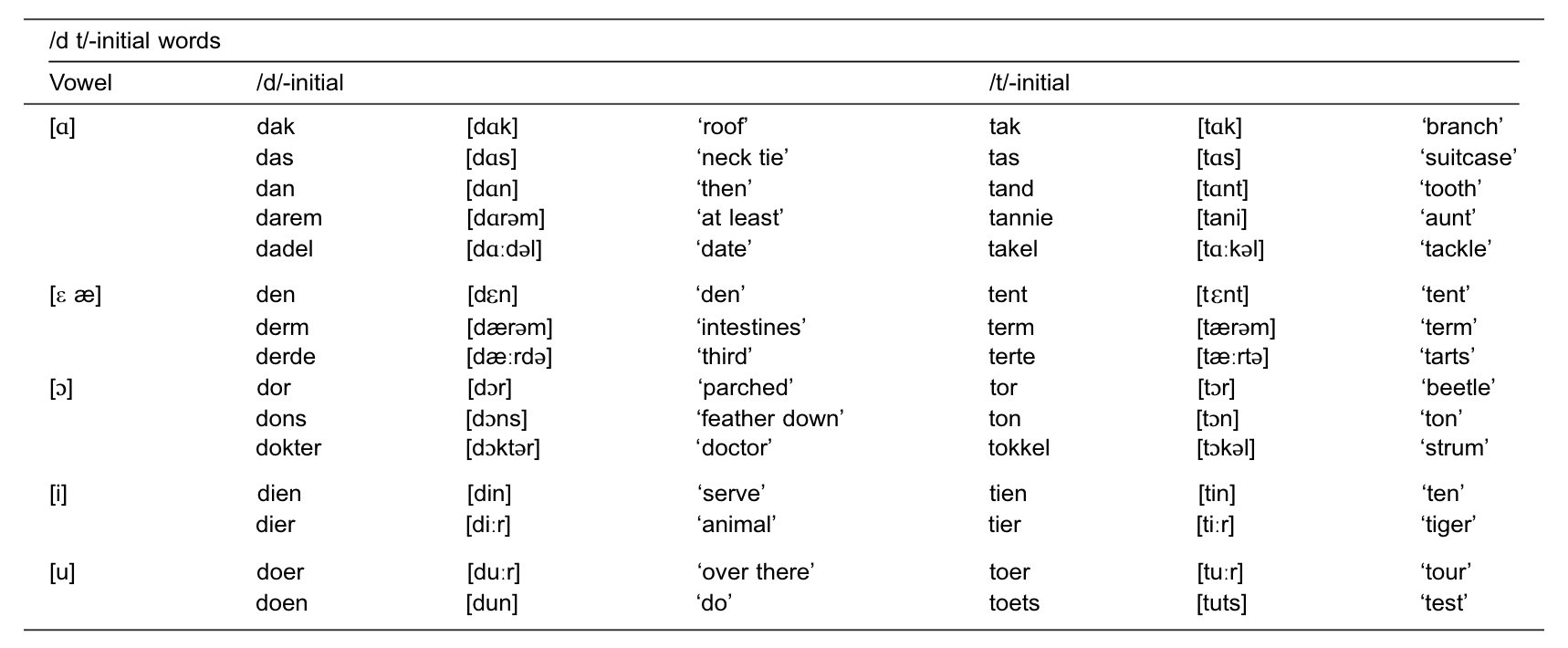}
  \end{subfigure}
    \begin{subfigure}{\textwidth}
  \includegraphics[width=0.86\textwidth]{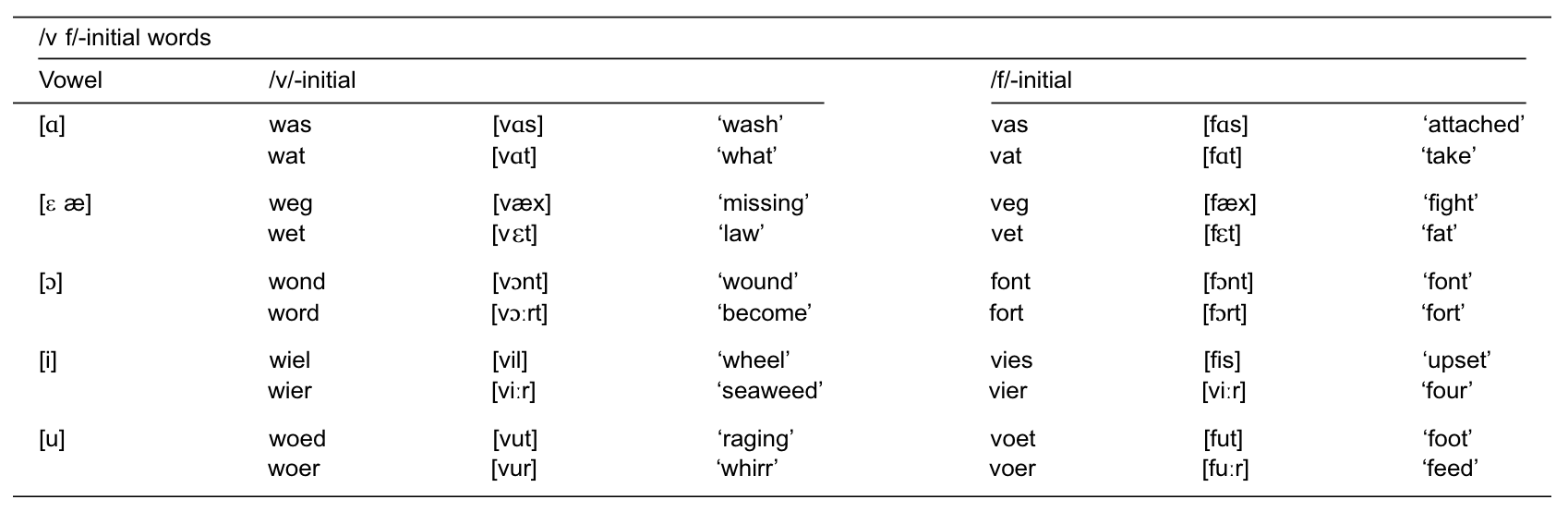}
  \end{subfigure}
    \begin{subfigure}{\textwidth}
  \includegraphics[width=0.86\textwidth]{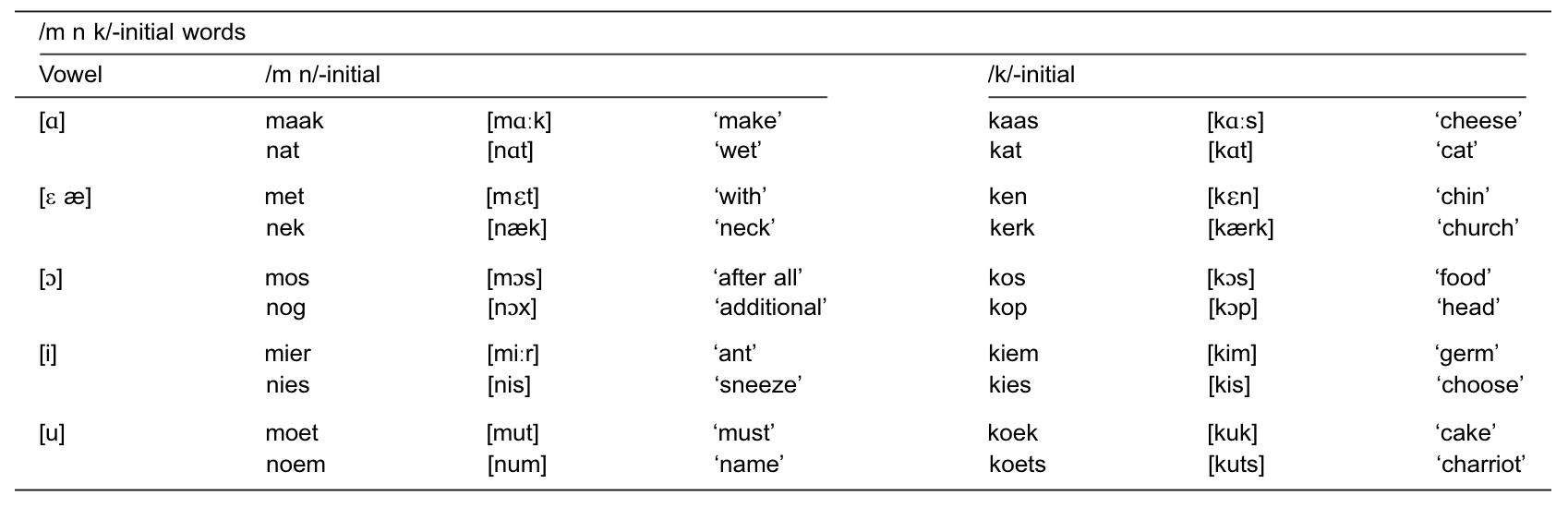}
  \end{subfigure}
  \caption{From Appendix A of \cite{coetzee2018plosive}}
  \end{figure}

\section{Word-word graphical models and inverse correlation estimation} \label{supplementVisualizationEdges}

\begin{table}[H] 
 \centering
 \begin{tabular}{|c|c|c| c|c|}
 \hline
     Word Group & Avg. node degree & \# edges & $\text{tr}(B) / \lVert B \rVert_F$ & $\lVert B \rVert_2$  \\ 
     \hline
     All words ($93$ words) & $9.3$ & $88$ &  $4.05$ & $5.0$ \\ 
     Labial ($26$ words) & $9.5$ & $90$ &  $4.05$ & $5.0$ \\ 
     Alveolar ($30$ words) & $9.8$ & $93$ &  $4.05$ & $5.0$ \\ 
     Nasal words ($20$ words) & $9.8$ & $93$ &  $4.07$ & $5.0$ \\ 
     vf words ($17$ words) & $8.4$ & $80$ &  $4.03$ & $5.1$ \\
     \hline
\end{tabular}
\caption{Metrics related to estimate of time-time correlation matrix.} \label{timeCorrMetrics}
\end{table} 

\subsection{Analyzing edges related to long vowels} \label{sec:SupplementLongVowels}

\begin{figure}[h!]
\includegraphics[width=\textwidth]{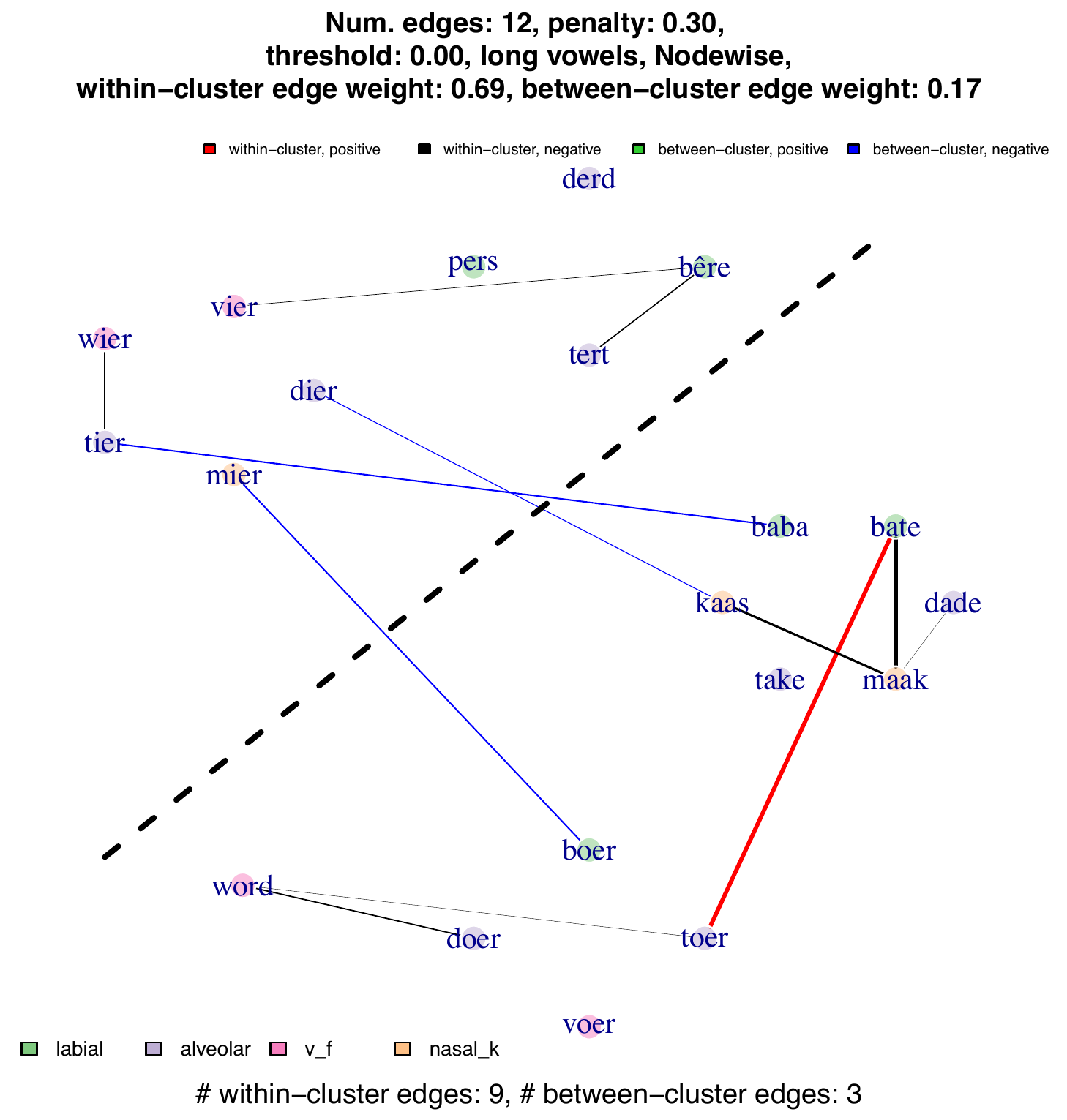} 
\caption{Inverse correlation edge graph for words with long vowels.  Based on the estimated effective sample ($n(r) = 3$, $n_{\text{eff}}(t)=3$ or $4$, $n(s) = 20$) and the theoretical guidance from ~\cite{Zhou14a},
we believe the theoretical penalty should be in the range of $[0.11, 0.13]$.  The words are organized by vowel, with each circle of words sharing a common vowel (``word'' is the only word with a long \textopeno \, vowel; in Afrikaans, it means ``become'').}  \label{OlympicVowelsWithLong_Nodewise_longThenShortVowels_pen_thresh_pen0p16_thresh0p08}
\end{figure} 



\begin{table}[H] 
\centering
\begin{tabular}{c|c|c}
     Word One & Word Two & Pearson Correlation  \\ 
     \hline
     bate &  maak & $0.50$  \\ 
     kaas & maak & $0.41$ \\ 
     baba & tier & $0.37$ \\
     bate & maak & $0.50$ \\
     bate & toer & $-0.48$ \\
     boer & kaas & $0.28$\\
     boer & mier & $0.36$ \\
     b\^{e}re & tert & $0.35$ \\
     b\^{e}re & vier & $0.33$ \\
     bate & kaas & $0.22$ \\
     dade & maak & $0.32$\\
     derd & wier & $0.27$ \\
     dier & kaas & $0.35$ \\
     doer & voer & $0.26$ \\
     doer & word & $0.36$ \\
     kaas & tert & $0.28$ \\
     tert & vier & $0.30$ \\
     wier & tier & $0.35$ \\
\end{tabular}
\caption{Word-word Pearson correlations.} \label{wordWordPearson}
\end{table} 

\begin{figure}[h!]
\includegraphics[width=\textwidth]{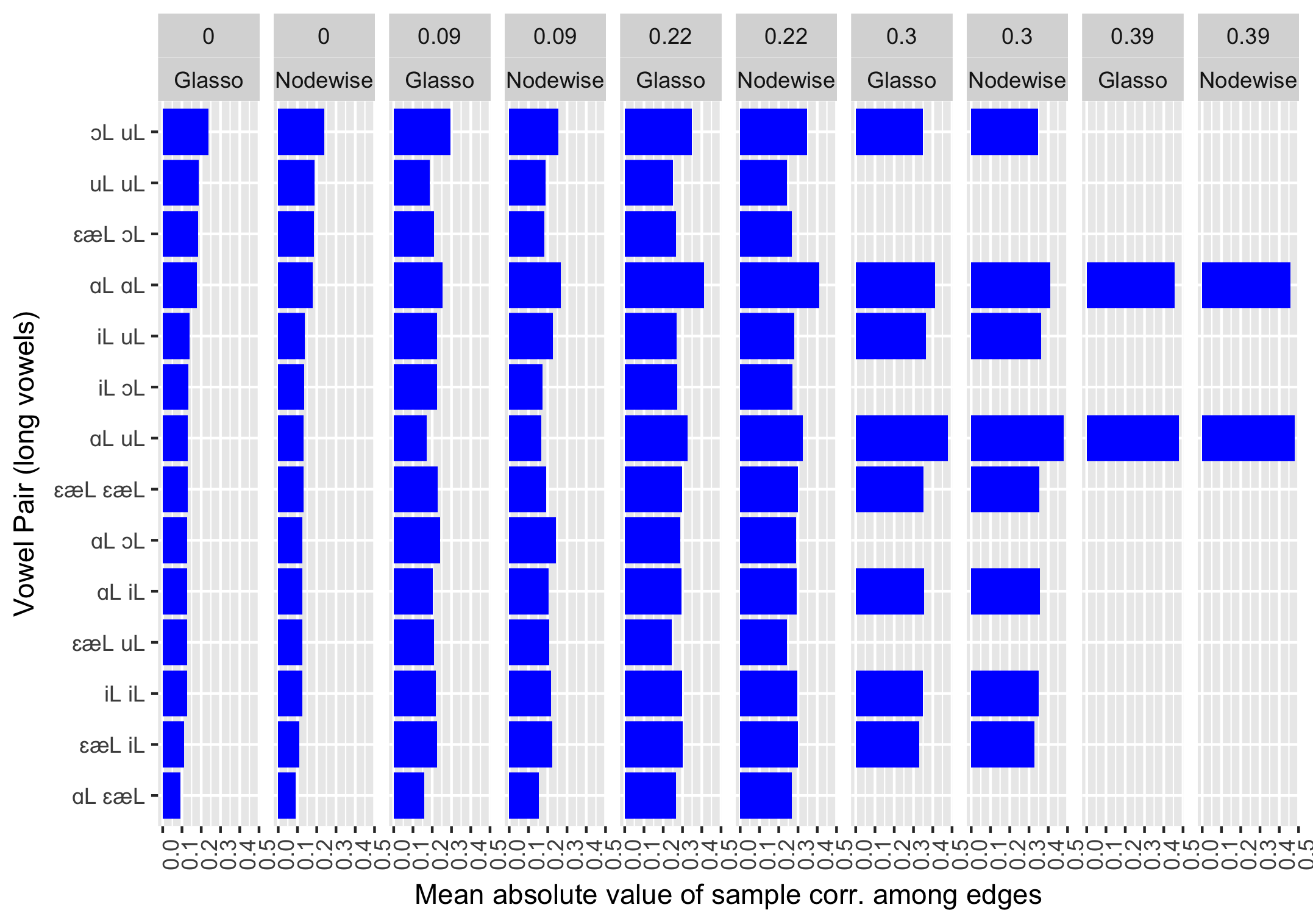} \caption{Bar chart of average sample correlation among edges for long vowels, estimated using Glasso and nodewise regression.}  \label{hist_vowelWithLongMerged_longVowels_glasso_nodewise_absSampleCorr}
\end{figure} 

\begin{figure}[h!]
\includegraphics[width=\textwidth]{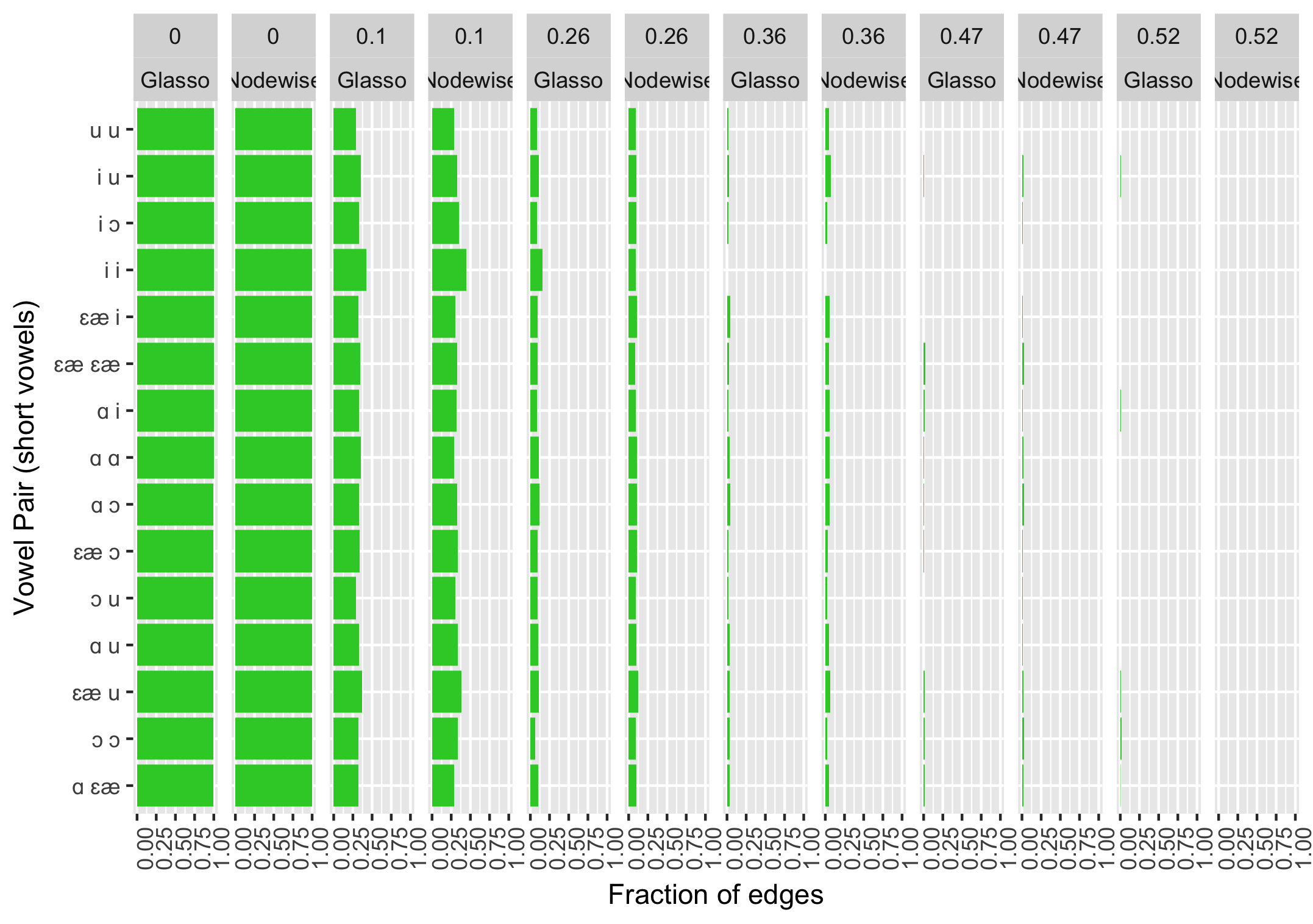} \caption{Bar chart of fraction of edges for short vowels, estimated using Glasso and nodewise regression.}  \label{hist_vowelWithLongMerged_shortVowels_glasso_nodewise_fracedges}
\end{figure} 

\begin{figure}[h!]
\includegraphics[width=\textwidth]{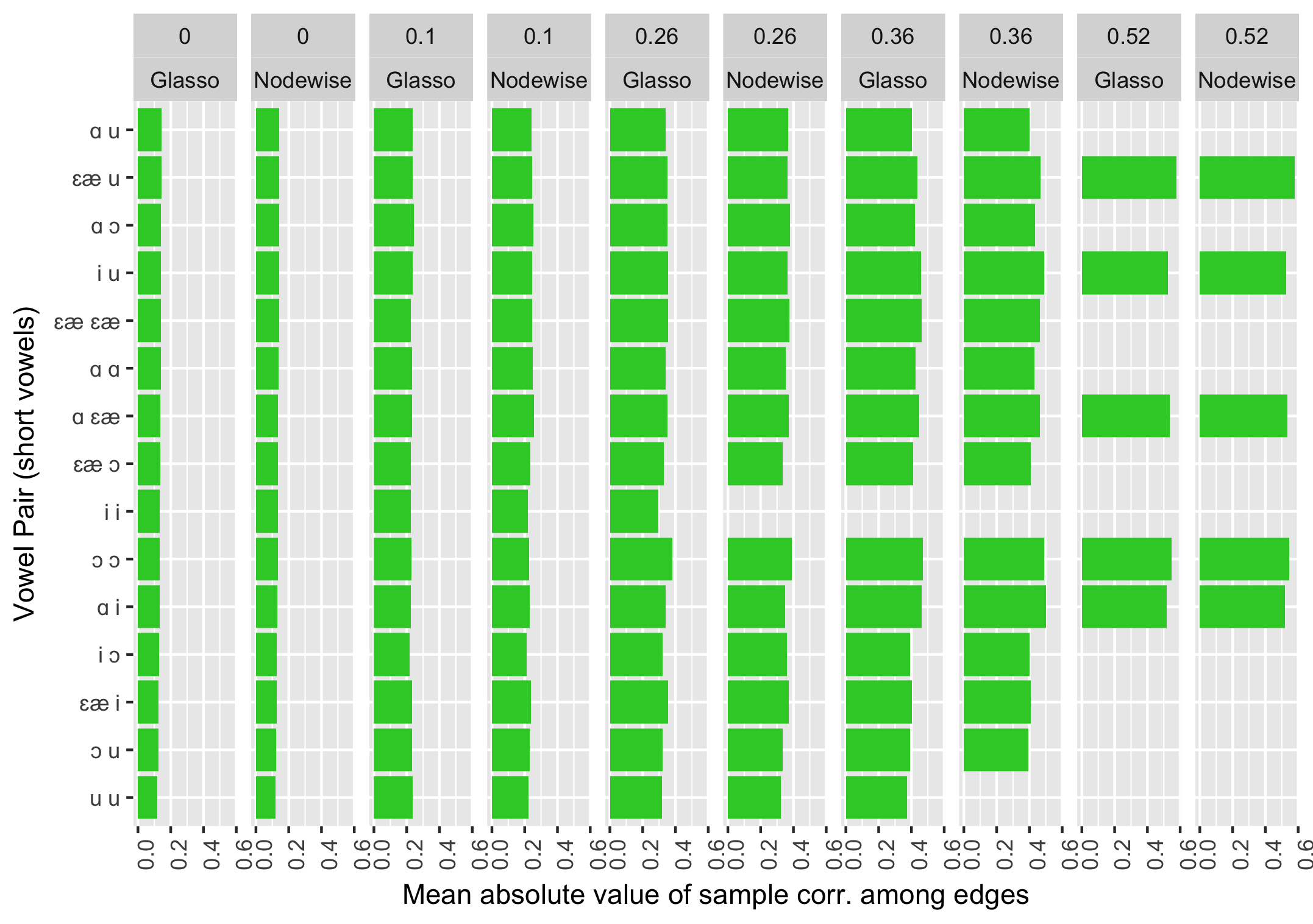} \caption{Bar chart of average sample correlation among edges for short vowels, estimated using Glasso and nodewise regression.}  \label{hist_vowelWithLongMerged_shortVowels_glasso_nodewise_absSampleCorr}
\end{figure}


\begin{figure}[h!]
\includegraphics[width=\textwidth]{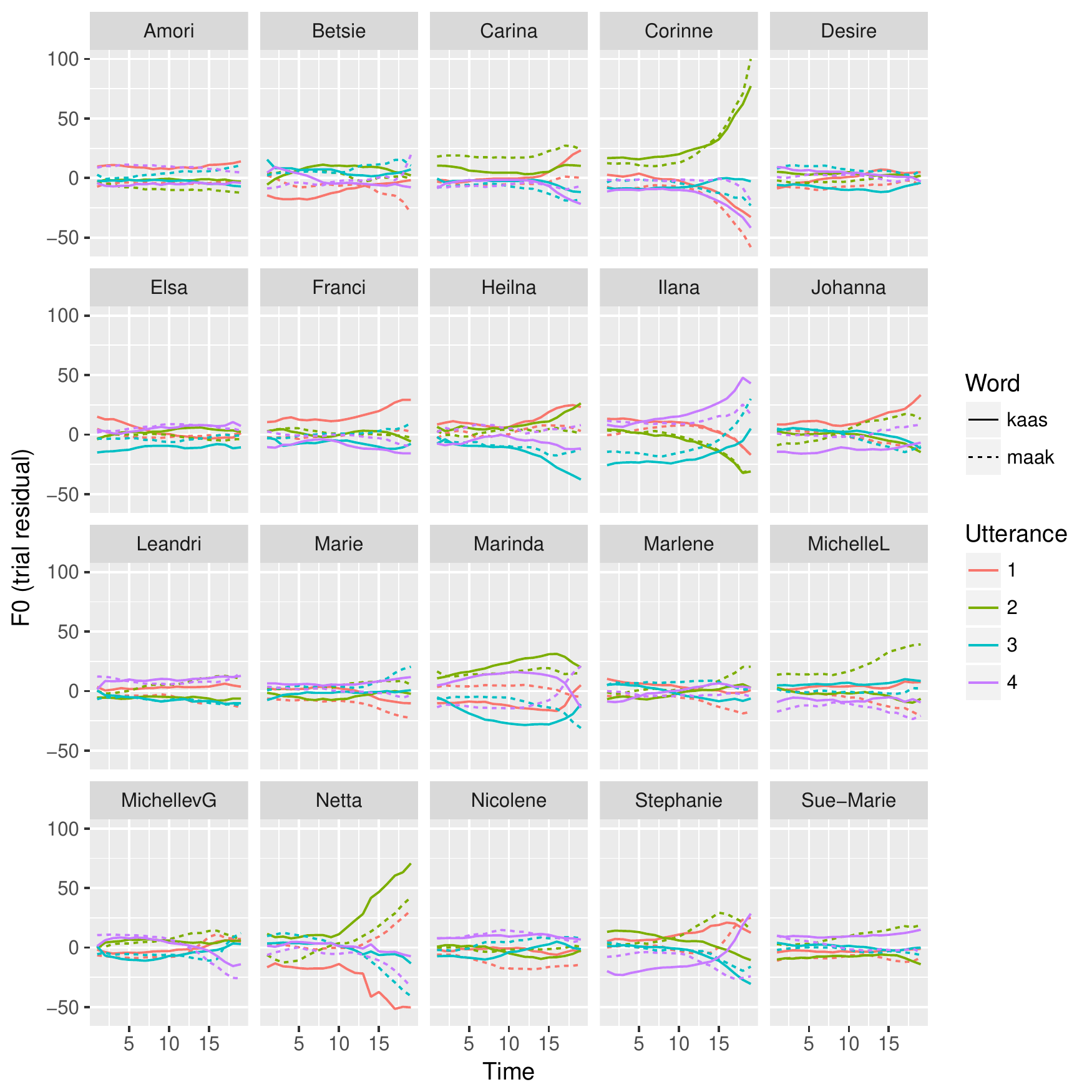} \caption{Trial residual pitch curves for the words maak and kaas.}  \label{pitchCurves_trialResid_maak_kaas}
\end{figure} 

\begin{figure}[h!]
\includegraphics[width=\textwidth]{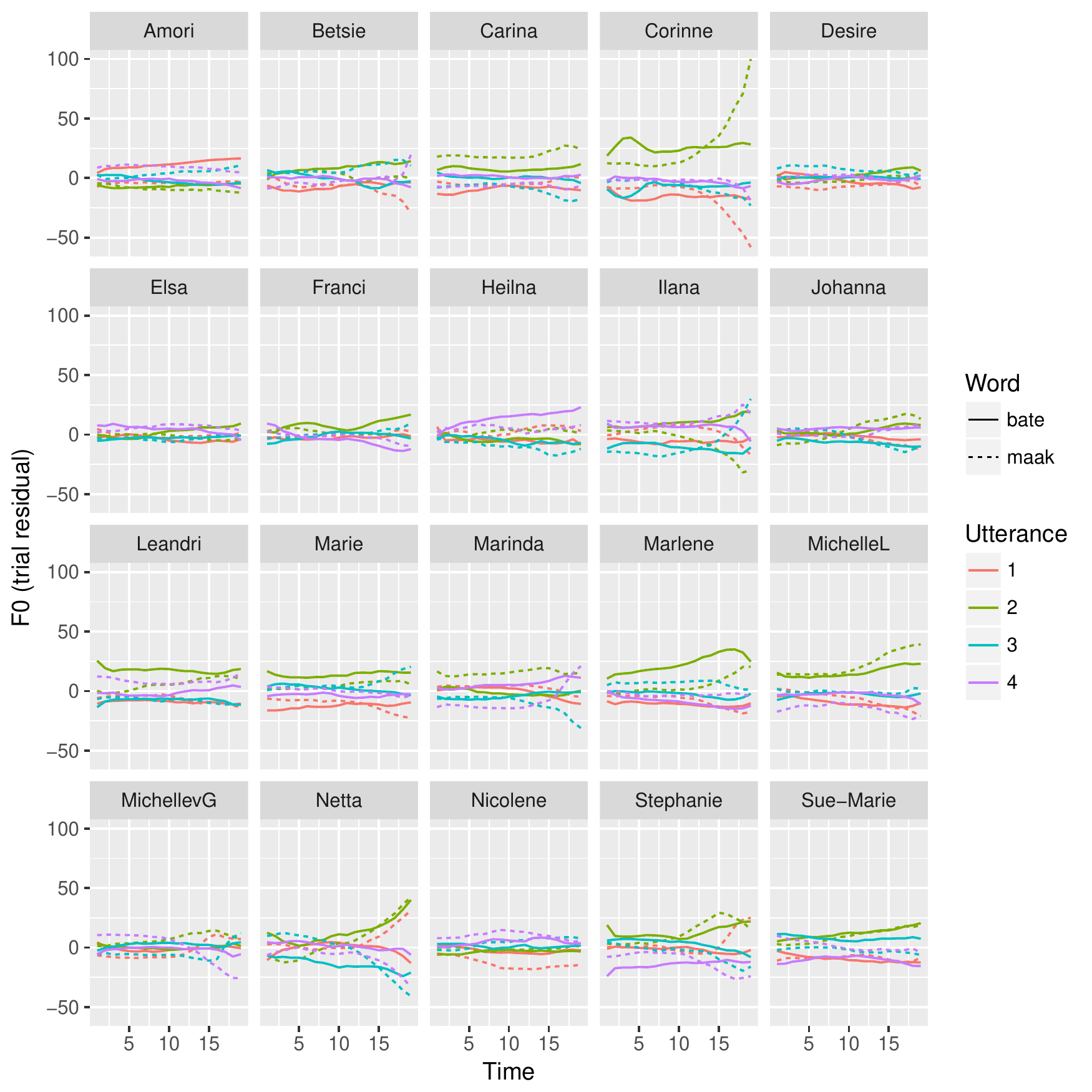} \caption{Trial residual pitch curves for the words bate and maak.}  \label{pitchCurves_trialResid_bate_maak}
\end{figure} 

\begin{figure}[h!]
\includegraphics[width=\textwidth]{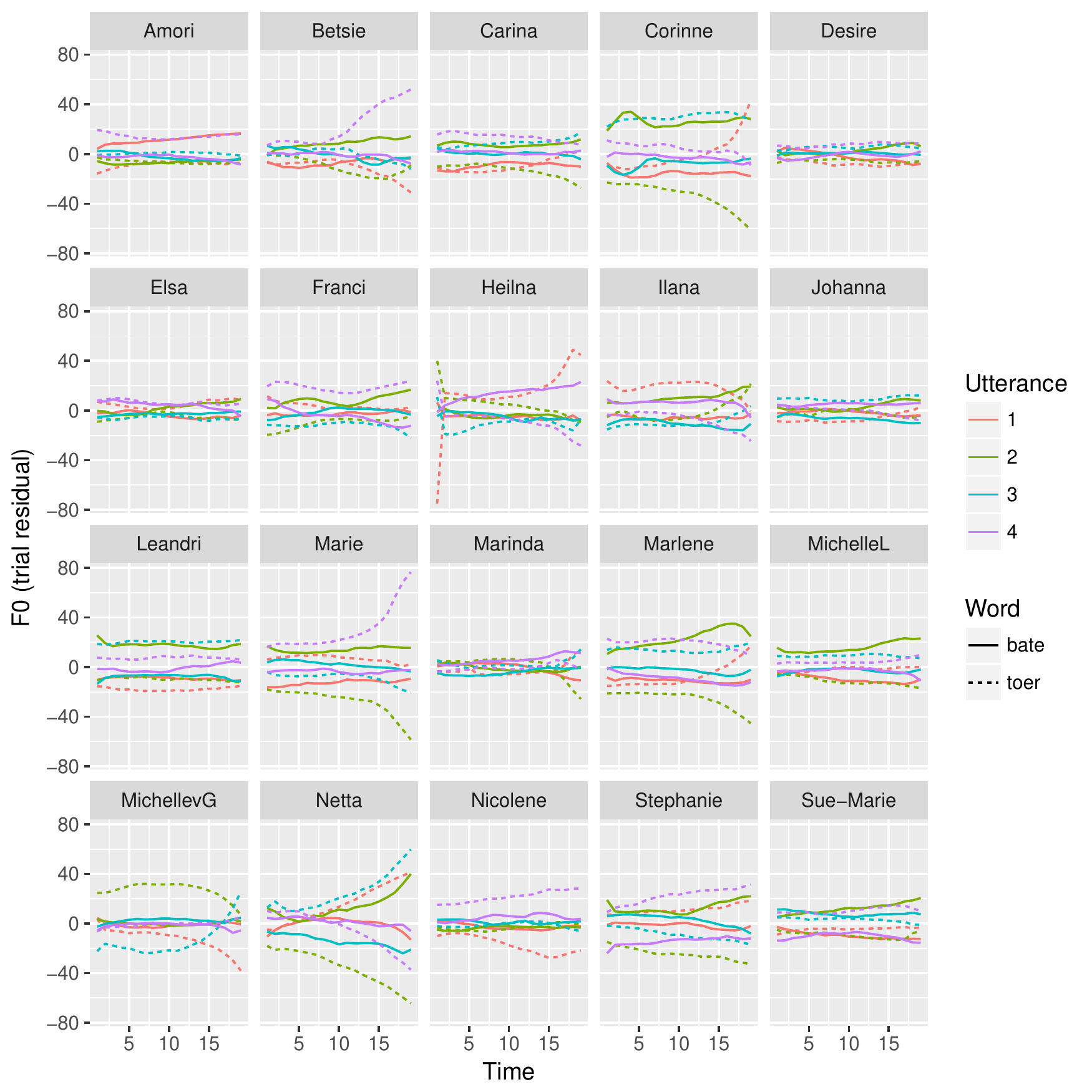} \caption{Trial residual pitch curves for the words bate and toer.}  \label{pitchCurves_trialResid_bate_toer}
\end{figure} 

\begin{figure}[h!]
\includegraphics[width=\textwidth]{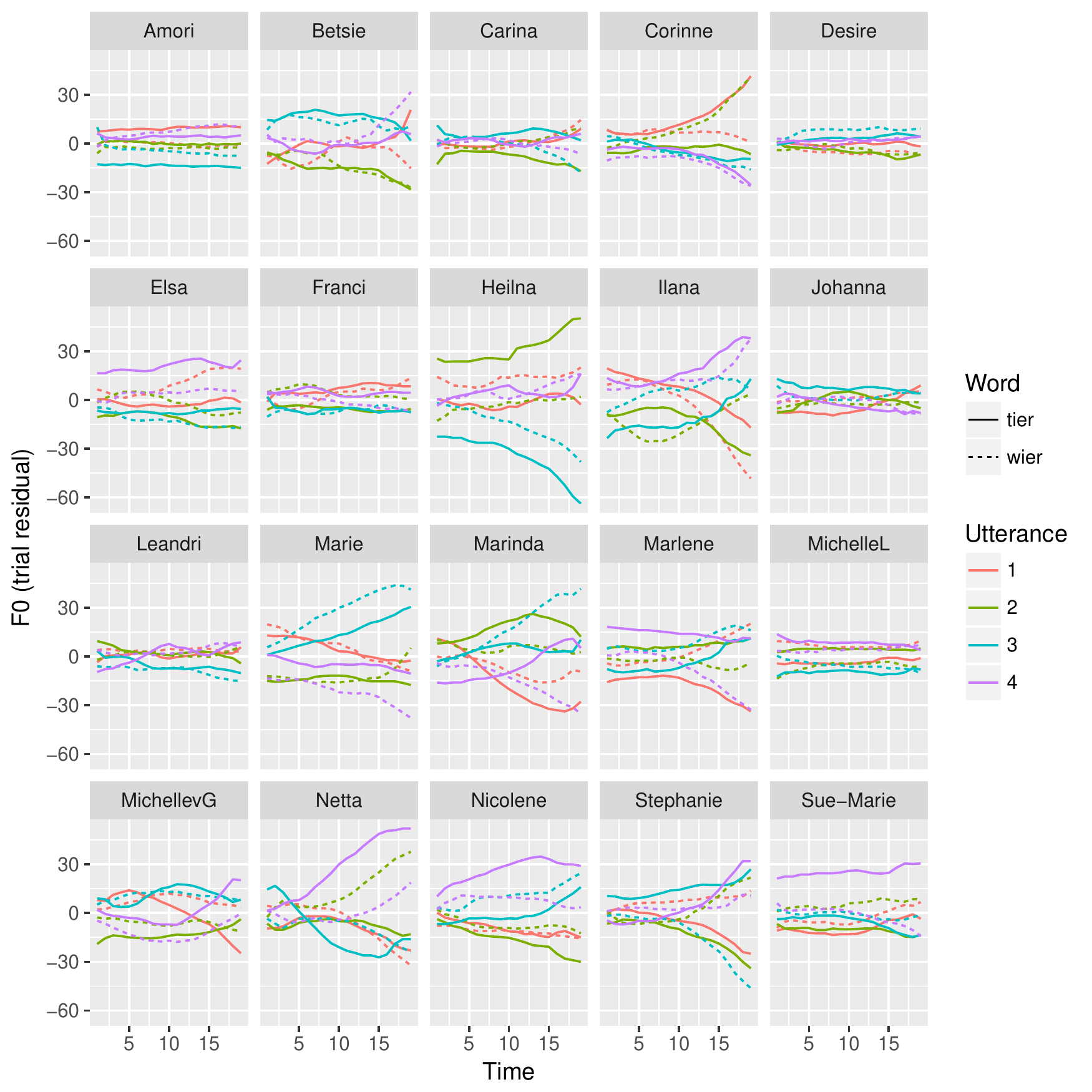} \caption{Trial residual pitch curves for the words wier and tier.}  \label{pitchCurves_trialResid_wier_tier}
\end{figure}

\clearpage
\subsection{Labial and alveolar words} \label{sec:LabialAlveolar}

\begin{figure}[h!]
\includegraphics[width=\textwidth]{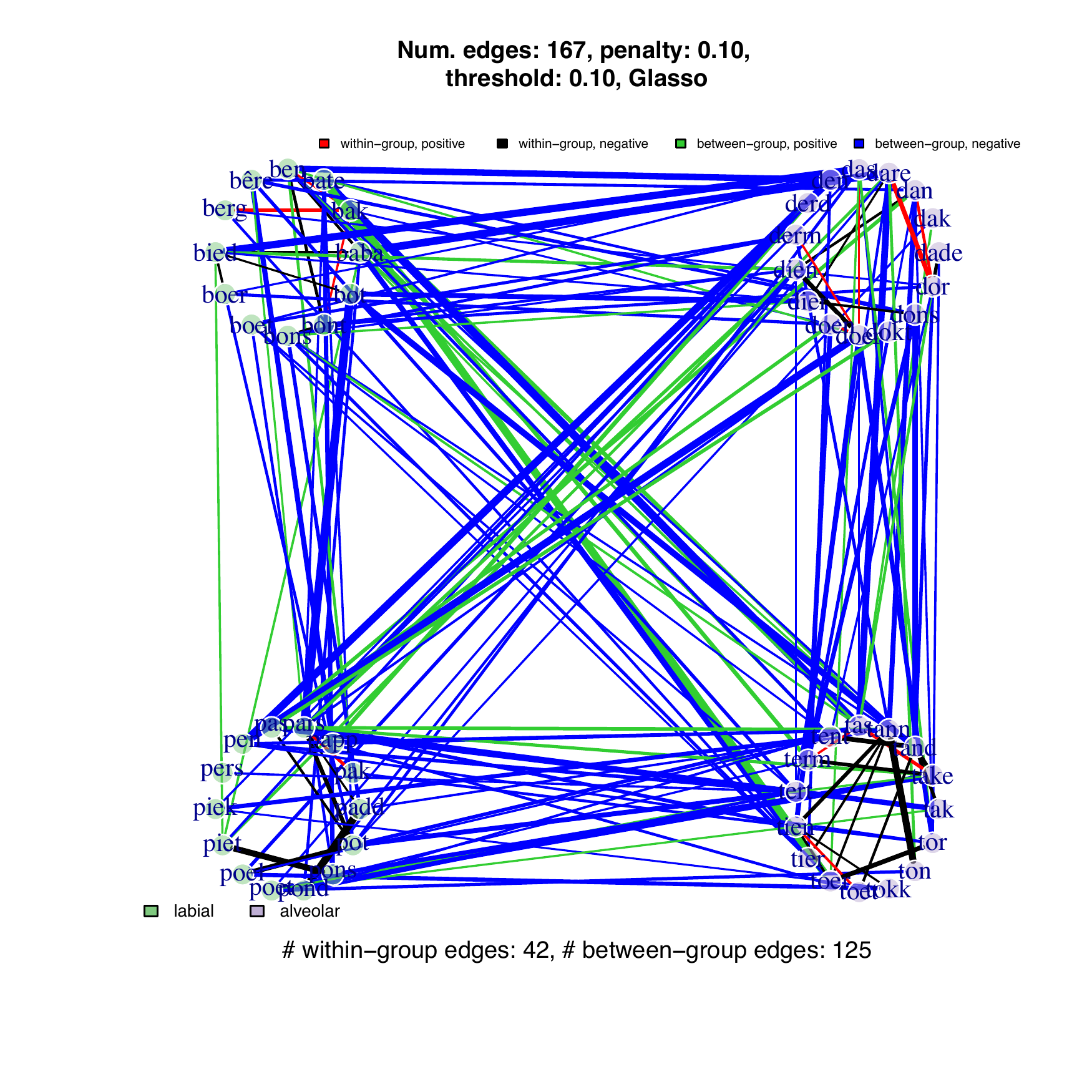} \caption{Inverse covariance graph of labial and alveolar words Glasso with a peanlty of $0.1$ and a threshold of $0.1$.}  \label{OlympicVowels_labialAlveolarMerged_initialConsonant_seqPen_pen0p10_thresh0p10}
\end{figure}

\begin{figure}[h!]
\includegraphics[width=\textwidth]{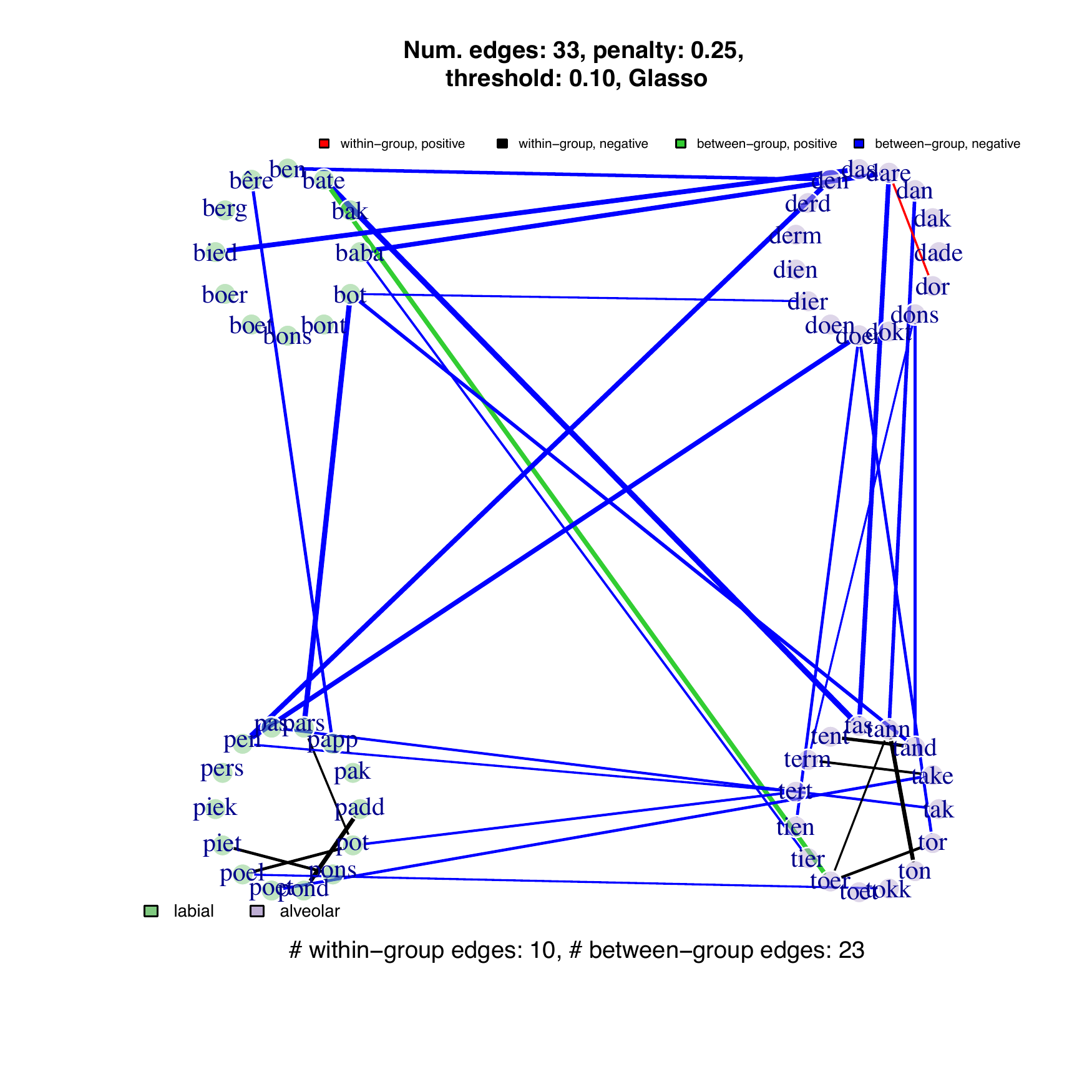} \caption{Inverse covariance graph of labial and alveolar words Glasso with a peanlty of $0.25$ and a threshold of $0.1$.}  \label{OlympicVowels_labialAlveolarMerged_initialConsonant_seqPen_pen0p25_thresh0p10}
\end{figure}

\begin{figure}[h!]
\includegraphics[width=\textwidth]{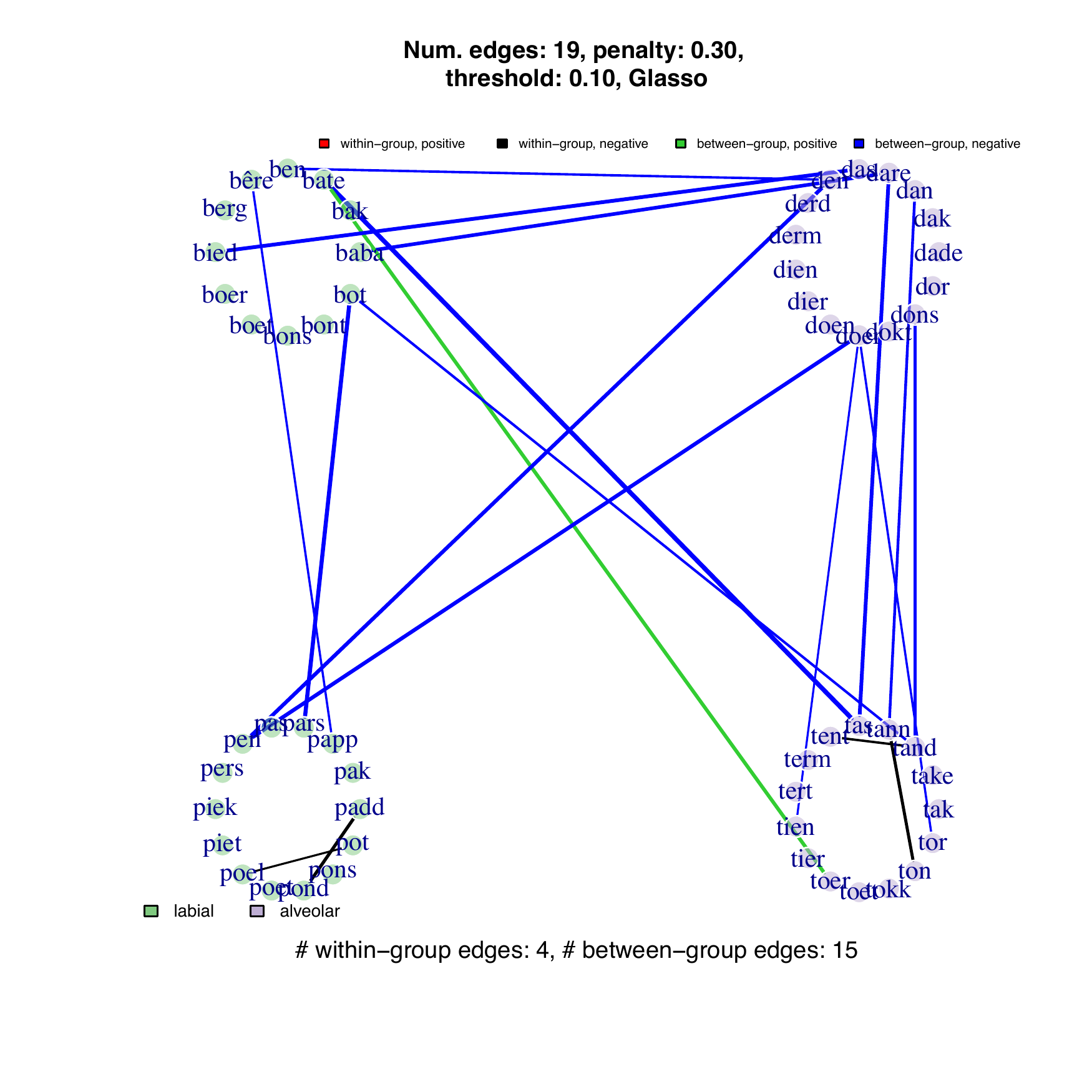} \caption{Inverse covariance graph of labial and alveolar words Glasso with a peanlty of $0.3$ and a threshold of $0.1$.}  \label{OlympicVowels_labialAlveolarMerged_initialConsonant_seqPen_pen0p30_thresh0p10}
\end{figure}

In Figures \ref{OlympicVowels_labialAlveolarMerged_initialConsonant_seqPen_pen0p10_thresh0p10}, \ref{OlympicVowels_labialAlveolarMerged_initialConsonant_seqPen_pen0p25_thresh0p10}, and \ref{OlympicVowels_labialAlveolarMerged_initialConsonant_seqPen_pen0p30_thresh0p10} we show the inverse covariance graph estimated using a sequence of Glasso penalty parameters, with a threshold of $0.1$.  For small penalty values, words of all four initial consonants (b, d, p, t) are densely connected.  As the penalty increases the edges between words beginning with p and t drop off.  

\clearpage
\subsection{Initial consonant connectivities} \label{supplementOnsetBarCharts}

\begin{figure}[h!]
\includegraphics[width=\textwidth]{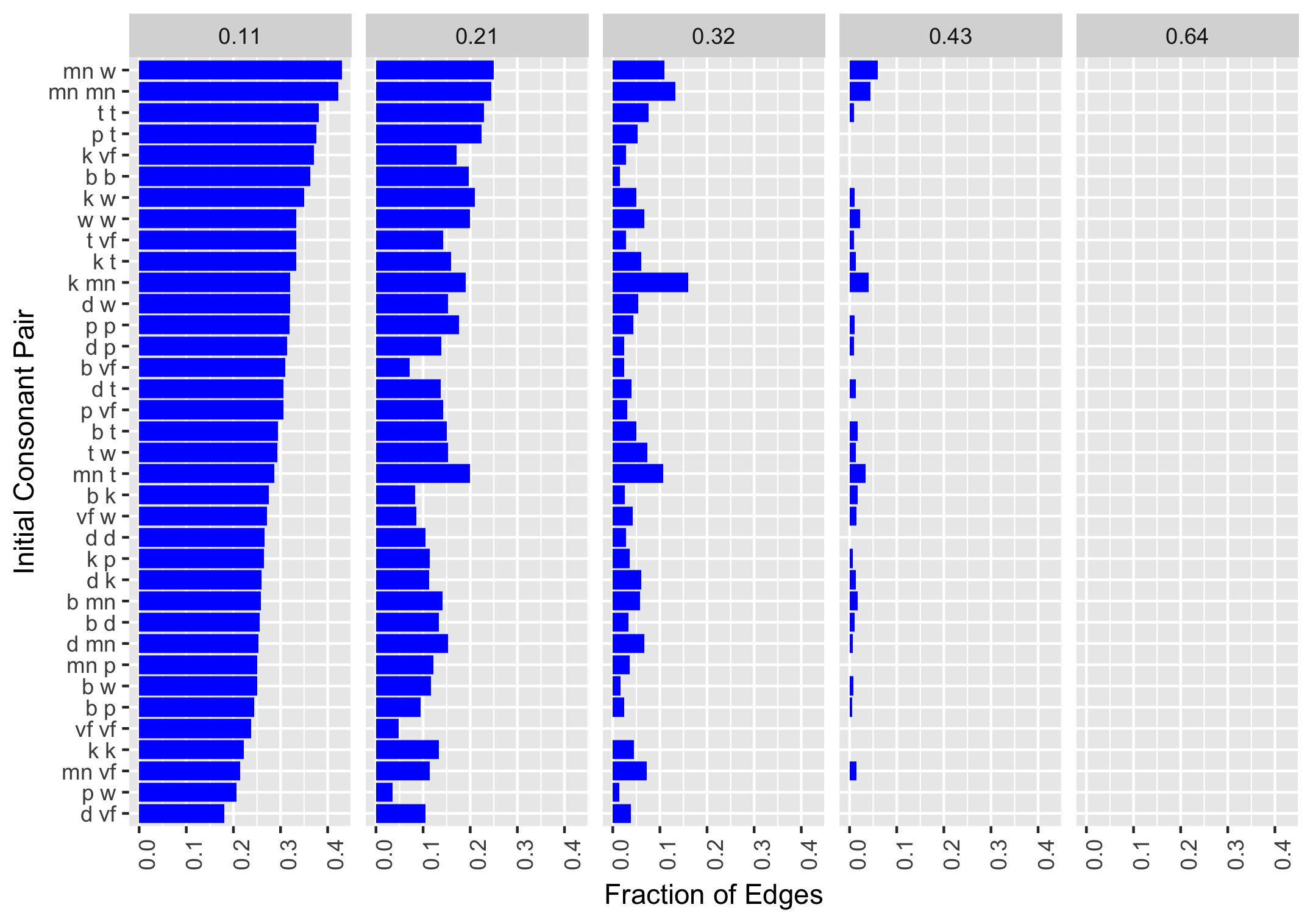} \caption{Fraction of edges between each pair of initial consonants as we vary the Glasso penalty.}  \label{hist_initialConsonantMerged_glasso_fracEdges}
\end{figure}

\begin{figure}[h!]
\includegraphics[width=\textwidth]{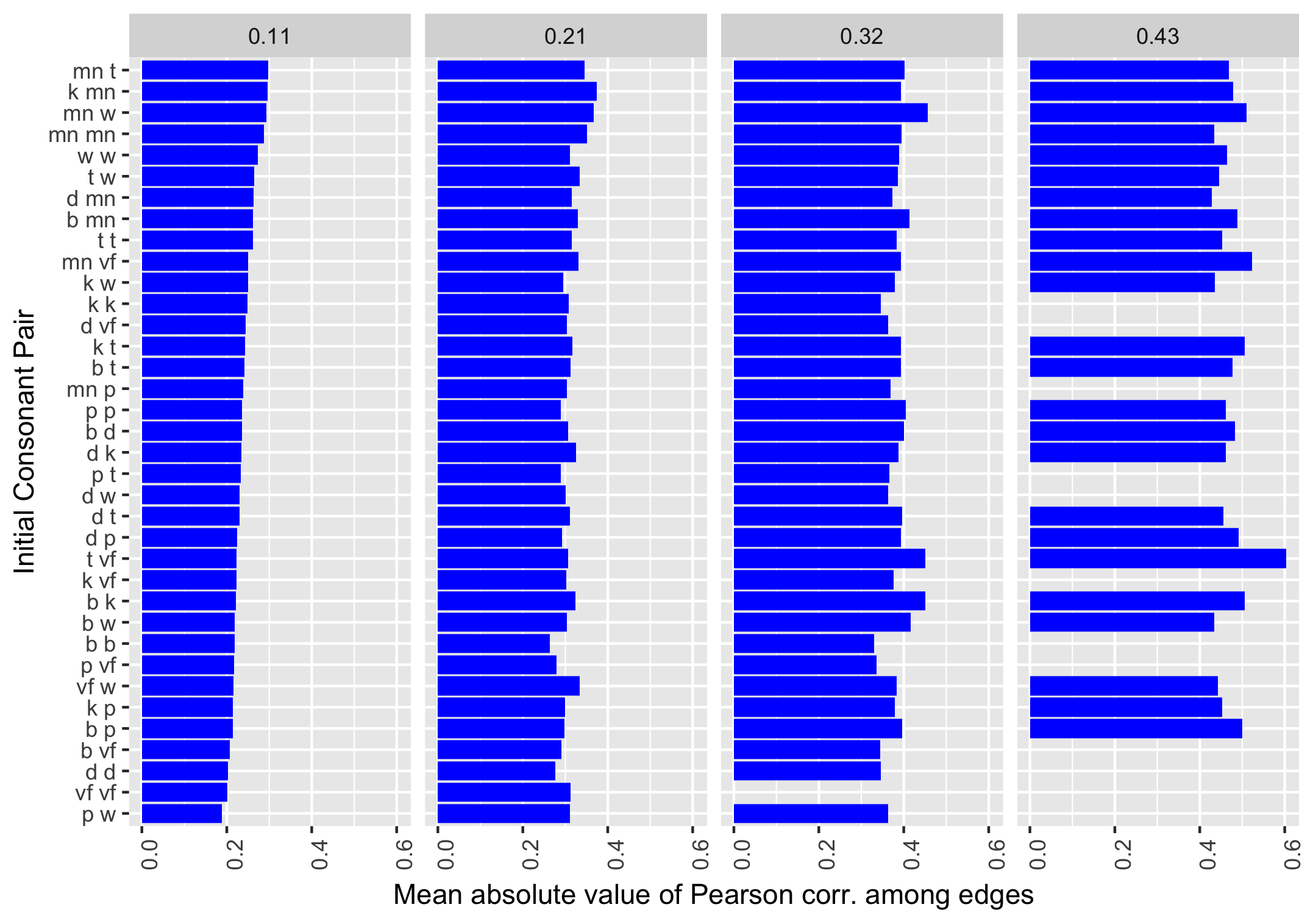} \caption{Mean absolute value of Pearson correlation among edges between each pair of initial consonants.}  \label{hist_initialConsonantMerged_glasso_meanAbsPearsonCor}
\end{figure}

  Figure \ref{hist_initialConsonantMerged_glasso_fracEdges} displays a bar chart of the fraction of edges between each pair of onsets (i.e.\ initial consonants), for a sequence of Glasso penalty parameters.  When counting edges, the ``m'' and ``n'' are treated as a single consonant, as are the consonants ``v'' and ``f''.   Figure \ref{hist_initialConsonantMerged_glasso_meanAbsPearsonCor} displays the mean Pearson correlation among edges in the for each consonant pair.  


\clearpage
 \subsection{Associations between short vowel edges and word attributes} \label{sec:ShortVowel}

 \begin{figure}[h]  
      \includegraphics[width=\textwidth]{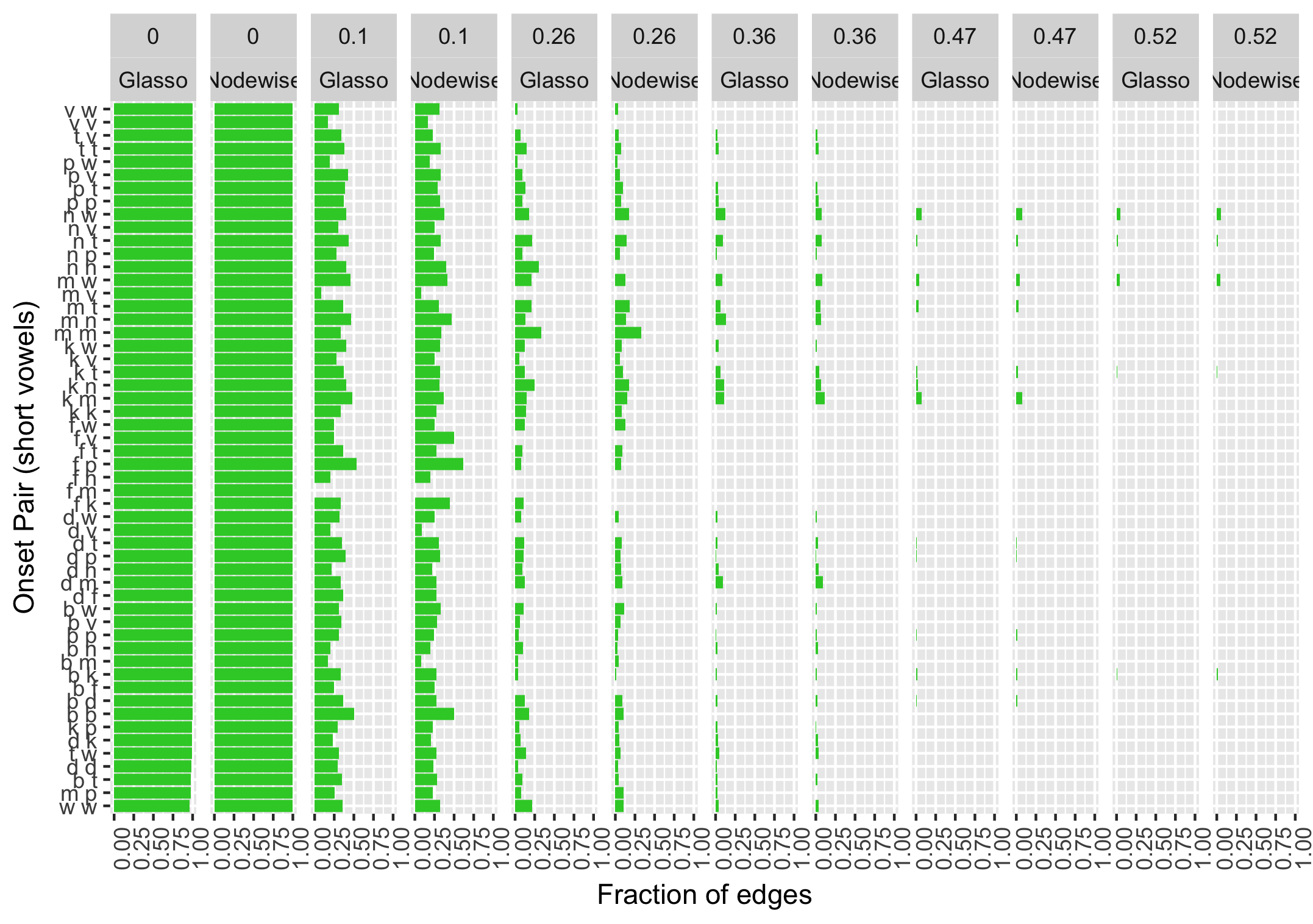}
  \caption{Bar chart of fraction of edges between each pair of onsets (initial consonants) for words with short vowels, estimated using Glasso and nodewise regression.}
\end{figure} 

\begin{figure}[h]
      \includegraphics[width=\textwidth]{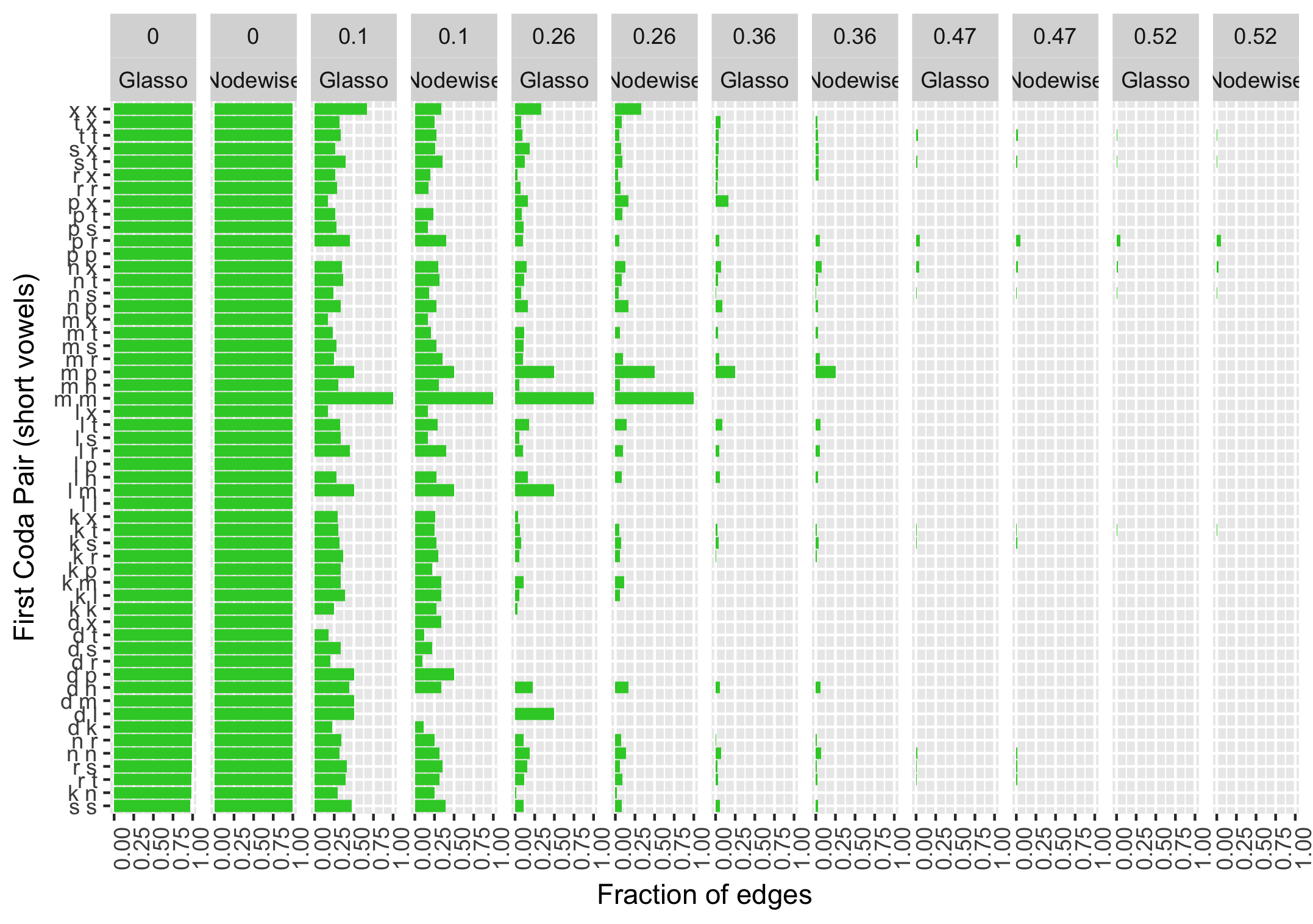}
  \caption{Bar chart of fraction of edges between each pair of first codas (consonant after the vowel) for words with short vowels, estimated using Glasso and nodewise regression.}
\end{figure}

  \begin{figure}[h] 
      \includegraphics[width=\textwidth]{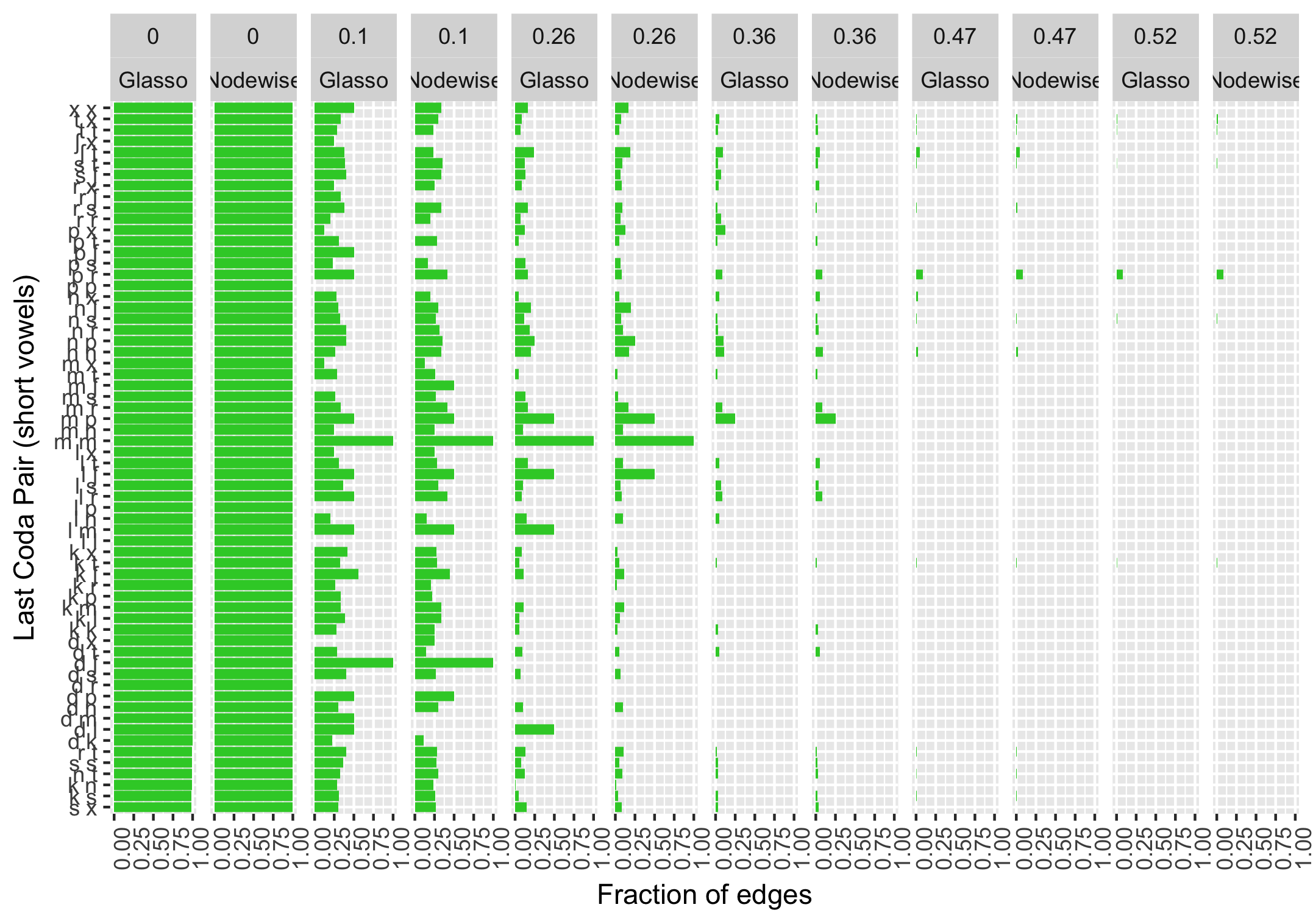}
  \caption{Bar chart of fraction of edges between each pair of last codas (last consonant of the syllable) for words with short vowels, estimated using Glasso and nodewise regression.}
\end{figure} 
  \begin{figure}[h]
  \includegraphics[width=\textwidth]{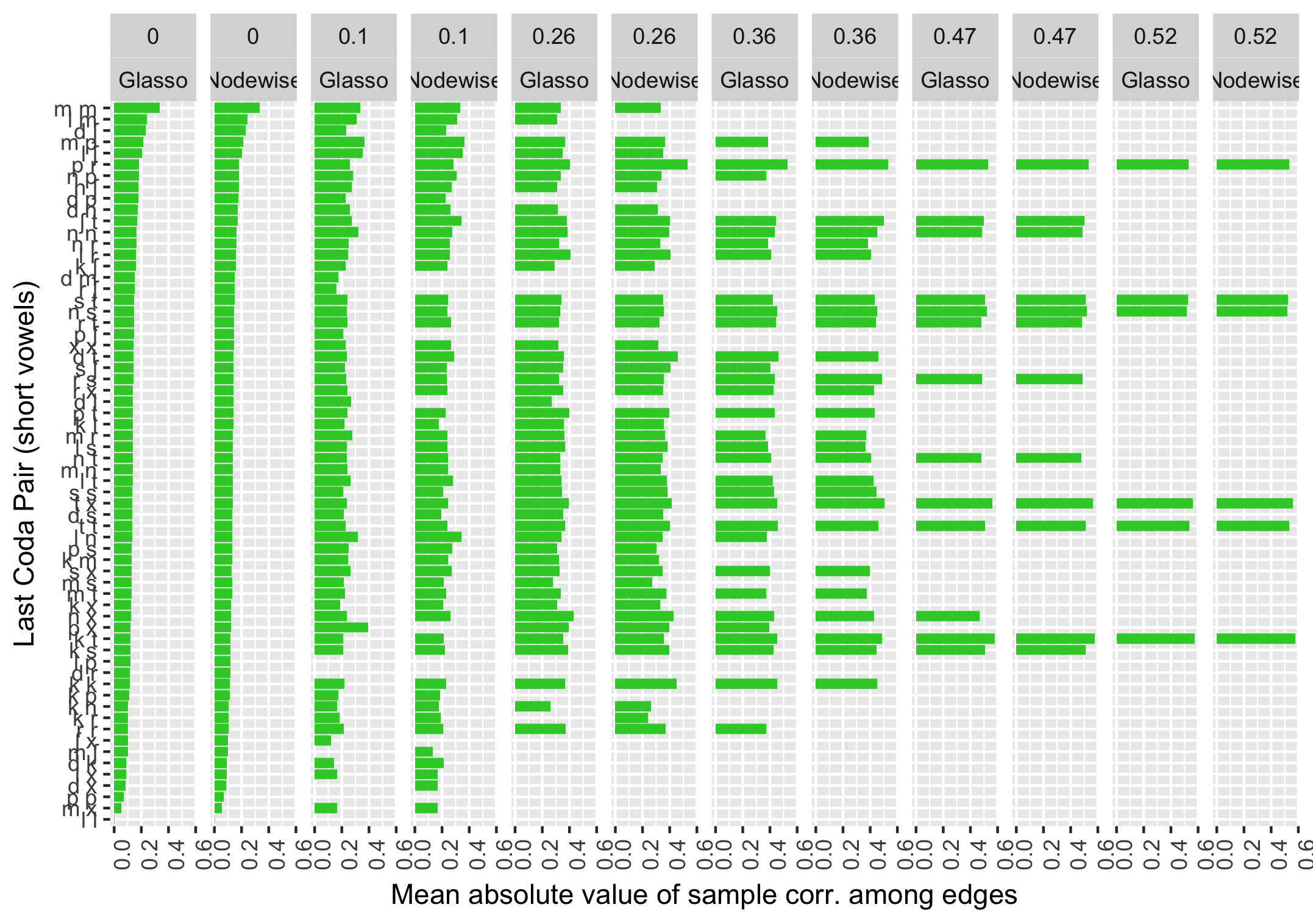}
  \caption{Bar chart of average sample correlation between words with edges for each pair of last codas, estimated using Glasso and nodewise regression.}
\end{figure}        

\begin{figure}[h]
\includegraphics[width=\textwidth]{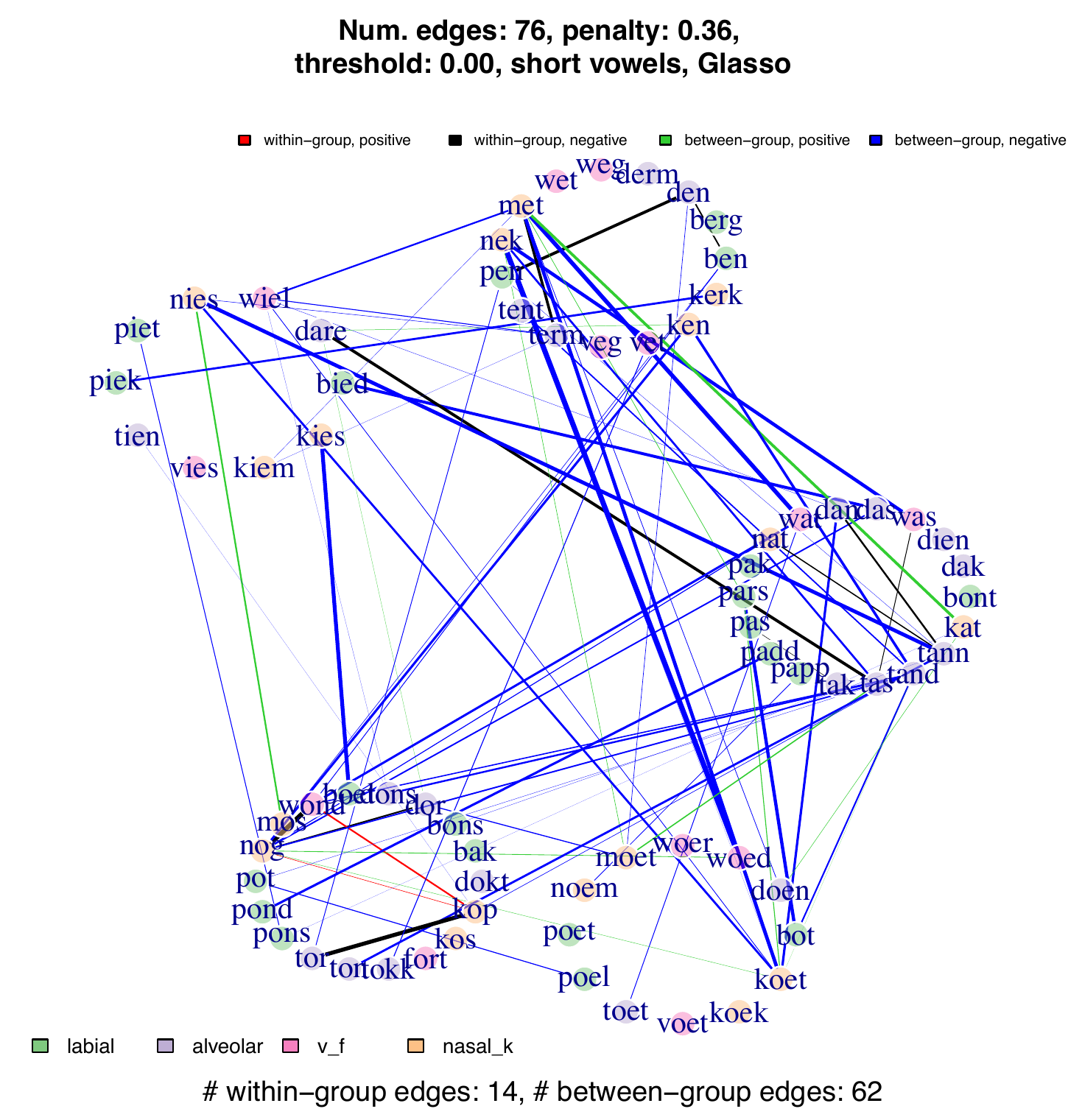}
\caption{Inverse covariance graph of words with short vowels, estimated using Glasso regression, organized by vowel.}
\label{OlympicVowelsWithLong_Glasso_shortVowel_pen0p36_thresh0}
\end{figure} 

\begin{figure}[h] 
\includegraphics[width=\textwidth]{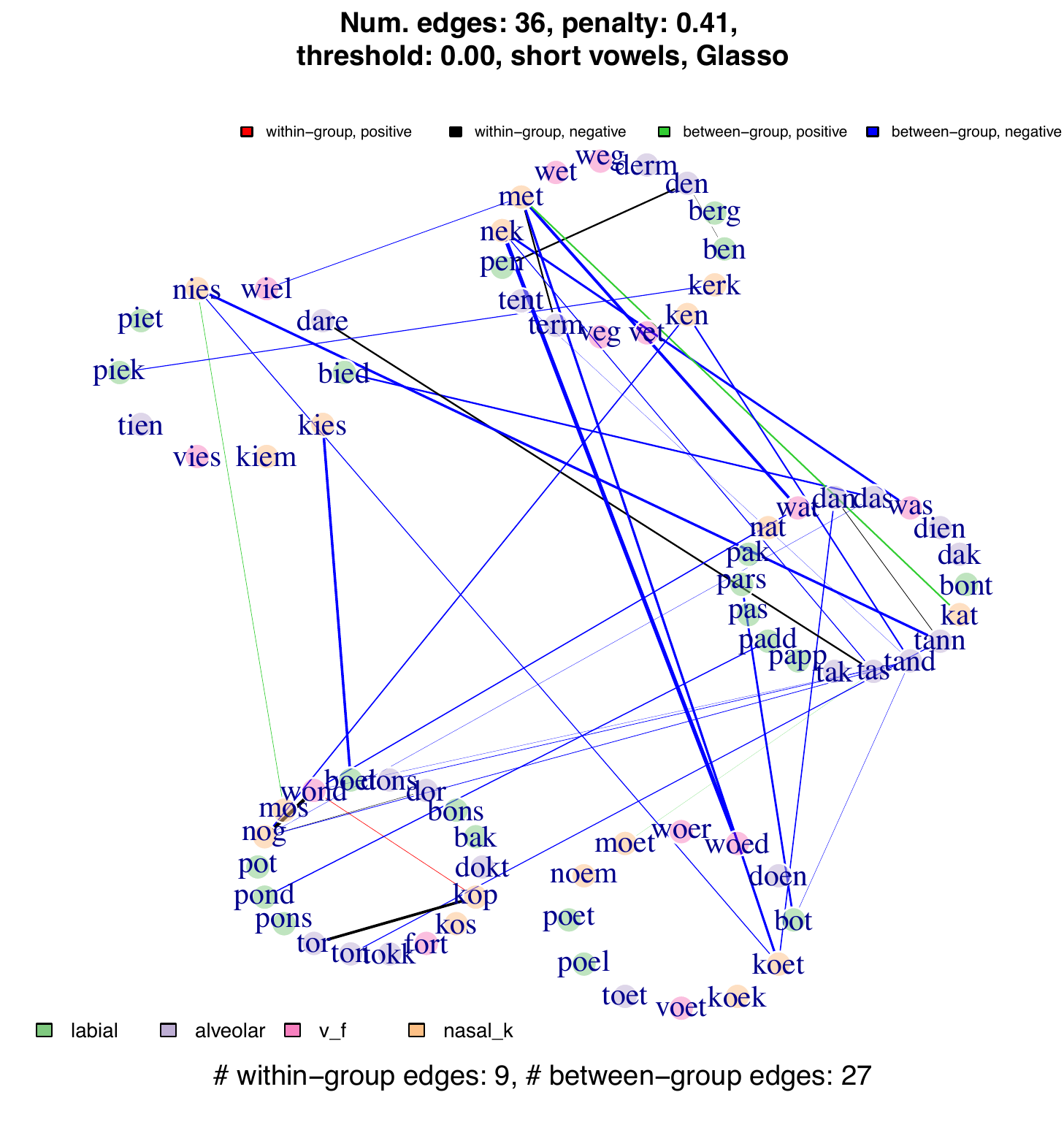}
\caption{Inverse covariance graph of words with short vowels, estimated using Glasso regression, organized by vowel.}
\label{OlympicVowelsWithLong_Glasso_shortVowel_pen0p41_thresh0}
\end{figure} 

\begin{figure}[h]
\includegraphics[width=\textwidth]{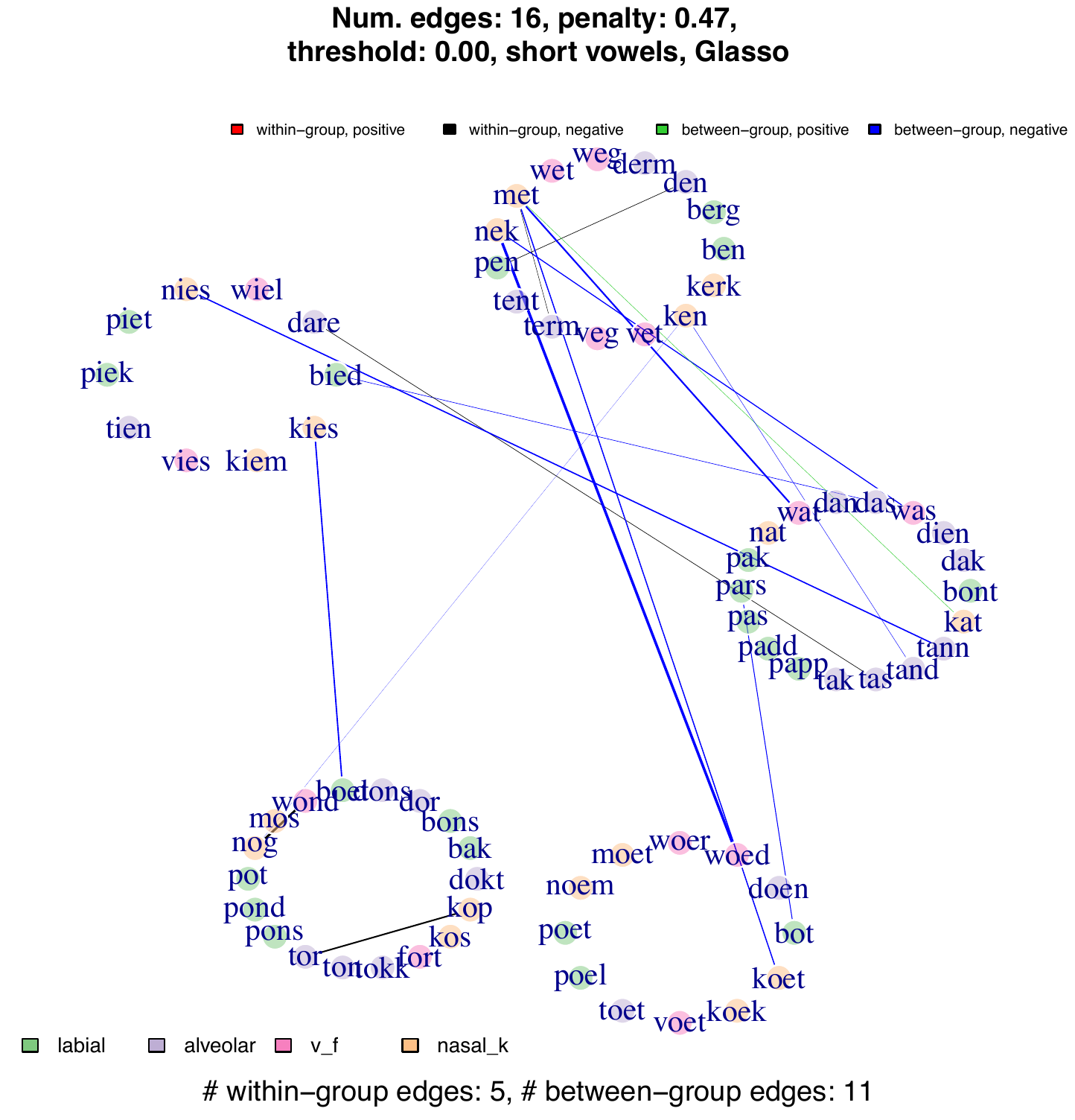}
\caption{Inverse covariance graph of words with short vowels, estimated using Glasso regression, organized by vowel.}
\label{OlympicVowelsWithLong_Glasso_shortVowel_pen0p47_thresh0}
\end{figure} 

\begin{figure}[h] 
\includegraphics[width=\textwidth]{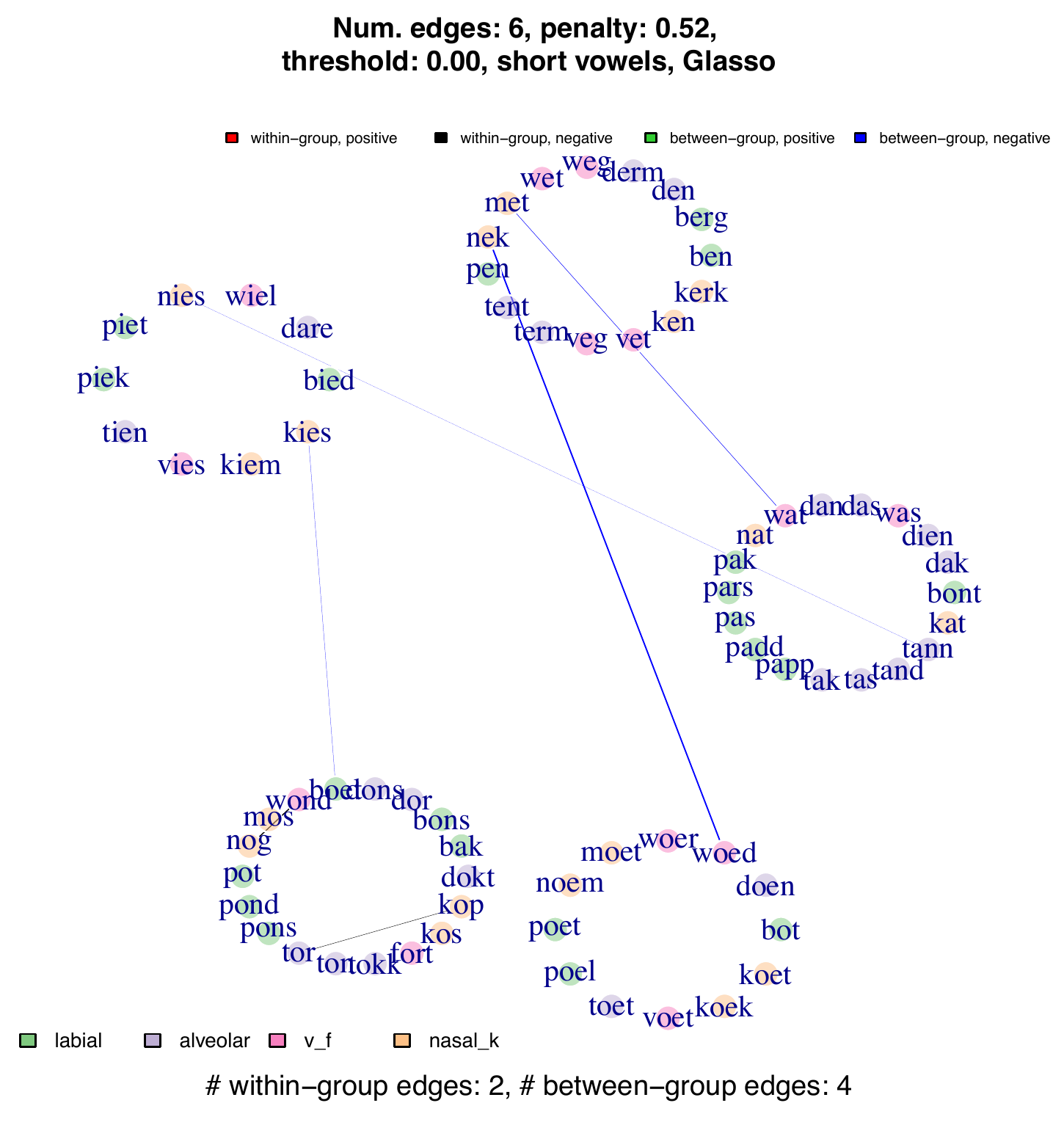}
\caption{Inverse covariance graph of words with short vowels, estimated using Glasso regression, organized by vowel.}
\label{OlympicVowelsWithLong_Glasso_shortVowel_pen0p52_thresh0}
\end{figure}

\clearpage
\subsection{Comparing Glasso and nodewise regression graphs for pairs of word groups} \label{sec:BicyclePlotsAllWordsSmallerPen}
We display inverse correlation graphs between each pair of word groups (labial, alveolar, nasal, and vf).  Glasso and nodewise regression were run on all the words; in the following figures, we visualize subgraphs of the full graph.  The line type indicates whether the edge appears in both the Glasso and nodewise regression graphs or in just one of the two.  Both methods are run with a penalty of $0.32$ and threshold of $0.16$.  We see that the edges are similar between the methods, but with more edges for Glasso than nodewise regression.  In Section \ref{bicyclePlotsAppendix} of the Appendix, we display analogous plots with a penalty of $0.26$ and threshold $0.08$.  

\begin{figure}[h!]
\includegraphics[width=\textwidth]{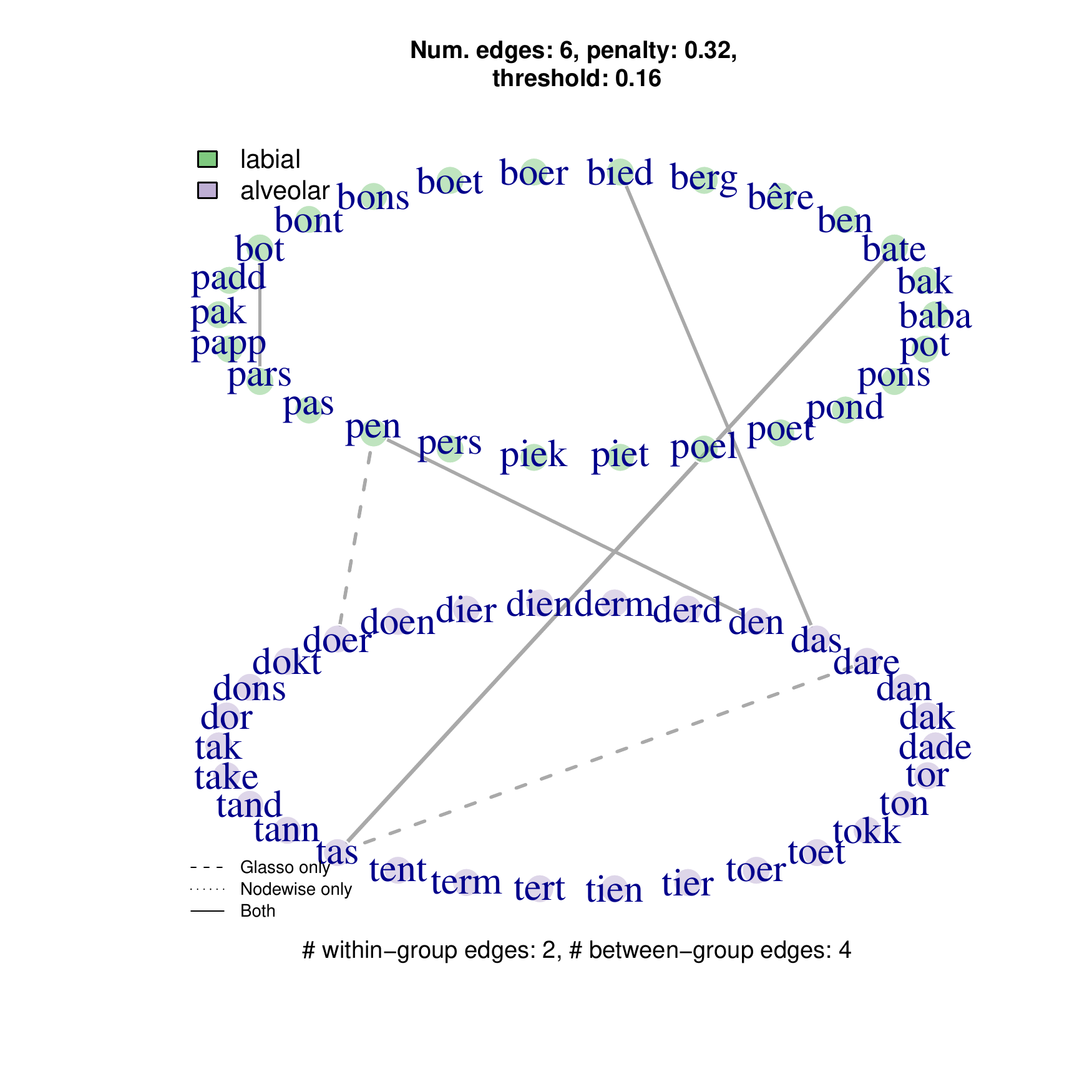} \caption{Inverse covariance graph of labial and alveolar words.  This graph displays a subgraph of a graph for all $93$ words, estimated using Glasso and nodewise regression with thresholding.}  \label{twoCircles_labial_alveolar_fromAllWords_pen0p32_thresh0p16_compare}
\end{figure} 
\begin{figure}[h!]
\includegraphics[width=\textwidth]{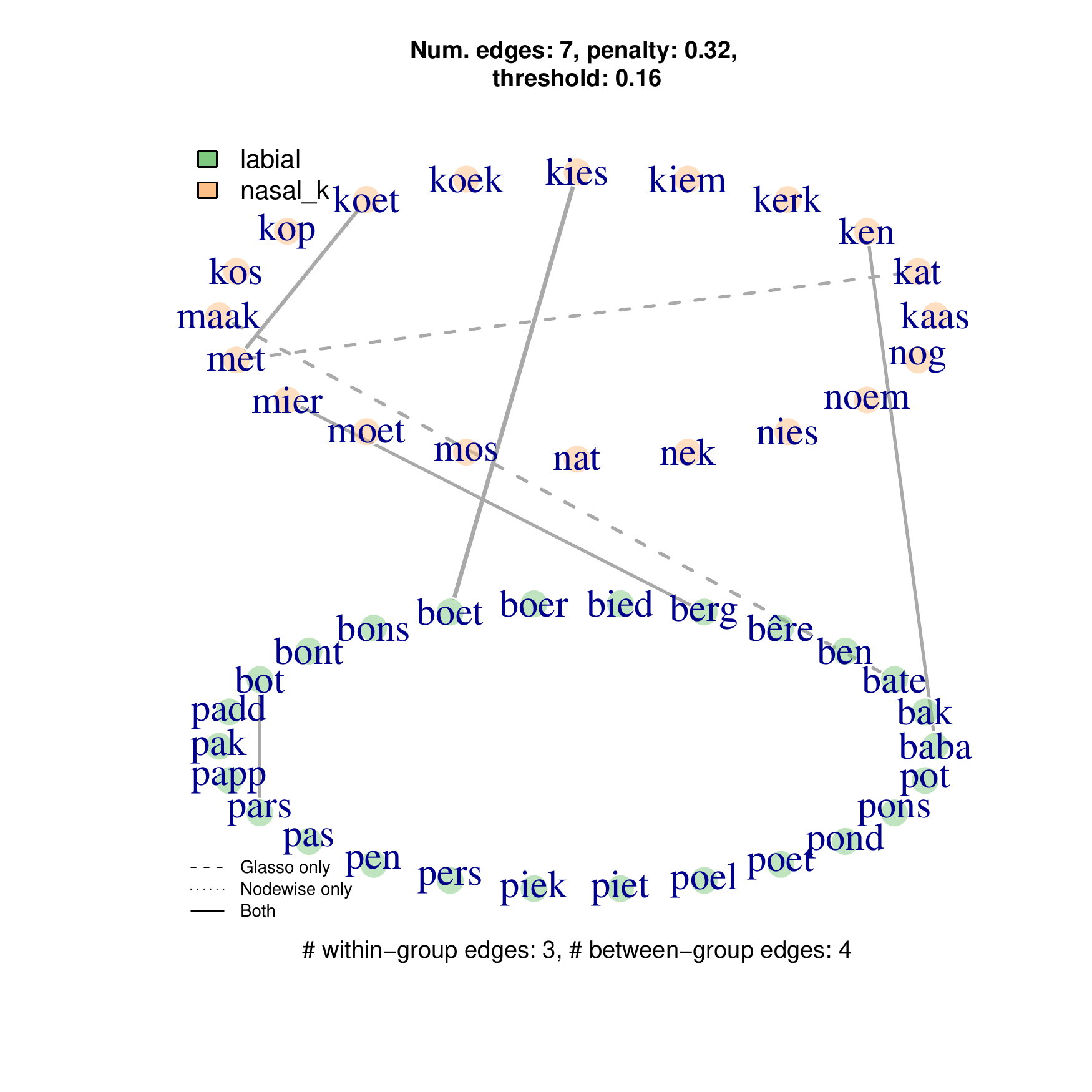} \caption{Inverse covariance graph of labial and nasal words.  This graph displays a subgraph of a graph for all $93$ words, estimated using Glasso and nodewise regression with thresholding.}  \label{twoCircles_labial_nasal_fromAllWords_pen0p32_thresh0p16_compare}
\end{figure} 
\begin{figure}[h!]
\includegraphics[width=\textwidth]{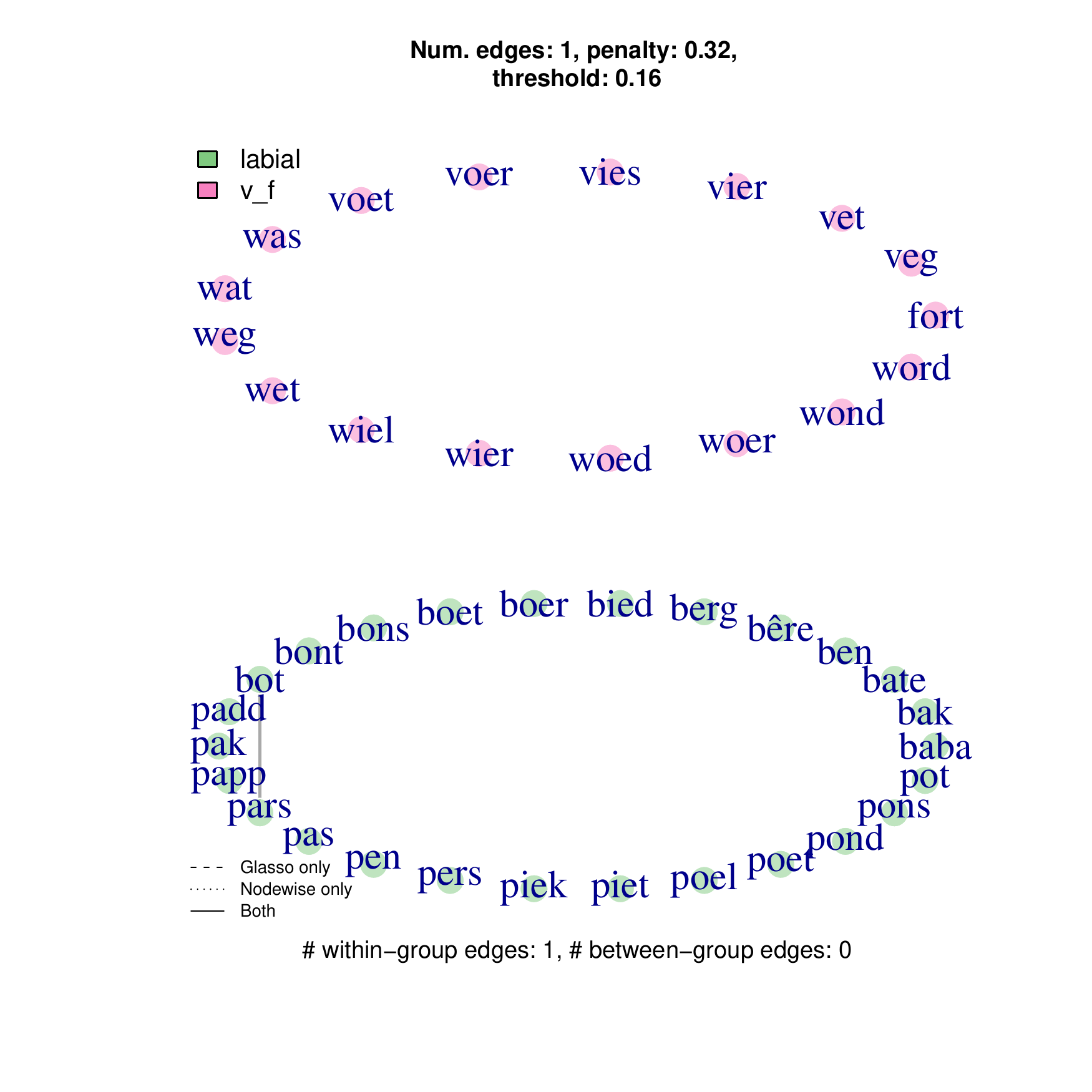} \caption{Inverse covariance graph of labial and vf words.  This graph displays a subgraph of a graph for all $93$ words, estimated using Glasso and nodewise regression with thresholding.}  \label{twoCircles_labial_vf_fromAllWords_pen0p32_thresh0p16_compare}
\end{figure} 
\begin{figure}[h!]
\includegraphics[width=\textwidth]{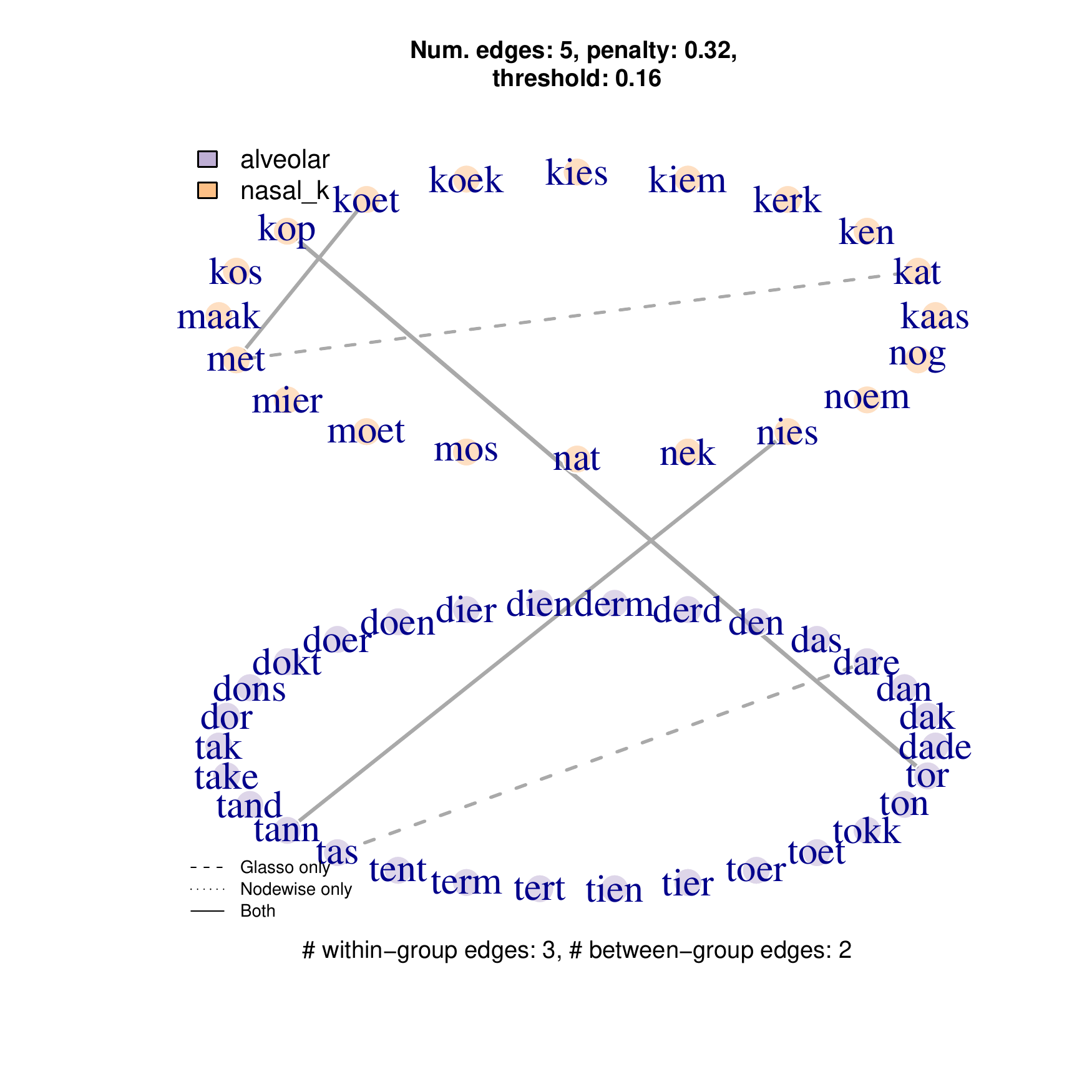} \caption{Inverse covariance graph of alveolar and nasal words.  This graph displays a subgraph of a graph for all $93$ words, estimated using Glasso and nodewise regression with thresholding.}  \label{twoCircles_alveolar_nasal_fromAllWords_pen0p32_thresh0p16_compare}
\end{figure} 
\begin{figure}[h!]
\includegraphics[width=\textwidth]{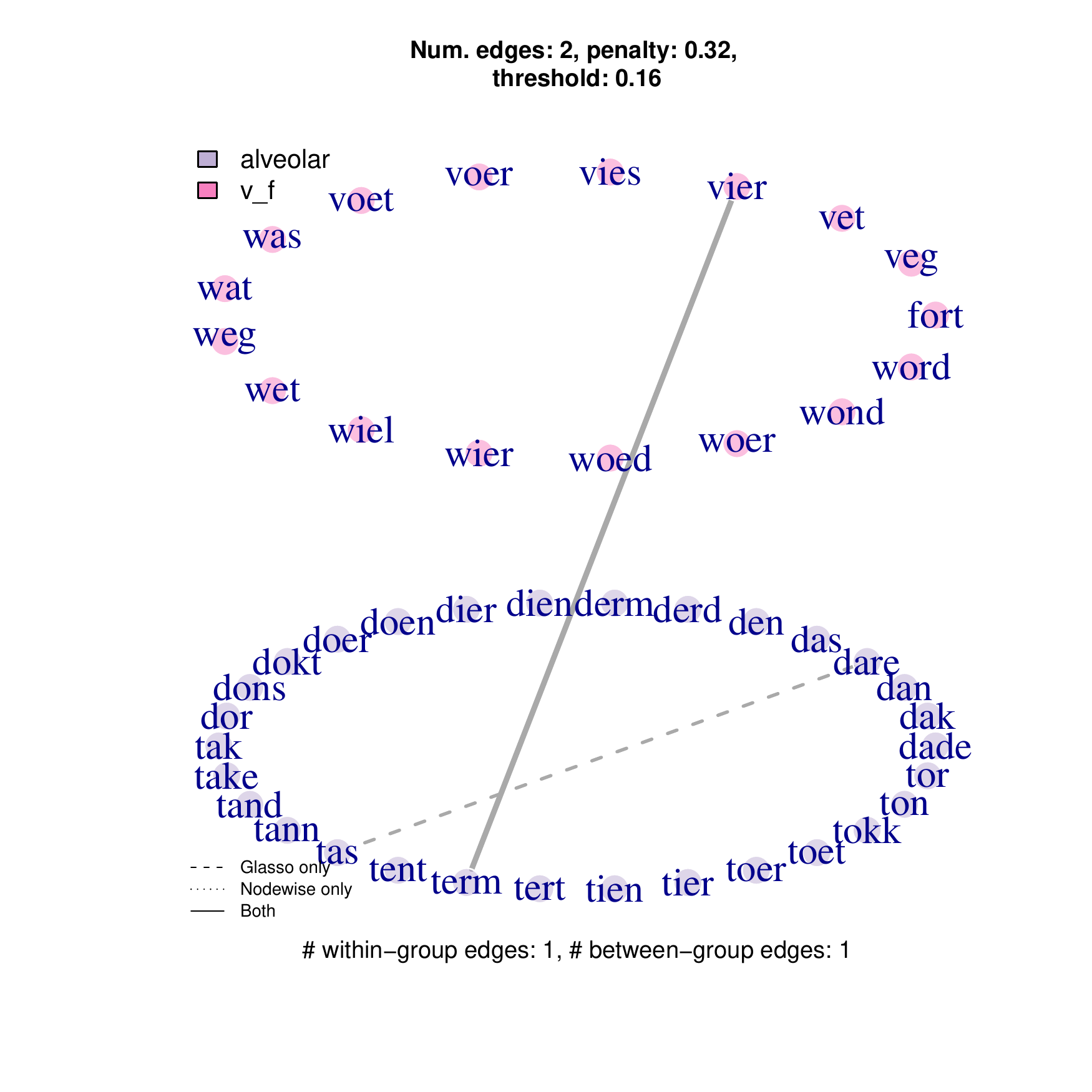} \caption{Inverse covariance graph of alveolar and vf words.  This graph displays a subgraph of a graph for all $93$ words, estimated using Glasso and nodewise regression with thresholding.}  \label{twoCircles_alveolar_vf_fromAllWords_pen0p32_thresh0p16_compare}
\end{figure} 
\begin{figure}[h!]
\includegraphics[width=\textwidth]{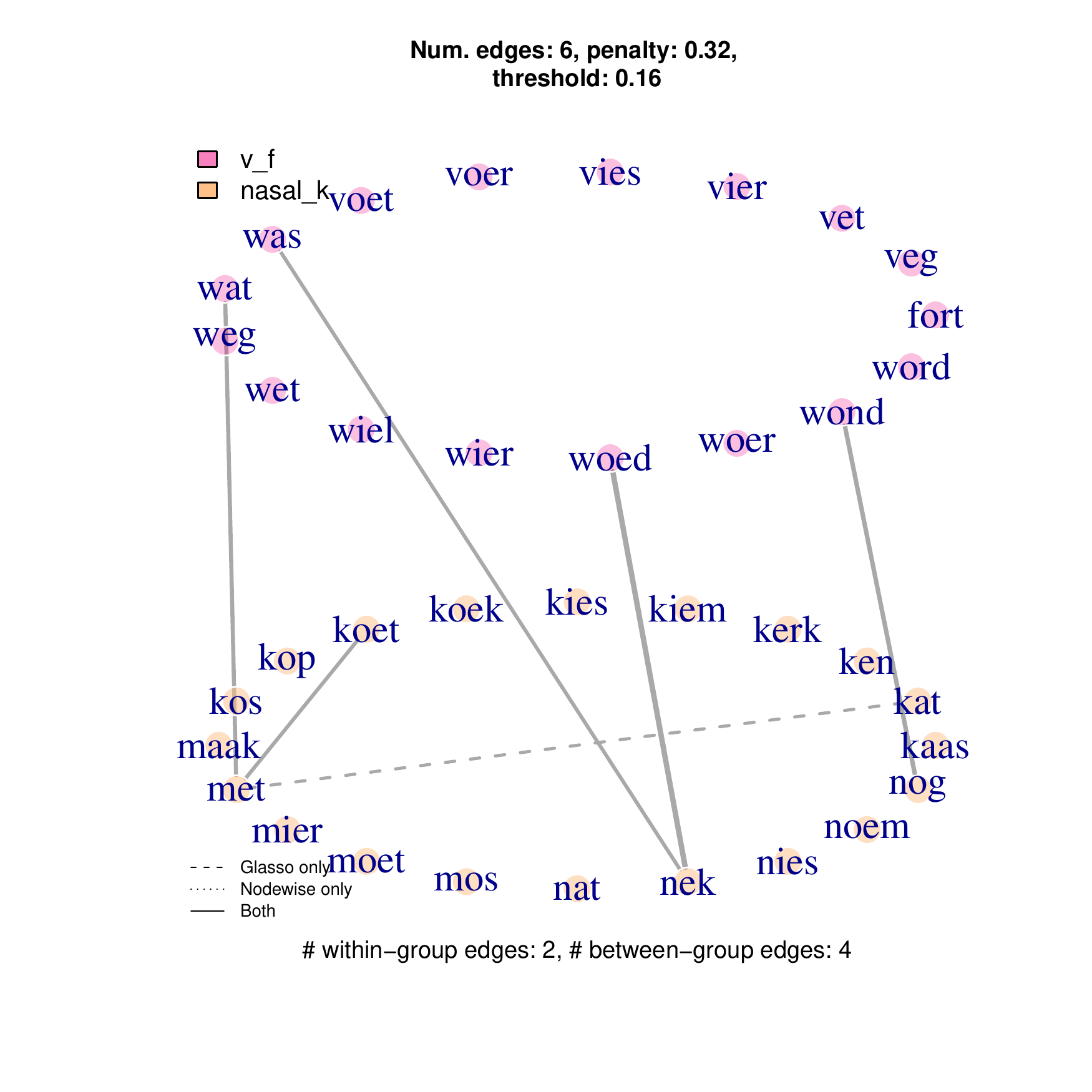} \caption{Inverse covariance graph of nasal and vf.  This graph displays a subgraph of a graph for all $93$ words, estimated using Glasso and nodewise regression with thresholding.}  \label{twoCircles_nasal_vf_fromAllWords_pen0p32_thresh0p16_compare}
\end{figure}

\clearpage
\subsection{Edge graphs comparing Glasso and nodewise regression, for each pair of word groups (labial, alveolar, nasal, vf)} \label{bicyclePlotsAppendix}

\begin{figure}[h!]
\includegraphics[width=\textwidth]{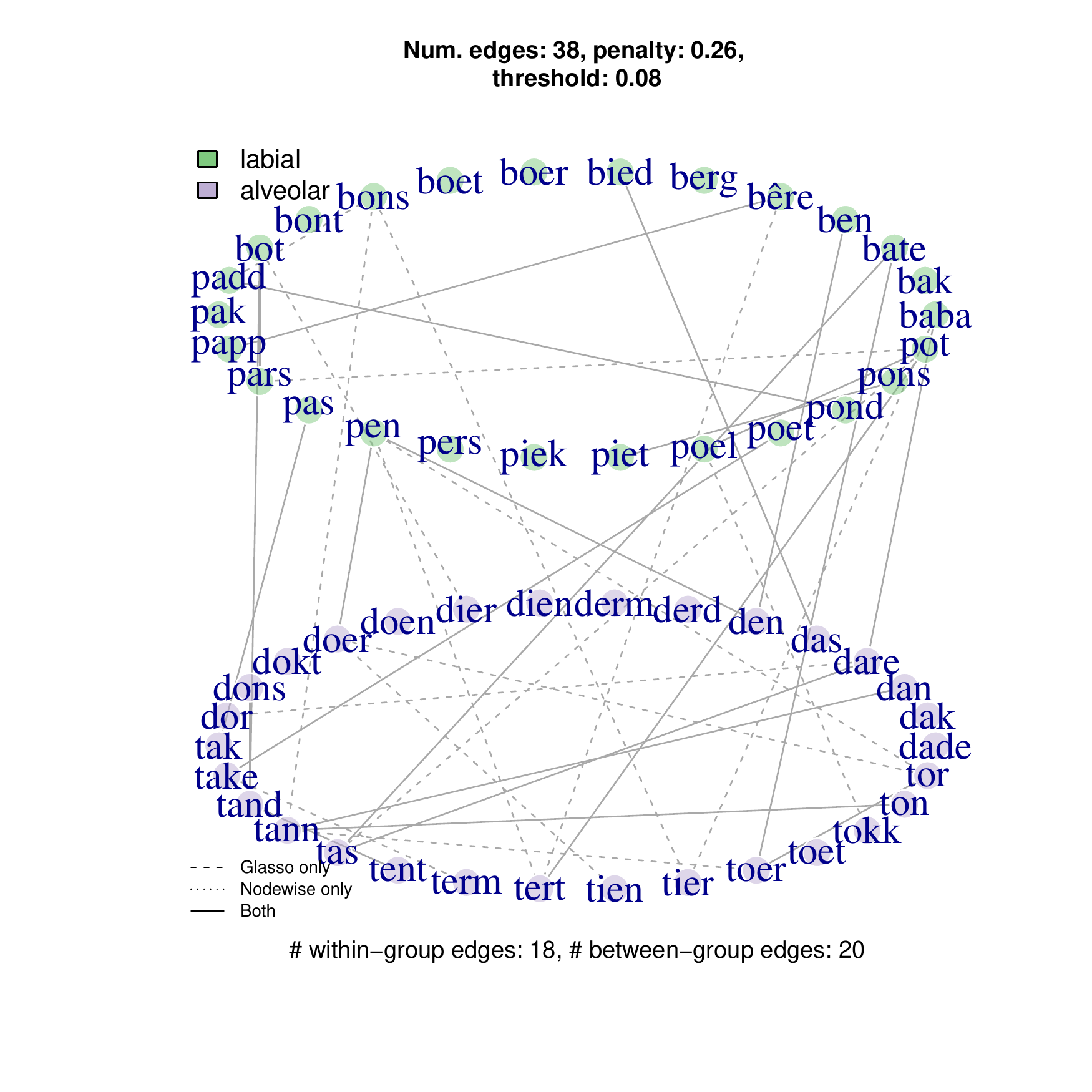} \caption{Inverse covariance graph of labial and alveolar words.  This graph displays a subgraph of a graph for all $93$ words, estimated using Glasso and nodewise regression with thresholding.}  \label{twoCircles_labial_alveolar_fromAllWords_pen0p26_thresh0p08_compare}
\end{figure} 
\begin{figure}[h!]
\includegraphics[width=\textwidth]{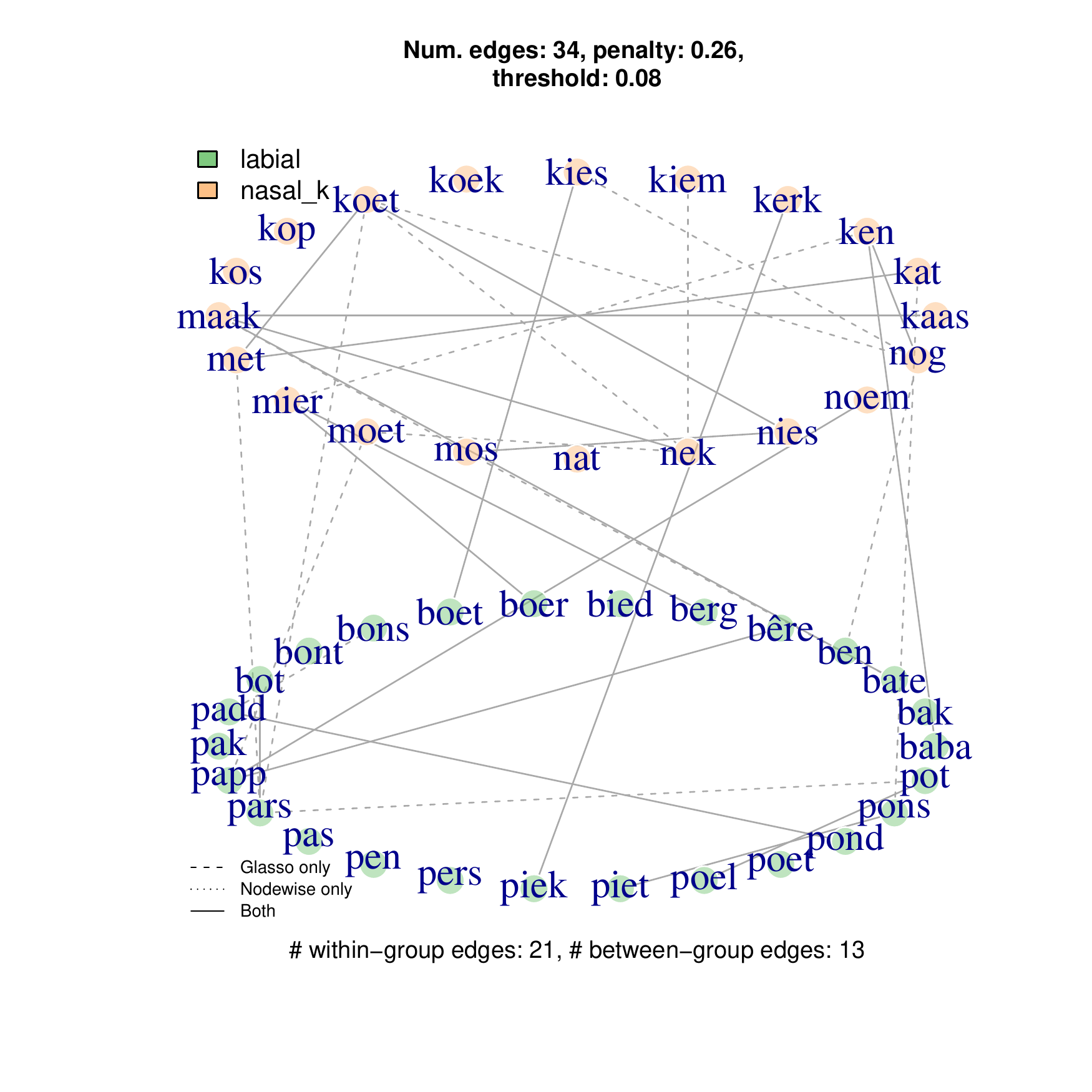} \caption{Inverse covariance graph of labial and nasal words.  This graph displays a subgraph of a graph for all $93$ words, estimated using Glasso and nodewise regression with thresholding.}  \label{twoCircles_labial_nasal_fromAllWords_pen0p26_thresh0p08_compare}
\end{figure} 
\begin{figure}[h!]
\includegraphics[width=\textwidth]{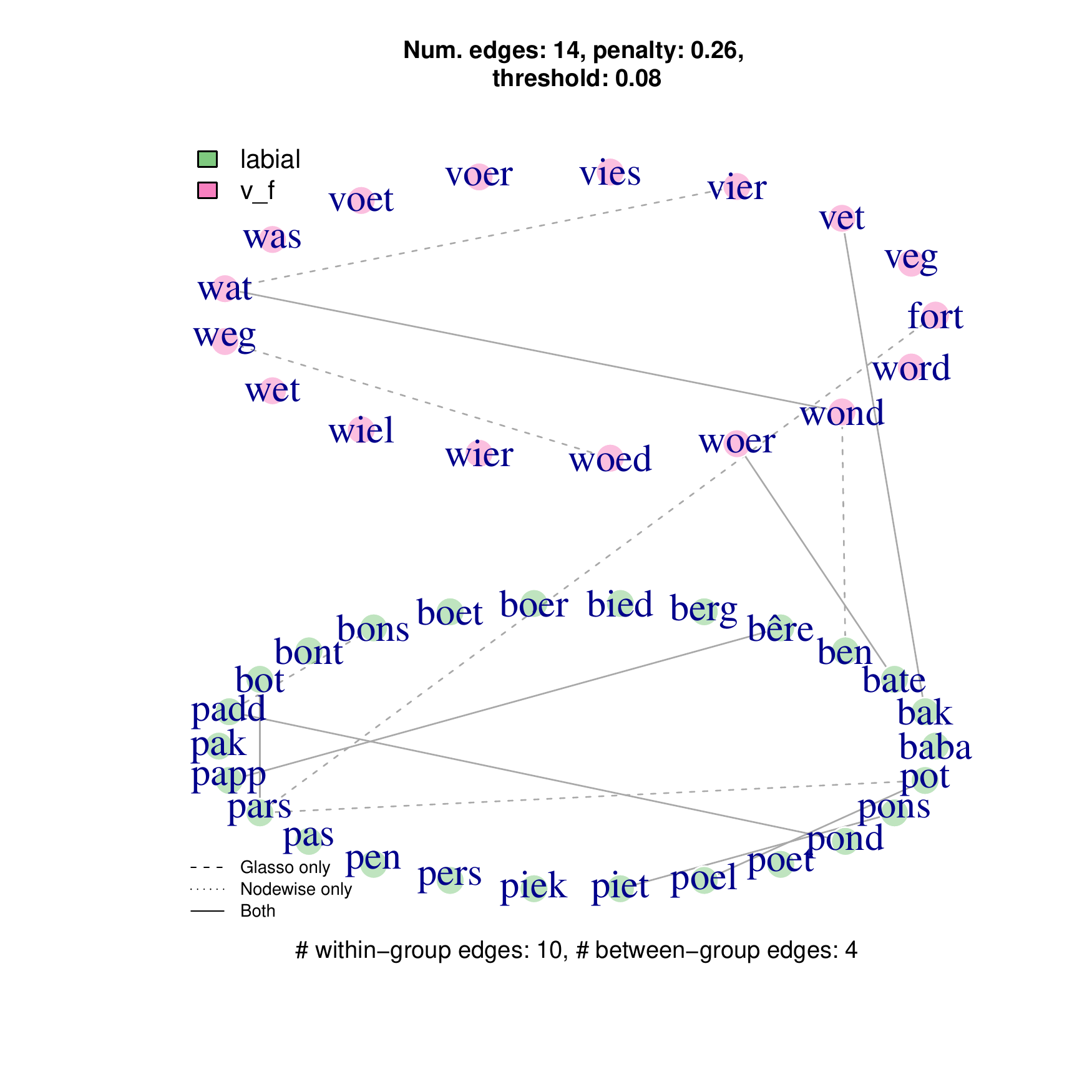} \caption{Inverse covariance graph of labial and vf words.  This graph displays a subgraph of a graph for all $93$ words, estimated using Glasso and nodewise regression with thresholding.}  \label{twoCircles_labial_vf_fromAllWords_pen0p26_thresh0p08_compare}
\end{figure} 
\begin{figure}[h!]
\includegraphics[width=\textwidth]{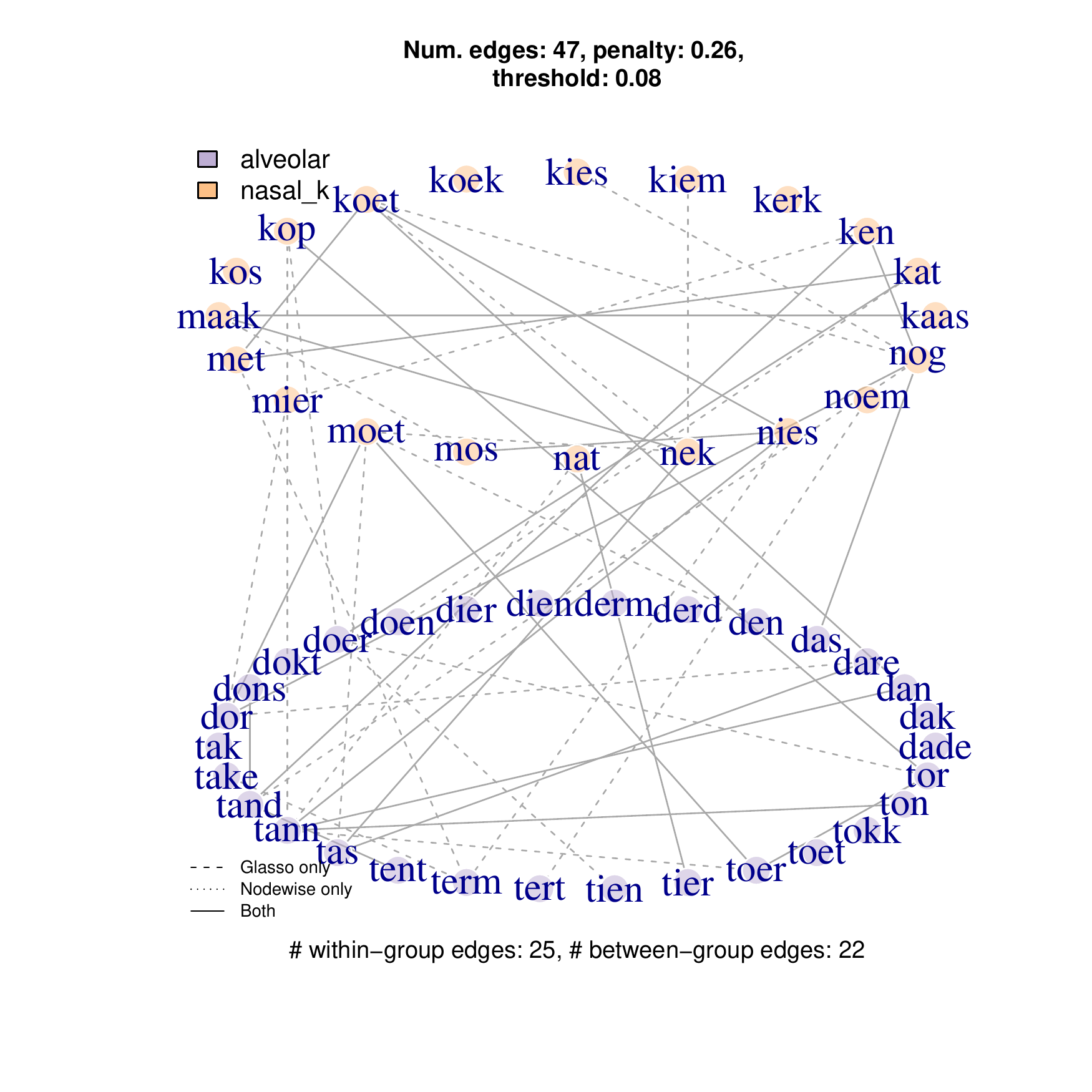} \caption{Inverse covariance graph of alveolar and nasal words.  This graph displays a subgraph of a graph for all $93$ words, estimated using Glasso and nodewise regression with thresholding.}  \label{twoCircles_alveolar_nasal_fromAllWords_pen0p26_thresh0p08_compare}
\end{figure} 
\begin{figure}[h!]
\includegraphics[width=\textwidth]{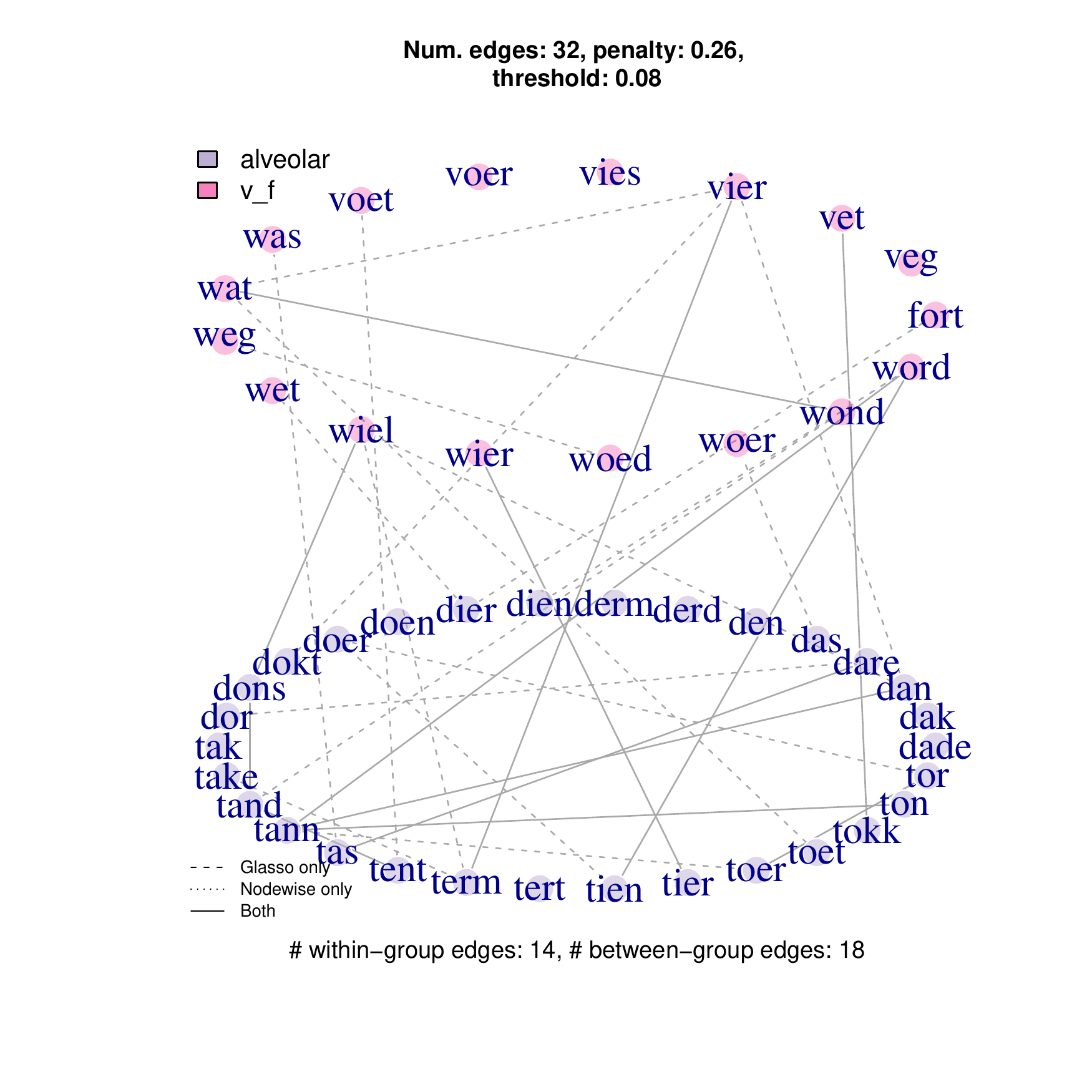} \caption{Inverse covariance graph of alveolar and vf words.  This graph displays a subgraph of a graph for all $93$ words, estimated using Glasso and nodewise regression with thresholding.}  \label{twoCircles_alveolar_vf_fromAllWords_pen0p26_thresh0p08_compare}
\end{figure} 
\begin{figure}[h!]
\includegraphics[width=\textwidth]{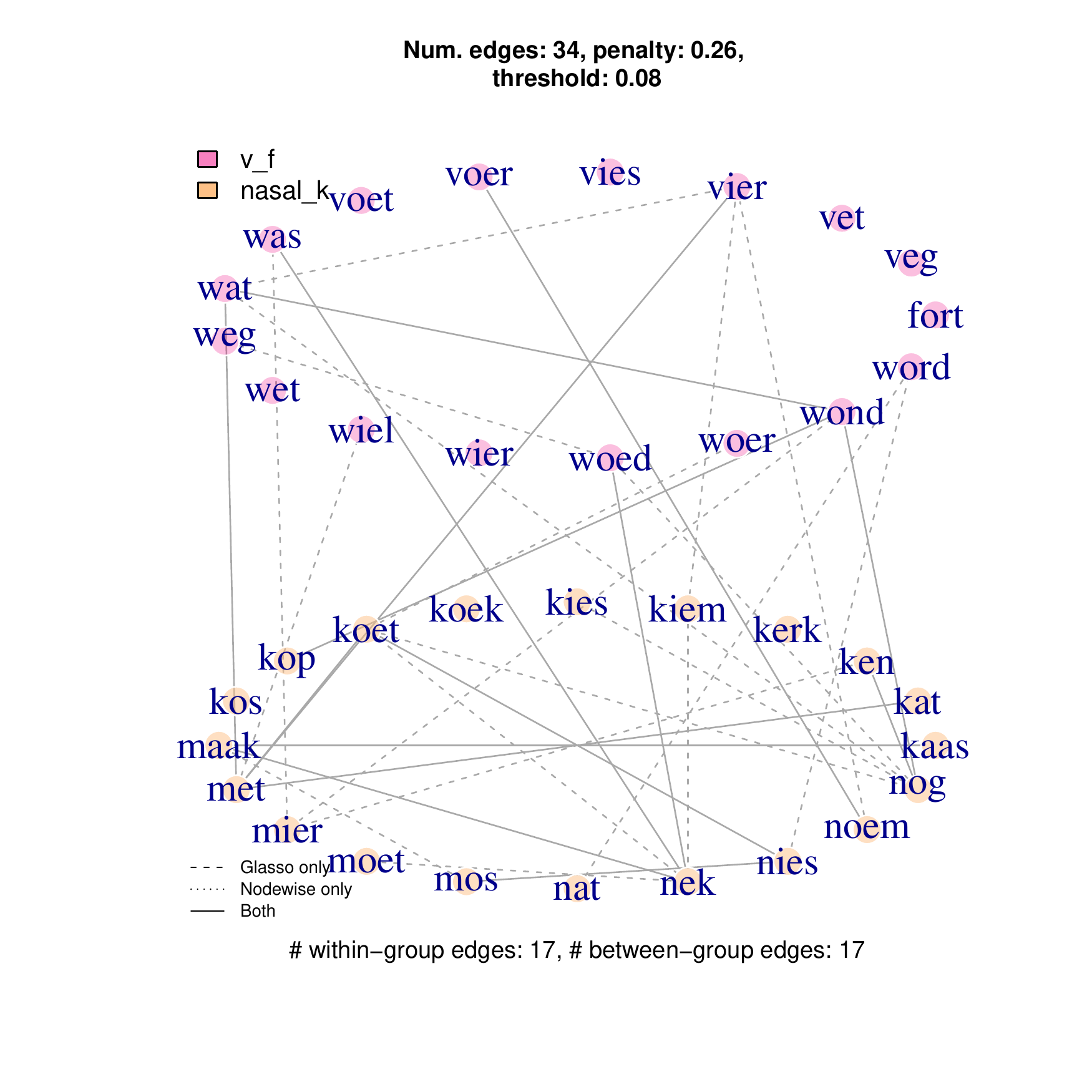} \caption{Inverse covariance graph of nasal and vf words.  This graph displays a subgraph of a graph for all $93$ words, estimated using Glasso and nodewise regression with thresholding.}  \label{twoCircles_nasal_vf_fromAllWords_pen0p26_thresh0p08_compare}
\end{figure}

\clearpage
\section{Time-time inverse covariance modeling}

\subsection{Time-time and word-word correlation and covariance}

\begin{figure}[h!]
\includegraphics[width=\textwidth]{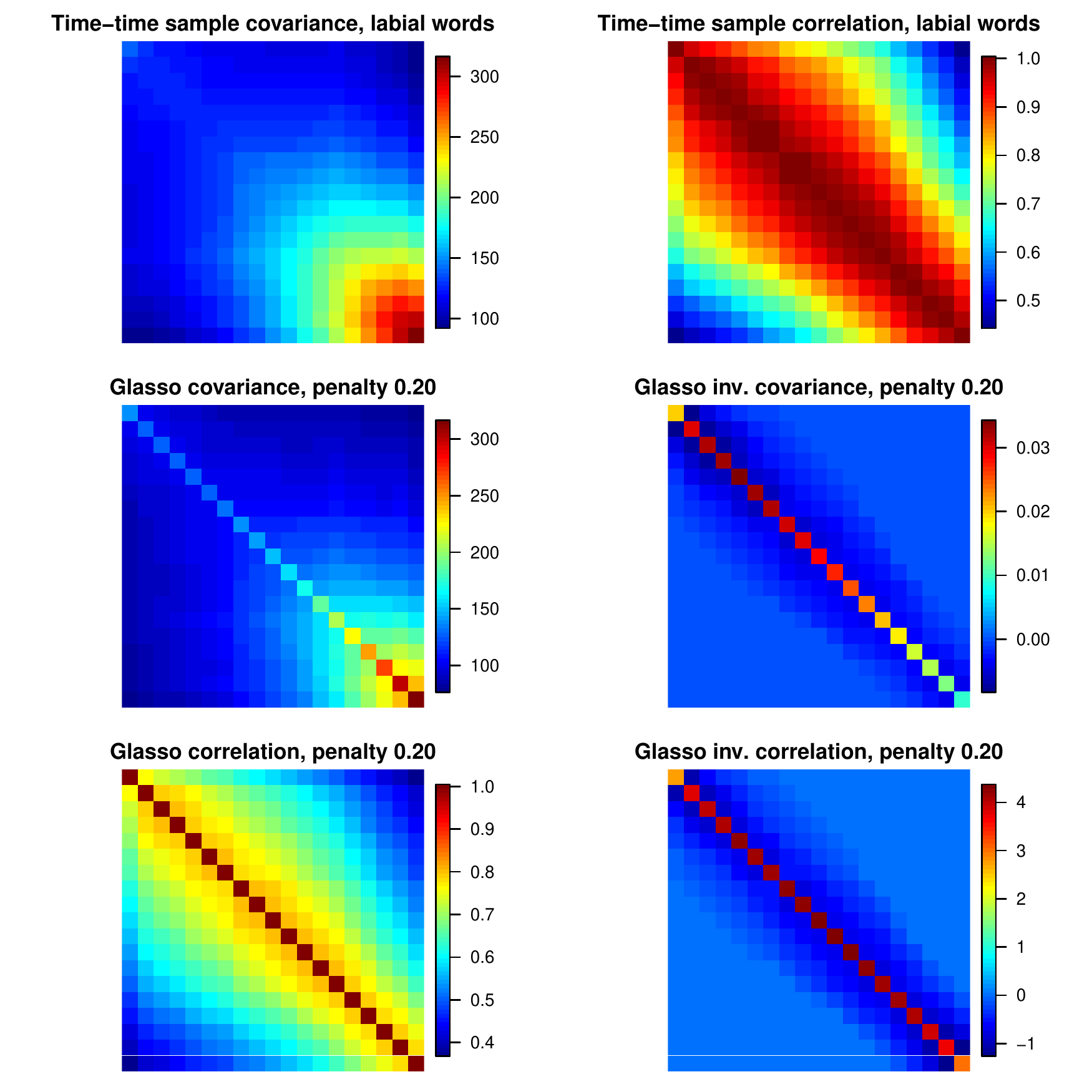} \caption{Time-time sample covariance (top left), sample correlation (top right), Glasso covariance (middle left), Glasso inverse covariance (middle right), Glasso correlation (bottom left), and Glasso inverse correlation (bottom right),  for words beginning with a labial consonant.  The sample covariance is calculated as in \eqref{sampleCovTimeTrialResid}, and the Glasso penalty parameter is chosen as five times the value of  \eqref{glassoWordTimeTuningParam}.}  \label{labial_Glasso_heatmaps_time_theoryPenTimesFive}
\end{figure}

Figure \ref{labial_Glasso_heatmaps_time_theoryPenTimesFive} displays sample covariance, sample correlation, Glasso covariance, Glasso inverse covariance, Glasso correlation, and Glasso inverse correlation for the labial words.  Glasso is run using a penalty five times that of of the theoretical value.  In Section \ref{timeGlassoOutput} of the Supplement, analogous figures are shown for the other word groups.

\subsection{Comparison of time inverse covariance graphs for each pair of word groups}

For each pair of word groups, we compare the time-time inverse correlation graphs, by taking intersections and set differences.  We threshold each graph down to $70$ edges.  In each graph, nodes are connected to approximately five nearest neighbors on each side.  The time-time edges are similar among the word groups; that is, most of the nodes are in the intersections of the graphs.  This suggests that we can consider using a combined time-time inverse covariance matrix pooling over the words to decorrelate along the time axis, potentially improving the word-word covariance estimates, discussed in \cite{Zhou14a}.    

\begin{figure}[h!]
\includegraphics[width=\textwidth]{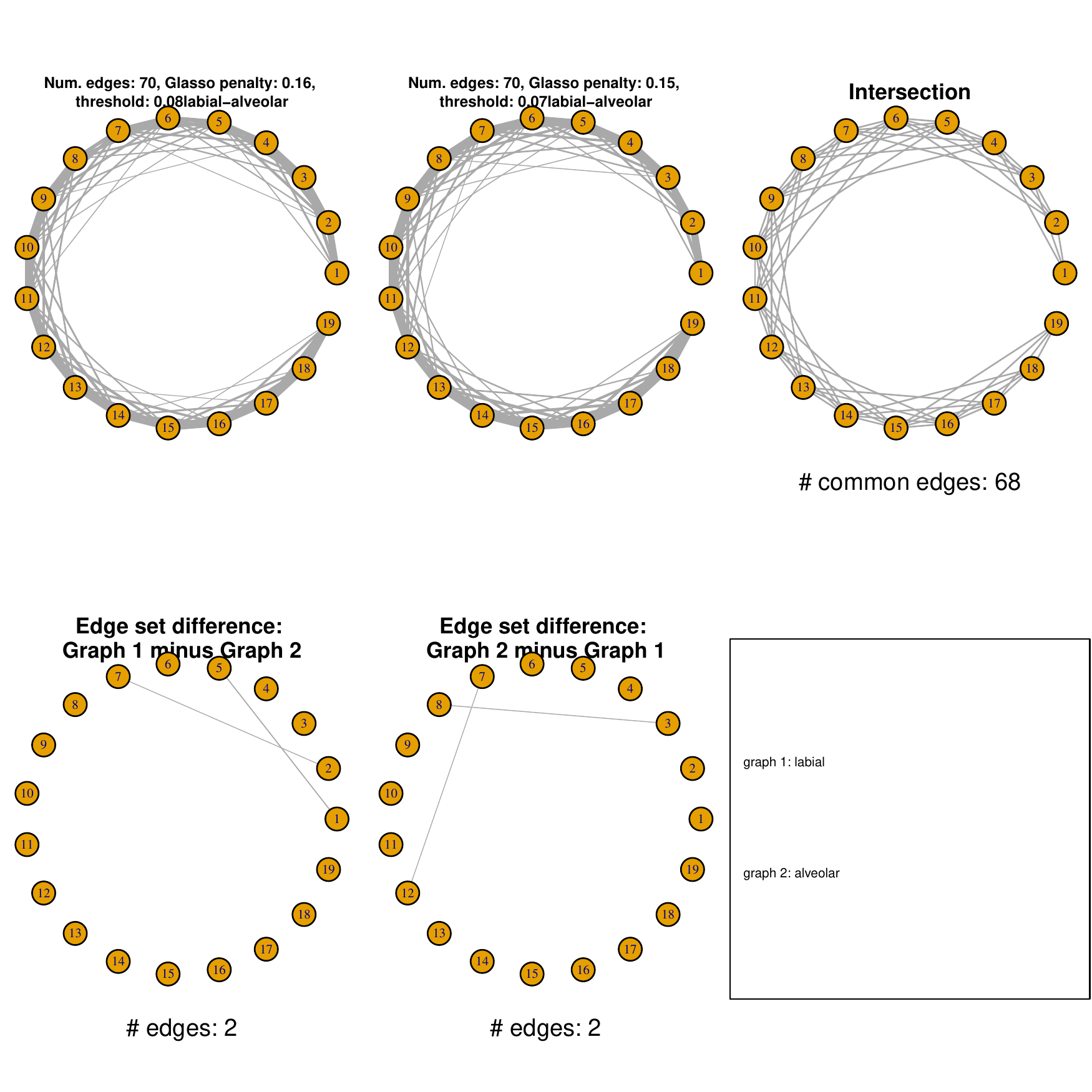} \caption{Time-time inverse covariance graphs for labial and alveolar words, as well as graph intersection and set differences.  The inverse correlation matrices are thresholded so that $70$ edges remain in each word group.}  \label{timeGraph_compareEdgesAmongWordGroups_labial_alveolar_70edges}
\end{figure} 
\begin{figure}[h!]
\includegraphics[width=\textwidth]{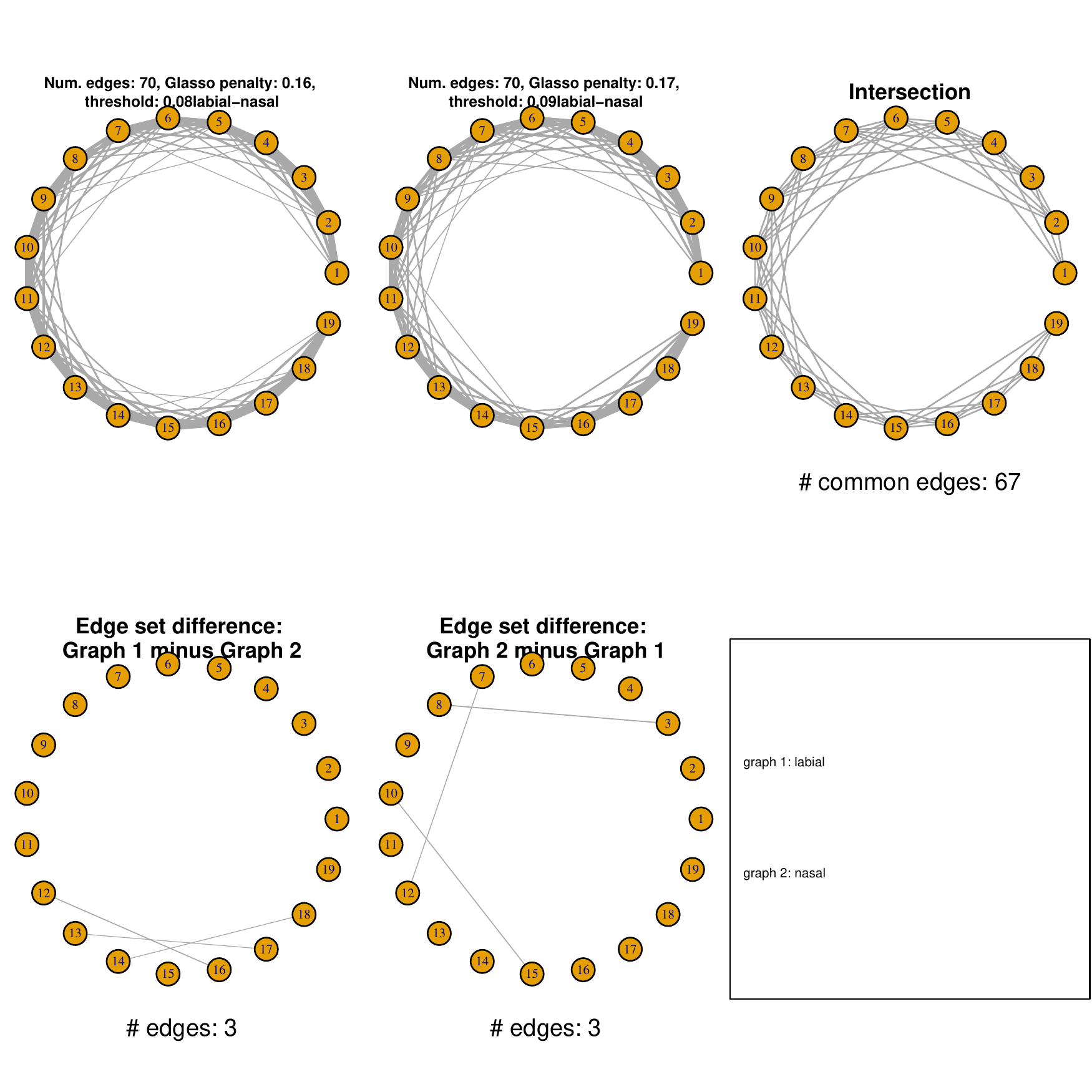} \caption{Time-time inverse covariance graphs for labial and nasal words, as well as graph intersection and set differences.}  \label{timeGraph_compareEdgesAmongWordGroups_labial_nasal_70edges}
\end{figure} 
\begin{figure}[h!]
\includegraphics[width=\textwidth]{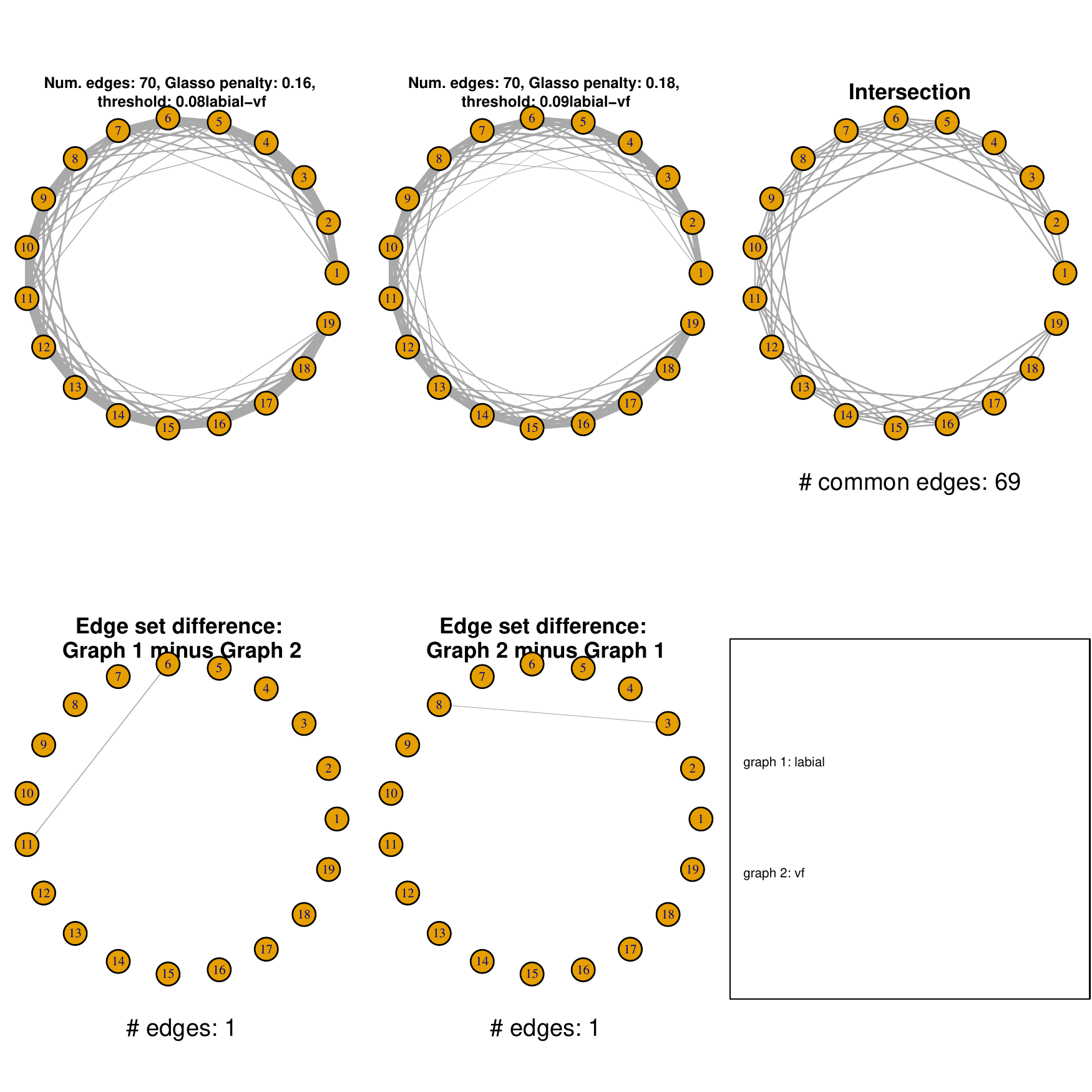} \caption{Time-time inverse covariance graphs for labial and vf words, as well as graph intersection and set differences.}  \label{timeGraph_compareEdgesAmongWordGroups_labial_vf_70edges}
\end{figure} 
\begin{figure}[h!]
\includegraphics[width=\textwidth]{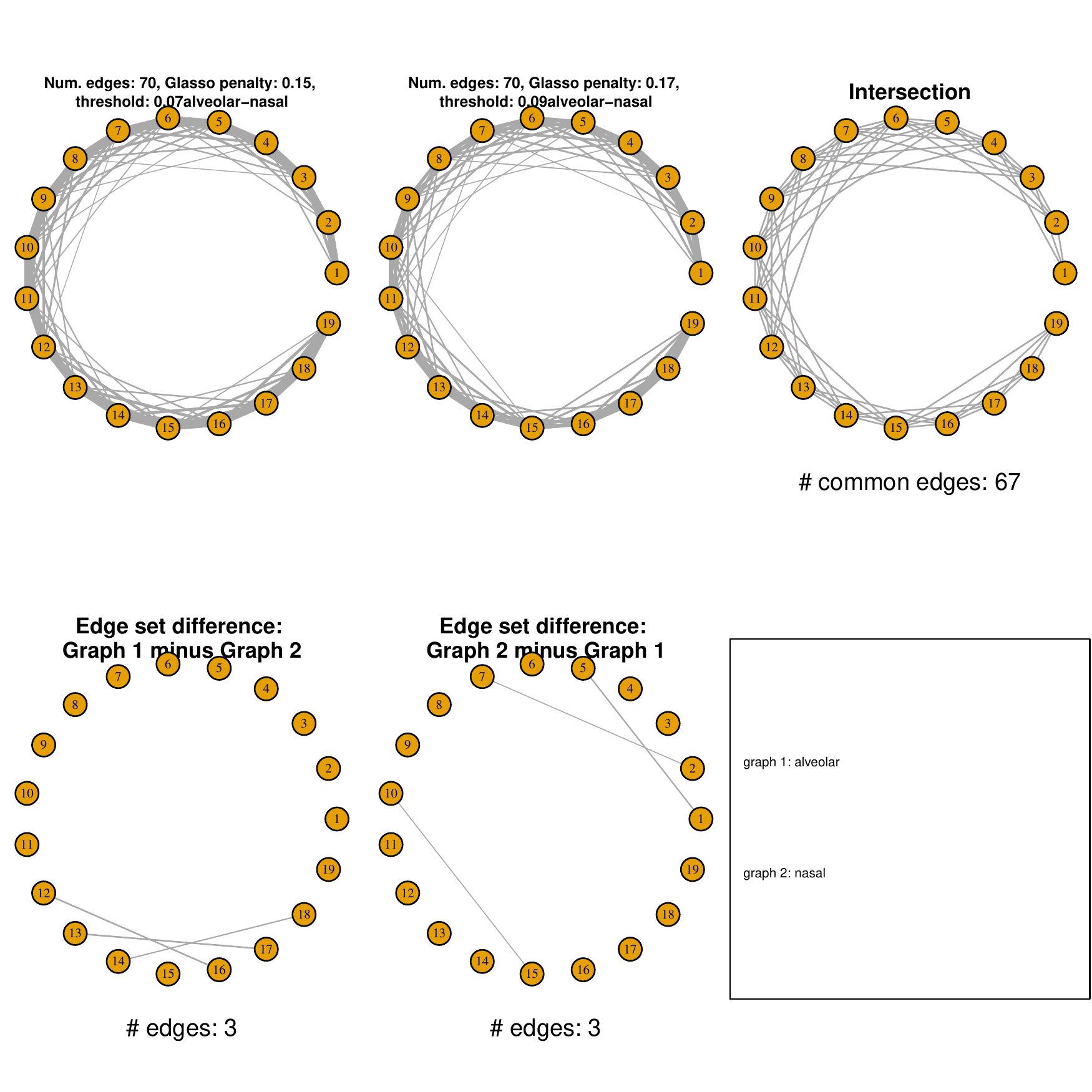} \caption{Time-time inverse covariance graphs for alveolar and nasal words, as well as graph intersection and set differences.}  \label{timeGraph_compareEdgesAmongWordGroups_alveolar_nasal_70edges}
\end{figure} 
\begin{figure}[h!]
\includegraphics[width=\textwidth]{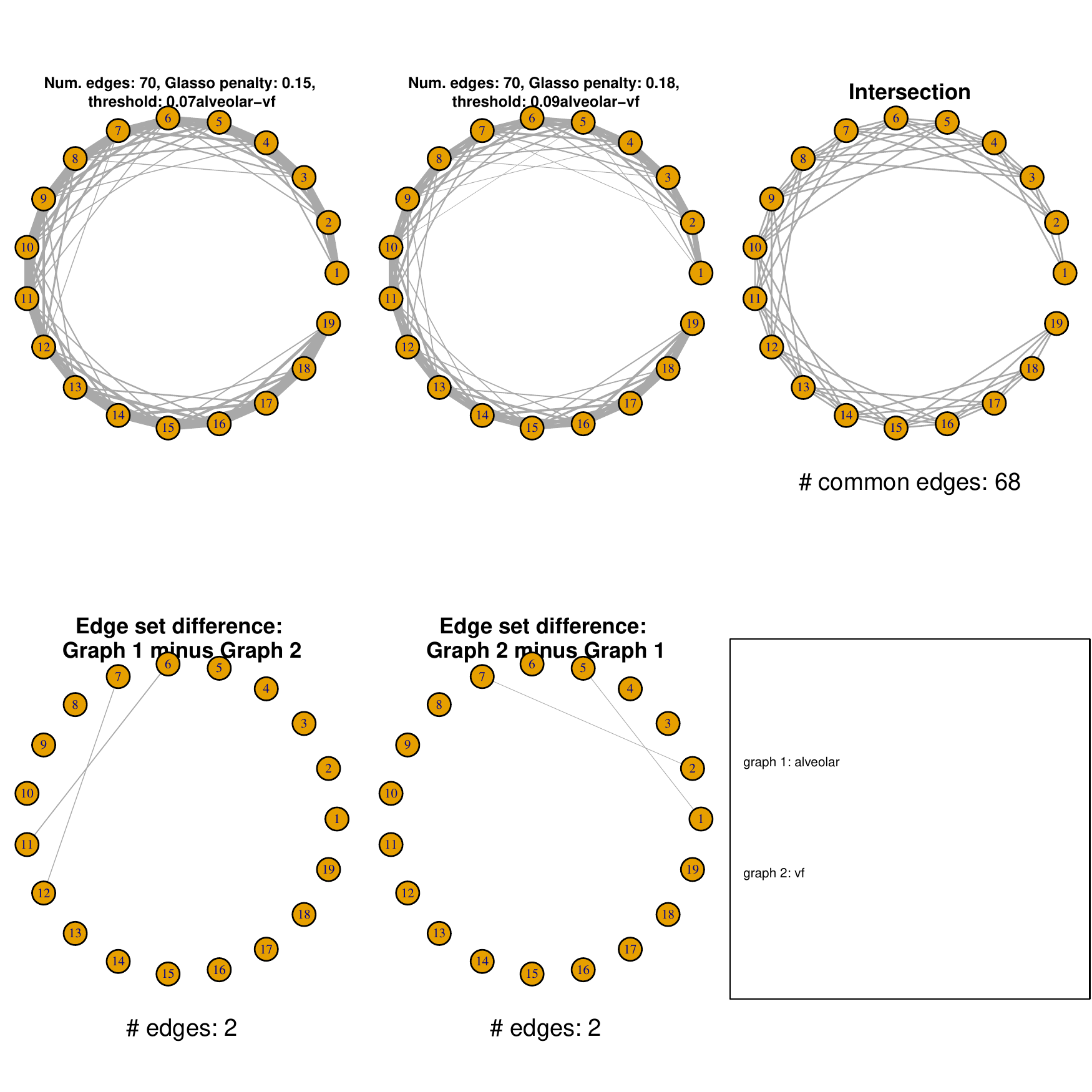} \caption{Time-time inverse covariance graphs for alveolar and vf words, as well as graph intersection and set differences.}  \label{timeGraph_compareEdgesAmongWordGroups_alveolar_vf_70edges}
\end{figure} 
\begin{figure}[h!]
\includegraphics[width=\textwidth]{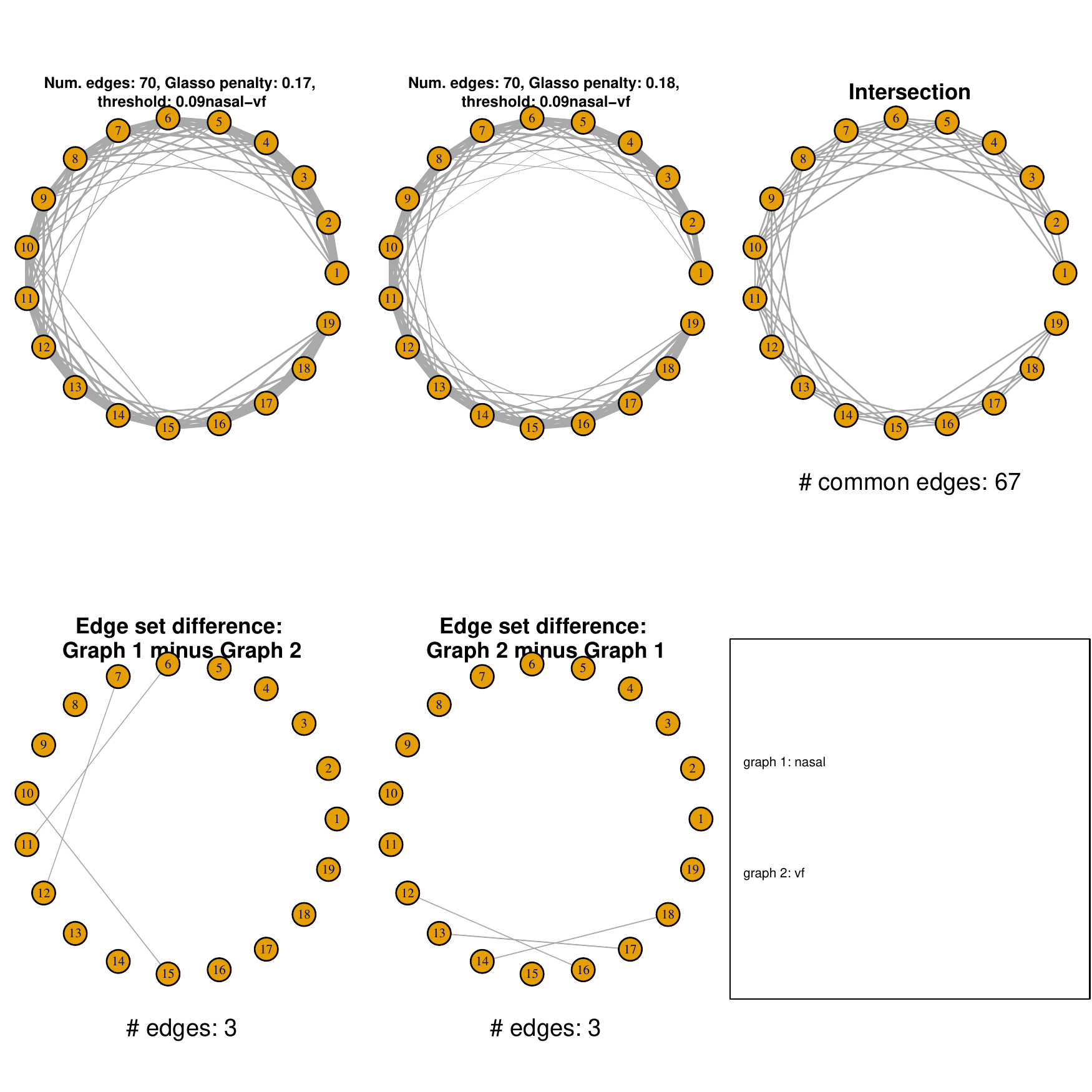} \caption{Time-time inverse covariance graphs for nasal and vf words, as well as graph intersection and set differences.}  \label{timeGraph_compareEdgesAmongWordGroups_nasal_vf_70edges}
\end{figure}

\clearpage
\subsection{Time-time covariance, correlation, inverse covariance, and inverse correlation} \label{timeGlassoOutput}

\begin{figure}[h!]
\includegraphics[width=\textwidth]{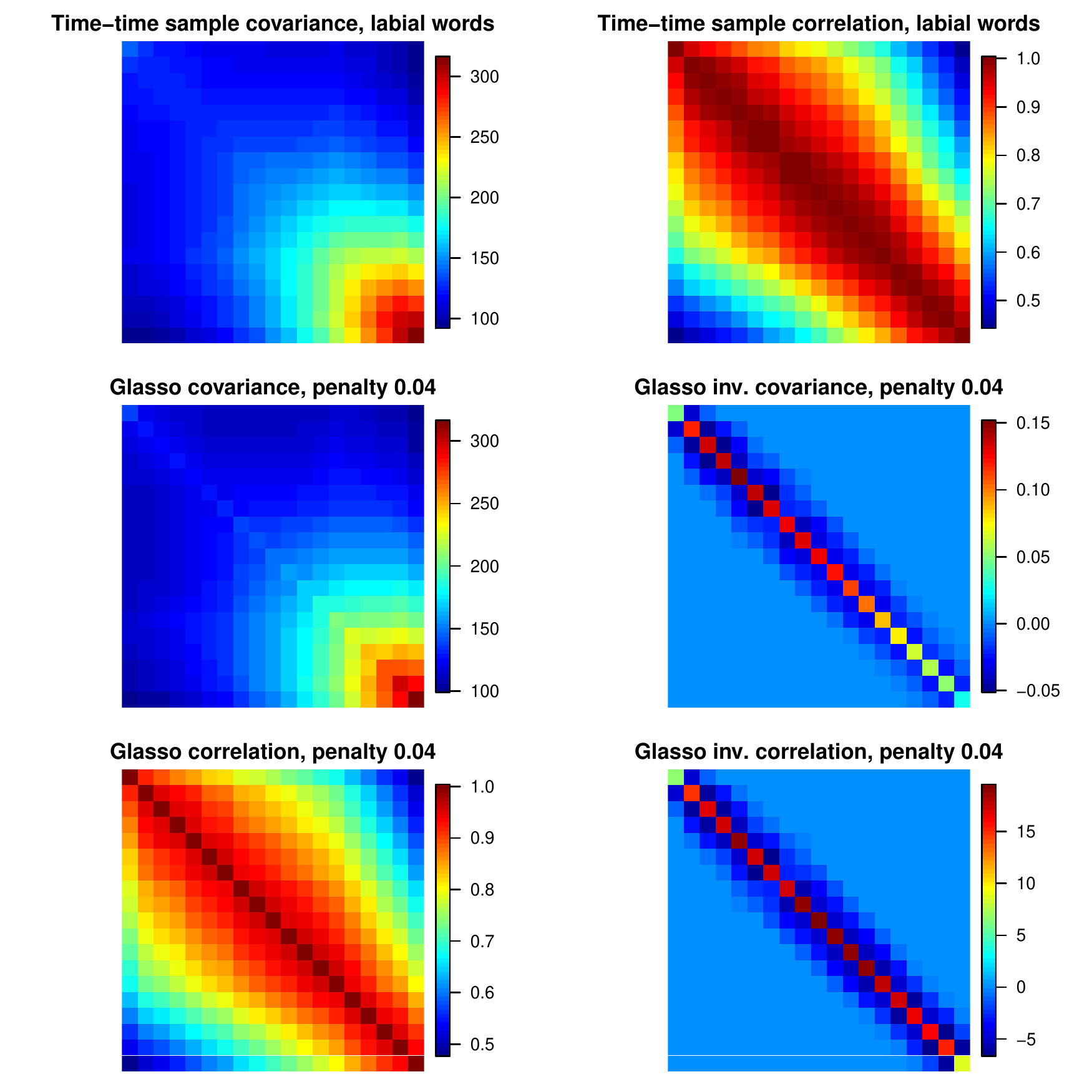} \caption{Time-time sample covariance (top left), sample correlation (top right), Glasso covariance (middle left), Glasso inverse covariance (middle right), Glasso correlation (bottom left), and Glasso inverse correlation (bottom right),  for words beginning with a labial consonant.  The sample covariance is calculated as in \eqref{sampleCovTimeTrialResid}, and the Glasso penalty parameter is chosen as in \eqref{glassoWordTimeTuningParam}.}  \label{labial_Glasso_heatmaps_time_theoryPen}
\end{figure}
\begin{figure}[h!]
\includegraphics[width=\textwidth]{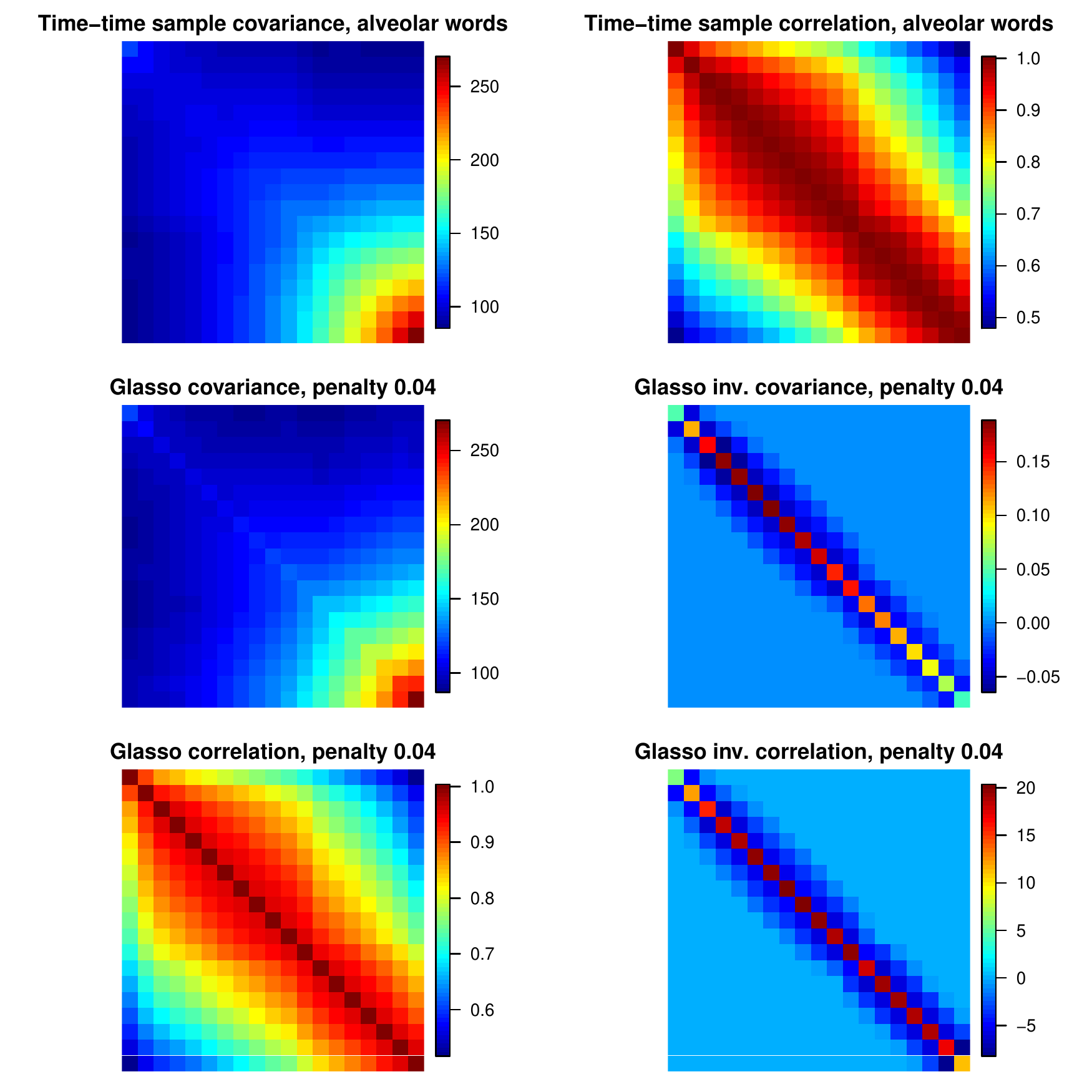} \caption{Time-time sample covariance (top left), sample correlation (top right), Glasso covariance (middle left), Glasso inverse covariance (middle right), Glasso correlation (bottom left), and Glasso inverse correlation (bottom right),  for words beginning with an alveolar consonant.  The sample covariance is calculated as in \eqref{sampleCovTimeTrialResid}, and the Glasso penalty parameter is chosen as in \eqref{glassoWordTimeTuningParam}.}  \label{alveolar_Glasso_heatmaps_time_theoryPen}
\end{figure}
\begin{figure}[h!]
\includegraphics[width=\textwidth]{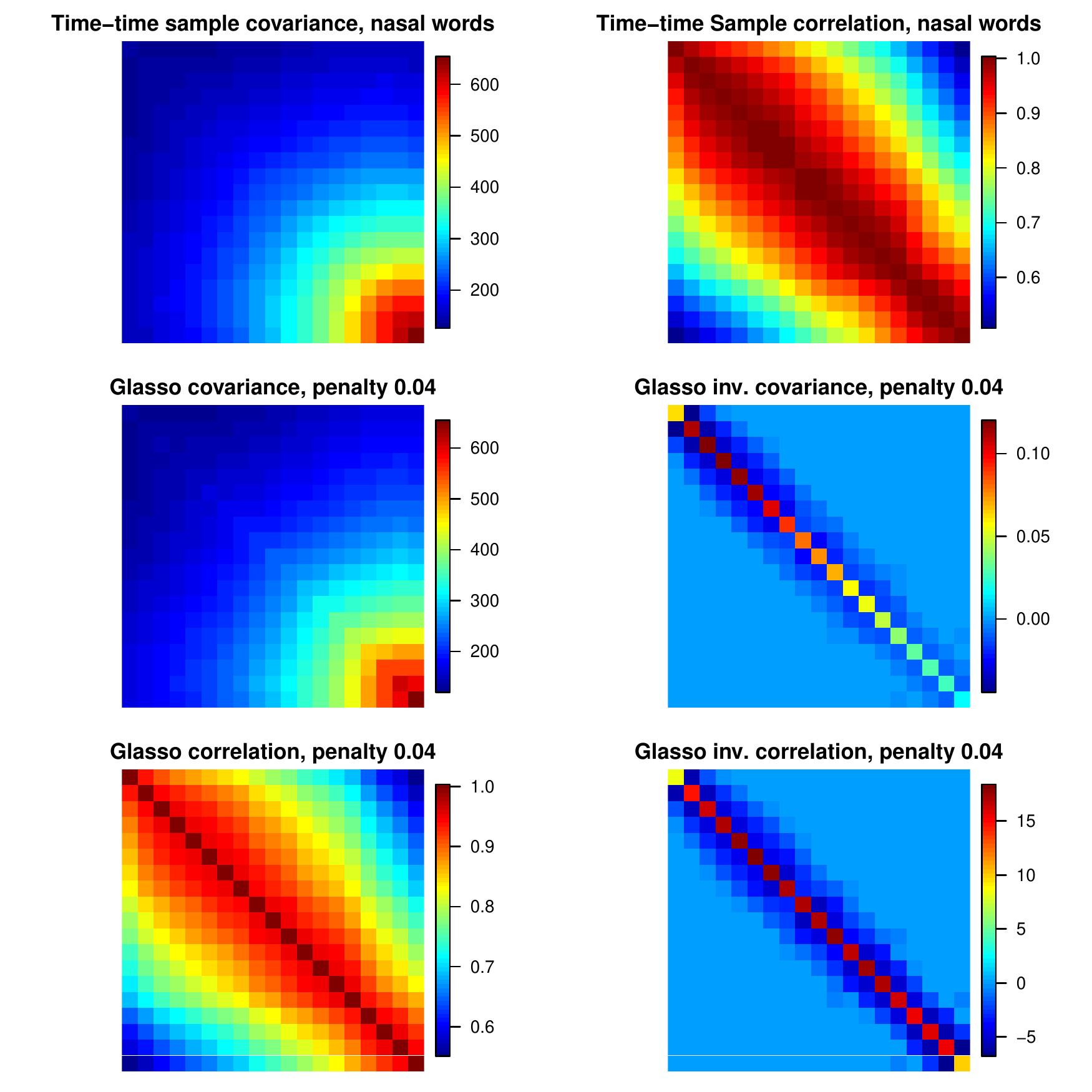} \caption{Time-time sample covariance (top left), sample correlation (top right), Glasso covariance (middle left), Glasso inverse covariance (middle right), Glasso correlation (bottom left), and Glasso inverse correlation (bottom right),  for words beginning with a nasal consonant.  The sample covariance is calculated as in \eqref{sampleCovTimeTrialResid}, and the Glasso penalty parameter is chosen as in \eqref{glassoWordTimeTuningParam}.}  \label{nasal_Glasso_heatmaps_time_theoryPen}
\end{figure}
\begin{figure}[h!]
\includegraphics[width=\textwidth]{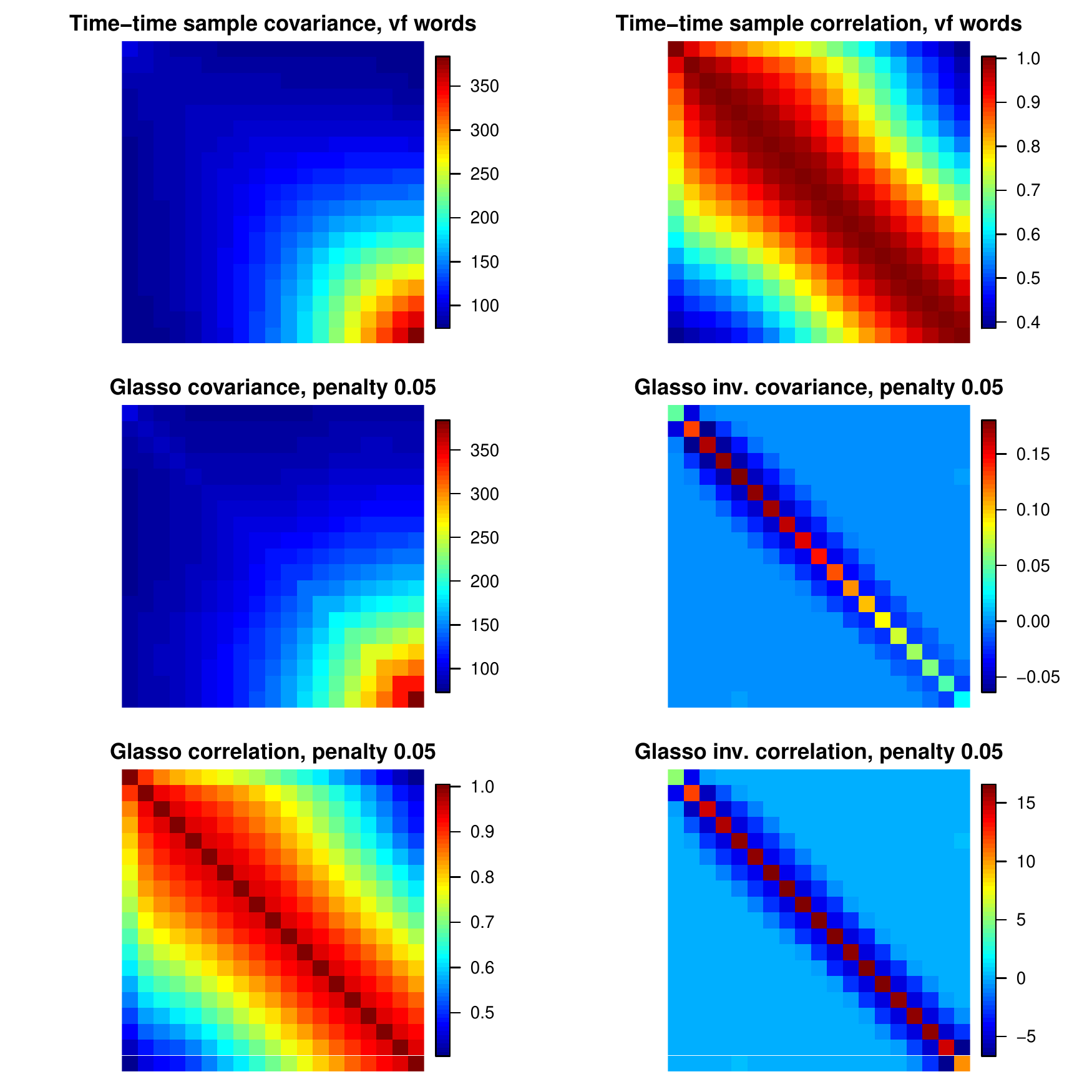} \caption{Time-time sample covariance (top left), sample correlation (top right), Glasso covariance (middle left), Glasso inverse covariance (middle right), Glasso correlation (bottom left), and Glasso inverse correlation (bottom right),  for words beginning with a vf consonant.  The sample covariance is calculated as in \eqref{sampleCovTimeTrialResid}, and the Glasso penalty parameter is chosen as in \eqref{glassoWordTimeTuningParam}.}  \label{vf_Glasso_heatmaps_time_theoryPen}
\end{figure}

\begin{figure}[h!]
\includegraphics[width=\textwidth]{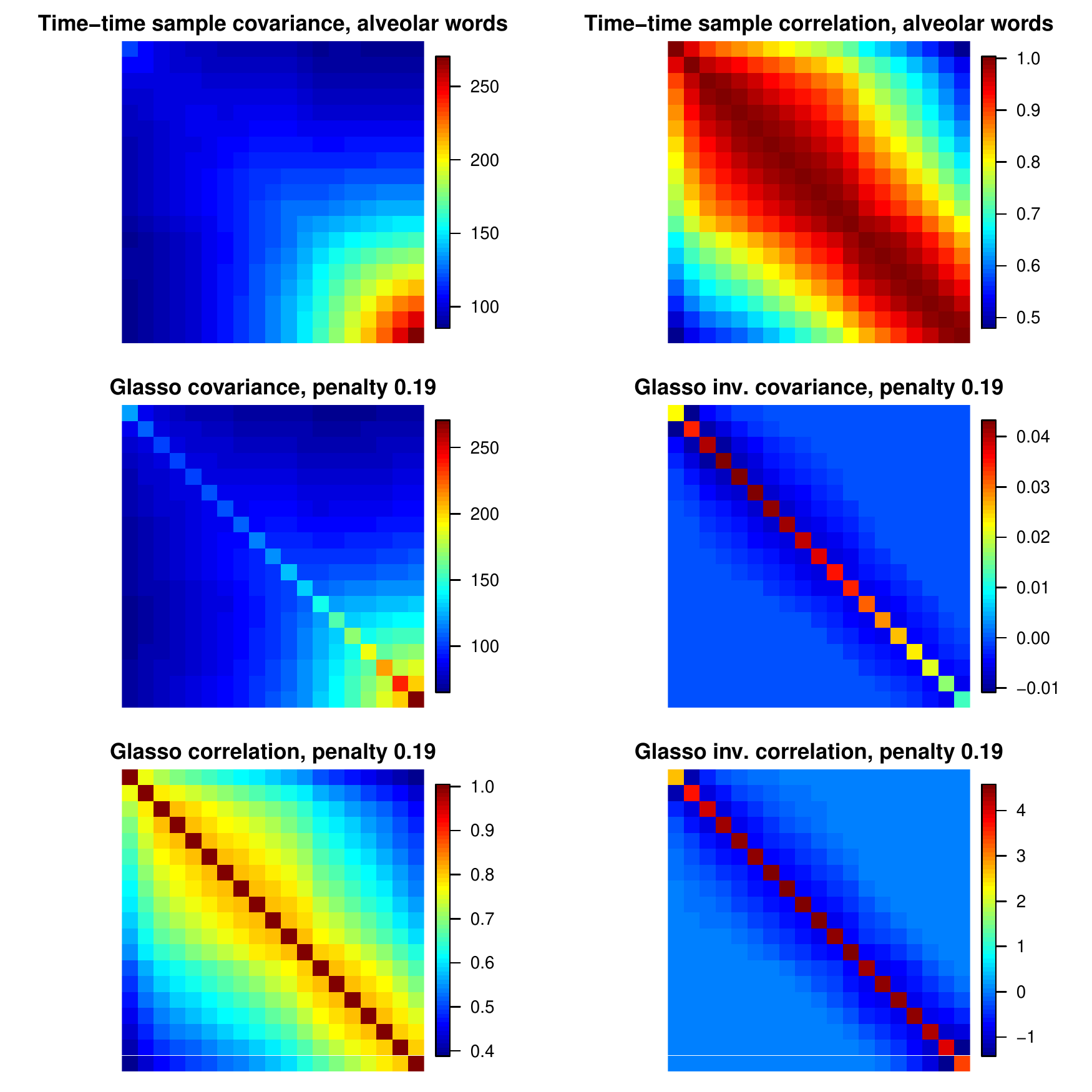} \caption{Time-time sample covariance (top left), sample correlation (top right), Glasso covariance (middle left), Glasso inverse covariance (middle right), Glasso correlation (bottom left), and Glasso inverse correlation (bottom right),  for words beginning with an alveolar consonant.  The sample covariance is calculated as in \eqref{sampleCovTimeTrialResid}, and the Glasso penalty parameter is chosen as five times the value of  \eqref{glassoWordTimeTuningParam}.}  \label{alveolar_Glasso_heatmaps_time_theoryPenTimesFive}
\end{figure}
\begin{figure}[h!]
\includegraphics[width=\textwidth]{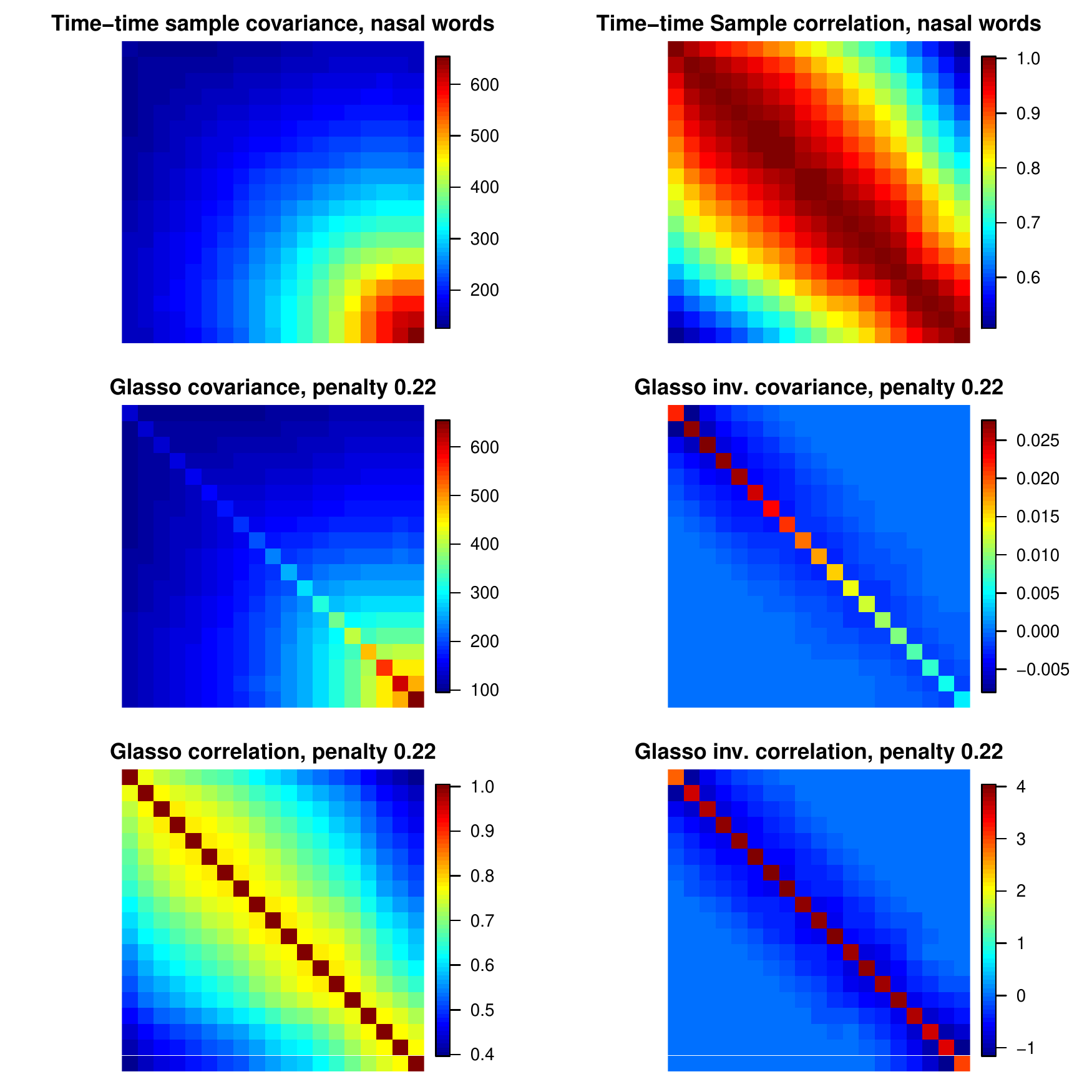} \caption{Time-time sample covariance (top left), sample correlation (top right), Glasso covariance (middle left), Glasso inverse covariance (middle right), Glasso correlation (bottom left), and Glasso inverse correlation (bottom right),  for words beginning with a nasal consonant.  The sample covariance is calculated as in \eqref{sampleCovTimeTrialResid}, and the Glasso penalty parameter is chosen as five times the value of  \eqref{glassoWordTimeTuningParam}.}  \label{nasal_Glasso_heatmaps_time_theoryPenTimesFive}
\end{figure}
\begin{figure}[h!]
\includegraphics[width=\textwidth]{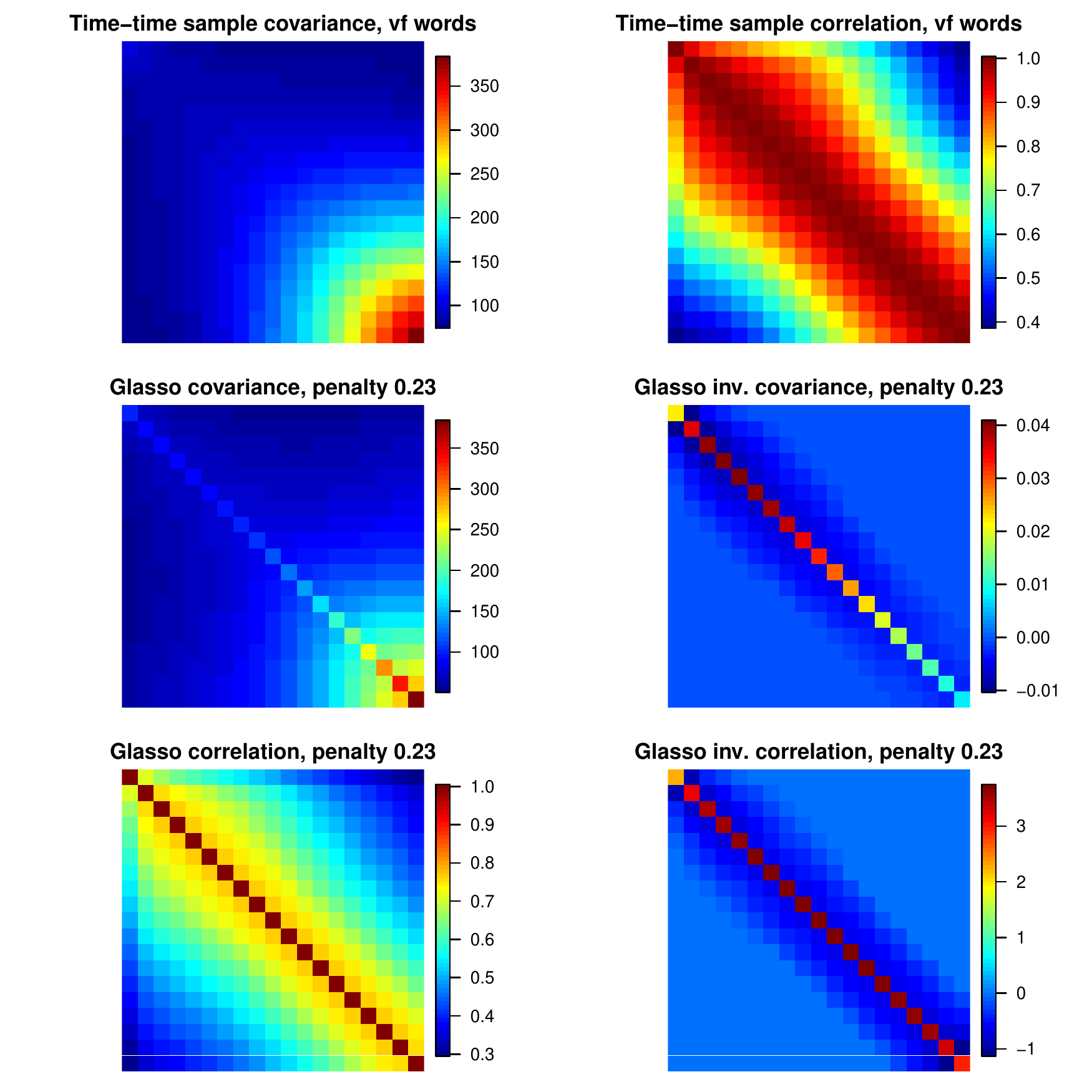} \caption{Time-time sample covariance (top left), sample correlation (top right), Glasso covariance (middle left), Glasso inverse covariance (middle right), Glasso correlation (bottom left), and Glasso inverse correlation (bottom right),  for words beginning with a vf consonant.  The sample covariance is calculated as in \eqref{sampleCovTimeTrialResid}, and the Glasso penalty parameter is chosen as five times the value of  \eqref{glassoWordTimeTuningParam}.}  \label{vf_Glasso_heatmaps_time_theoryPenTimesFive}
\end{figure}



\end{document}